\documentclass[floats,floatfix,amssymb,prd,twocolumn,superscriptaddress,
nofootinbib]{revtex4-1}

\usepackage{subcaption}
\usepackage{ragged2e}
\DeclareCaptionJustification{justified}{\justifying}
\makeatletter
\newcommand{\subsetsim}{\mathrel{\mathpalette\subset@sim\relax}}
\newcommand{\subset@sim}[2]{%
  \vtop{\offinterlineskip\m@th
    \ialign{\hfil##\cr
      $#1\subset$\cr\noalign{\kern0.5pt}\scalebox{0.9}{$#1\sim$}\cr
    }%
  }%
}
\makeatother

\usepackage{amssymb,amsmath,verbatim,mathtools,needspace,enumitem,etoolbox,graphicx,physics,microtype,afterpage,bm}
\usepackage[dvipsnames, usenames]{xcolor}
\definecolor{linkcolor}{rgb}{0.0,0.3,0.5}
\usepackage{booktabs}

\definecolor{oucrimsonred}{rgb}{0.6, 0.0, 0.0}
\definecolor{persianblue}{rgb}{0.11, 0.22, 0.73}
\definecolor{forestgreen}{rgb}{0.13,0.35,0.13}

\usepackage[unicode, 
colorlinks=true, 
linkcolor=persianblue, 
citecolor=forestgreen, 
filecolor=persianblue,
urlcolor=oucrimsonred, 
pdfusetitle]{hyperref}

\usepackage[all]{hypcap}
\usepackage[T1]{fontenc}
\usepackage[utf8]{inputenc}
\usepackage{tabularx}
\usepackage{adjustbox}
\usepackage{float}
\usepackage{ulem}
\usepackage{amsmath}
\usepackage{xfrac}
\usepackage{orcidlink}

\definecolor{deepfuchsia}{rgb}{0.76, 0.33, 0.76}
\definecolor{VioletRed4}{rgb}{0.55, 0.13, .32}
 
\definecolor{harvardcrimson}{rgb}{0.79, 0.0, 0.09}
\definecolor{oceanboatblue}{rgb}{0.0, 0.47, 0.75}
\definecolor{persianblue}{rgb}{0.11, 0.22, 0.73}
\definecolor{egyptianblue}{rgb}{0.06, 0.2, 0.65}
\definecolor{navyblue}{rgb}{0.0, 0.0, 0.5}

\usepackage{multirow}
\usepackage{pifont}
\usepackage{fontawesome}
\usepackage{lmodern}

\usepackage{multirow}

\allowdisplaybreaks
\usepackage{tikz}
\usepackage{color}
\usepackage{framed}

\definecolor{rossos}{cmyk}{0,1,1,0.55}
\definecolor{bluscuro}{rgb}{0.15, 0.2, .85}
\definecolor{bluchiaro}{cmyk}{1,.3,0.,0.1}
\definecolor{ForestGreen}{rgb}{0.13, 0.55, 0.13}

\definecolor{azure}{rgb}{0.0, 0.5, 1.0}

\newcommand{\MPl}{\bar{M}_{\textrm{\tiny{Pl}}}}

\def\nn{\nonumber}

\newcommand{\MT}{M_{\textrm{\tiny{T}}}}

\def\bea{\begin{eqnarray}}
\def\eea{\end{eqnarray}}

\newcommand{\bs}{\begin{subequations}}
\newcommand{\es}{\end{subequations}}

\newcommand{\be}{\begin{equation}}
\newcommand{\ee}{\end{equation}}

\def\lsim{\mathrel{\rlap{\lower4pt\hbox{\hskip0.5pt$\sim$}}
    \raise1pt\hbox{$<$}}}         
\def\gsim{\mathrel{\rlap{\lower4pt\hbox{\hskip0.5pt$\sim$}}
    \raise1pt\hbox{$>$}}}         

\makeatletter
\def\l@subsubsection#1#2{}
\makeatother

\begin{document}
\title{The tidal gap: causality bound on exotic compact objects \\ with applications in the solar and sub-solar mass range}

\author{Benedetta Russo}
\thanks{{\scriptsize Email}: \href{mailto:russo.1914359@studenti.uniroma1.it}{russo.1914359@studenti.uniroma1.it}.}
\affiliation{Dipartimento di Fisica, ``Sapienza'' Universit\`a di Roma, Piazzale Aldo Moro 5, 00185, Roma, Italy}

\author{Alfredo Urbano\orcidlink{0000-0002-0488-3256}}
\thanks{{\scriptsize Email}: \href{mailto:alfredo.urbano@uniroma1.it}{alfredo.urbano@uniroma1.it}.}
\affiliation{Dipartimento di Fisica, ``Sapienza'' Universit\`a di Roma, Piazzale Aldo Moro 5, 00185, Roma, Italy}
\affiliation{INFN sezione di Roma, Piazzale Aldo Moro 5, 00185, Roma, Italy}

\date{\today}

\begin{abstract}\noindent 
In this work, we highlight the existence of a lower limit on the tidal deformability parameter $\Lambda$, determined by the requirement of relativistic causality. Additionally, by considering the upper bound set on compactness, we identify the region within the parameter space of compactness versus tidal deformability, where physically motivated exotic compact objects (ECOs) could potentially reside.
Our analysis reveals the presence of a {\it tidal gap} between black holes, characterized by vanishing tidal deformability, and physically motivated ECOs. Prompted by this finding, we investigate the possibility that a population of maximally compact exotic objects, described by a linear equation of state (EoS), may simultaneously inhabit the lower mass gap and the sub-solar region, thus qualifying as (primordial) black hole mimickers while distinguishing themselves from the latter by their non-zero tidal deformability.  
Finally, considering the case of solitonic boson stars as proxies for ECOs described by a linear EoS, 
 we discuss how it is possible to further reduce the lower limit on $\Lambda$, provided that the strong energy condition is violated (but not the dominant energy condition, and therefore causality).
\end{abstract}

\maketitle


\noindent
\section{Introduction and motivations}

Exotic compact objects (ECOs) in astrophysics are hypothetical objects that deviate from the typical characteristics of known astrophysical objects like stars, neutron stars, and black holes. ECOs represent intriguing possibilities that could challenge our understanding of astrophysics, gravity, and the fundamental nature of the universe. 
Although speculative and not directly observed, ECOs play a significant role in gravitational wave astronomy because they have the potential to produce unique gravitational wave signatures distinct from those generated by more conventional astrophysical objects, such as binary neutron stars or binary black holes.

As suggested by its name, the characteristic property of an ECO is
the compactness $\mathcal{C}$, defined---in terms of  
the gravitational constant $G_N$, the speed of light $c$, the object's mass $M$, and its radius $R$---by the ratio
\begin{align}
\mathcal{C} = \frac{G_N M}{c^2 R}\,.\label{eq:Compa}
\end{align}
This dimensionless quantity encapsulates the degree to which mass is concentrated within a given volume, with higher compactness typically correlating with stronger gravitational fields near the object's surface.

The Buchdahl bound on compactness is a theoretical limit that imposes constraints on the maximum possible compactness of a spherical object without it collapsing into a black hole\,\cite{Buchdahl:1959zz}. 
Specifically, self-gravitating, isotropic (or mildly anisotropic), spherically symmetric, perfect fluid solutions of General Relativity (GR) adhere to the condition  $\mathcal{C} \leqslant 4/9$. 
This bound is attained by stars with isotropic pressure and constant density. However, such objects violate relativistic causality. Taking causality into account further restricts the maximal compactness of fluid stars to
 $\mathcal{C} \leqslant 0.354$\,\cite{1984ApJL,Glendenning:1992dr,Lattimer:2006xb,Urbano:2018nrs}.

The tidal deformability parameter 
is a measure of how easily an object can be deformed by tidal forces.
When a compact object is subjected to tidal forces, such as those arising from a nearby companion star or a gravitational wave passing through, it can undergo deformation. This deformation depends on the object's internal structure and its response to the external gravitational field. 
More formally, the tidal deformability $\lambda_{\textrm{Tidal}}$ establishes the relationship between the star's induced quadrupole moment $Q_{ij}$ in response to a time-independent external quadrupolar tidal field $\mathcal{E}_{ij}$, $Q_{ij} = -\lambda_{\textrm{Tidal}}\mathcal{E}_{ij}$. 

In Newtonian gravity, 
the quadrupole moment is given by 
$Q_{ij} = \int d^3\vec{x}
\rho(\vec{x})(x_ix_j - \frac{1}{3}r^2\delta_{ij})$
where $\rho$ is the mass density, $x_i$ is the $i$-th coordinate of a point in  space with
respect to the reference frame sets in the center of mass of the body, and $r$ is defined
through $r^2= {x_ix_j\delta_{ij}}$. 
The quadrupole moment has dimension $[Q_{ij}] = [M L^2]$. 
On the other hand, 
if we consider a quadrupolar tidal
field, its effect is characterized by the tidal momentum $\mathcal{E}_{ij} =  
\partial_{ij}\Phi_{\textrm{ext}}$, where
$\Phi_{\textrm{ext}}$ is the newtonian external potential sourced by the companion body and evaluated (after differentiation with respect to the spatial coordinates) at the center of mass of the body subject to the tidal field. The tidal momentum has dimension
$[\mathcal{E}_{ij}] = [G_N M/L^3]$. 
Consequently, 
the tidal deformability has dimensions 
$[\lambda_{\textrm{Tidal}}] = [G_N^{-1} L^5]$.

In GR, the above definitions of $Q_{ij}$ and $\mathcal{E}_{ij}$ are no longer valid. The quadrupole moment and the tidal momentum are implicitly defined through the asymptotic expansion of the $tt$-component of the metric. In
asymptotically Cartesian and mass-centred coordinates, one can write\,\cite{Thorne:1980ru}
\begin{align}
-\frac{(1+g_{tt})}{2} = 
&-\frac{G_N M}{r} \nn\\
&
- \frac{3G_N Q_{ij}}{
2r^3
}\left(
n_i n_j - \frac{1}{3}\delta_{ij}
\right) + O\left(\frac{1}{r^4}\right)\nn\\
&+\frac{1}{2}
\mathcal{E}_{ij}r^2 n_in_j + O(r^3)\,,
\label{eq:AsyMe}
\end{align}
where $n_i\equiv x_i/r$. 
The definition of $\lambda_{\textrm{Tidal}}$ is always given by the proportionality $Q_{ij} = -\lambda_{\textrm{Tidal}}\mathcal{E}_{ij}$, and it can be seen that the dimensional analysis carried out previously remains valid in the GR framework.
It is, therefore, convenient to rewrite the tidal deformability 
according to 
\begin{align}
\lambda_{\textrm{Tidal}} 
= \frac{2}{3}G_N^{-1}R^5k_2\,\label{eq:LoveNumber}
\end{align}
where the factor $2/3$ is a convention 
while the dimension $[L^5]$ is entirely expressed in terms of the radius of the object subjected to the tidal deformation;
the dimensionless coefficient 
$k_2$ 
is the so-called quadrupolar tidal Love number.  
The latter, in turn,  
is also commonly expressed
in dimensionless form in terms of the  tidal deformability parameter $\Lambda$
\begin{align}
\Lambda \equiv 
\frac{2}{3}k_2\left(
\frac{G_N M}{c^2 R}
\right)^{-5}
= \frac{2k_2}{3\mathcal{C}^5}
\,.\label{eq:Tidal}
\end{align}

In the plane $(\mathcal{C},\Lambda)$, 
black holes occupy a privileged position that renders them unique: they have compactness $\mathcal{C} = 1/2$ (since the Schwarzschild radius is given by $R = 2G_N M/c^2$) and vanishing tidal deformability. 
This second property ultimately arises as a consequence of a hidden symmetry which governs the dynamics of black hole perturbations in the near-zone approximation, cf. ref.\,\cite{Charalambous:2022rre,Combaluzier-Szteinsznaider:2024sgb}.

In the case of an ECO, both compactness and tidal deformability crucially depend on its internal structure. 
This is not very different from what happens in the case of neutron stars, where the behavior of nuclear matter inside the star under extreme gravitational pressures is described by means of an Equation of State (EoS). 
Understanding the EoS of neutron star matter is crucial for predicting their properties, such as their maximum mass, radius, and internal structure.

In the case of ECOs, it is possible to formulate the problem similarly by following two different approaches. {\it i)} On one hand, starting from a specific model of microphysics, it is feasible to compute the corresponding EoS from first principles and then investigate its macroscopic predictions.
{\it ii)} On the other hand, it is also possible to start from a specific ansatz for the EoS---possibly motivated by fundamental principles such as, for example, positivity, stability and causality---in order to obtain model-independent properties that can be valid for a broad class of models (and to understand which assumptions can or cannot be relaxed in order to modify them).
This is, for example, the case of the compactness limits mentioned below eq.\,(\ref{eq:Compa}).
In this work, we will tackle both these perspectives.

In particular, concerning approach {\it ii)}, we investigate whether there exists the possibility of deriving causality constraints on the tidal Love number $k_2$ and the tidal deformability $\Lambda$.
If this were true, it would be possible to constrain both directions of the planes $(\mathcal{C},\Lambda)$ and $(\mathcal{C},k_2)$, 
delineating the region where solutions describing ECOs could potentially reside.
Regarding approach {\it i)},  if such causality limits were to exist, it would be natural to ask  which realistic ECOs could potentially saturate them and what their phenomenological significance would be.

At the conceptual level, these are the main motivations of this work.
In the remainder of this paper, we proceed as follows. In section\,\ref{section:LinearEoS}, we introduce the theoretical framework underlying our analysis, reviewing the relevant energy conditions, discussing the implications of causality for barotropic equations of state, and examining in detail the properties of the linear EoS. In section\,\ref{sec:Pheno}, we explore the compactness–tidal deformability plane, deriving causality-driven lower limits on the tidal deformability and identifying the region in parameter space accessible to physically motivated ECO models. In section\,\ref{sec:BS}, we consider solitonic boson stars as proxies for ECOs described by a linear EoS, assessing the extent to which violations of the strong energy condition affect the minimal allowed tidal deformability. Finally, section\,\ref{sec:Conclusion} summarizes our conclusions and outlines potential directions for future work.

Throughout this paper, we will work in natural units where Planck’s constant $\hbar$, the speed of light $c$, and Newton’s constant $G_N$ are set to one.  On occasion, we will reintroduce $G_N$ to enhance the transparency of certain equations, particularly from the perspective of particle physics. Finally, our convention for the metric signature is mostly positive, denoted as $(-,+,+,+)$.

\noindent
\section{Causality and the linear EoS}\label{section:LinearEoS}

For completeness, we begin with a brief discussion introducing the definitions and notations used in this work.

The general form of the static spherically symmetric line element $ds^2 = g_{\mu\nu}dx^{\mu}dx^{\nu}$ in Boyer-Lindquist type coordinates
$(t, r,\theta,\varphi)$ reads
\begin{align}
ds^2 = -e^{\nu(r)}dt^2 + 
e^{\lambda(r)}dr^2 + r^2 (
d\theta^2 + \sin^2\theta d\varphi^2)\,.\label{eq:SphMetric}
\end{align}
The stress-energy tensor $T_{\mu\nu}$ for an anisotropic fluid is given by
\begin{align}
T_{\mu\nu} = 
(\epsilon + P_t)u_{\mu}u_{\nu} + 
P_t g_{\mu\nu} + 
(P_r - P_t)\chi_{\mu}\chi_{\nu}\,,\label{eq:AnisoTMUNU}
\end{align}
where $\epsilon = \epsilon(r)$ is the energy density, $u_{\mu}$ is the four-velocity of the fluid, 
$\chi_{\mu}$ is a unit spacelike vector in the radial direction, 
$P_r=P_r(r)$ and $P_t=P_t(r)$ are the radial and tangential pressures, respectively. 
For a static, spherically symmetric spacetime, the fluid is not moving in the spatial directions, so its four-velocity points only in the time direction. 
We write $u^{\mu} = (e^{-\nu(r)/2},0,0,0)$
with $u^{\mu}u_{\mu} = -1$. 
The unit spacelike vector $\chi^{\mu}$ 
defines the radial direction and is orthogonal to the fluid's four-velocity $u^{\mu}$. 
We write 
$\chi^{\mu} = (0,e^{-\lambda(r)/2},0,0)$ with 
$\chi^{\mu}\chi_{\mu} = 1$. 
The Einstein field equations are given by
$G_{\mu\nu} = 8\pi T_{\mu\nu}$, where 
$G_{\mu\nu} = R_{\mu\nu}-\frac{1}{2}g_{\mu\nu}R$ is the Einstein tensor, with 
$R_{\mu\nu}$ being the Ricci curvature tensor and 
$R=g^{\mu\nu}R_{\mu\nu}$ the Ricci scalar.
The continuity equation for the stress-energy tensor $T_{\mu\nu}$ is given by 
$\nabla^{\mu}T_{\mu\nu} = 0$. 
Here, $\nabla^{\mu}$ denotes the covariant derivative. The continuity equation is a direct consequence of the Bianchi identity in Riemaniann geometry applied to the Einstein field equations.
Following the redefinition 
$e^{\lambda(r)} = [1-2m(r)/r]^{-1}$, the system composed of Einstein's field equations and the continuity equation is equivalent to the two equations
\begin{align}
\frac{dm(r)}{dr} & = 4\pi r^2 \epsilon(r)\,,\label{eq:Anisotro1}\\
\frac{dP_r(r)}{dr} & = 
-\frac{
[\epsilon(r) + P_r(r)][4\pi r^3 P_r(r) + m(r)]
}{
r^2[1-2m(r)/r]
} - \frac{2\Delta(r)}{r}\,,\label{eq:Anisotro2}
\end{align}
where we defined the quantity 
$\Delta(r) \equiv P_r(r) - P_t(r)$ 
that represents the anisotropy parameter at a given radial coordinate $r$.
The isotropic limit is defined by the condition 
$P_r(r) = P_t(r) \equiv P(r)$, such that $\Delta(r)= 0$. 
We remain general in this introductory discussion because, throughout this work, we will be interested in both the isotropic limit and the anisotropic case (albeit limited to a specific model). 
Eq.\,(\ref{eq:Anisotro2}) describes the balance between gravitational forces and pressure gradients in a spherically symmetric, static configuration of a relativistic object.
Note the role of $\Delta$. 
If $\Delta < 0$ ($P_r < P_t$), the tangential pressure exceeds the radial pressure, reducing the effective gravitational pull and allowing the star to support more mass and achieve greater compactness.

It is possible to impose that the stress-energy tensor satisfies the so-called energy conditions\,\cite{Kontou:2020bta,Poisson:2009pwt}.
Energy conditions represent various generalizations of the principle that the energy density must not be negative, extended to the entire stress-energy tensor. 
Specifically, when applied to the anisotropic case, we have
\begin{itemize}
    \item[$\circ$] Null Energy Condition (NEC). 
    
    The NEC asserts that $T_{\mu\nu}k^{\mu}k^{\nu} \geqslant 0$ for all null vectors $k^{\mu}$ (i.e., $k_{\mu}k^{\mu} = 0$).\footnote{
Since the truth value of $T_{\mu\nu}k^{\mu}k^{\nu} \geqslant 0$ remains unchanged when the nonzero vector $k^{\mu}$ is replaced by any scalar multiple of $k^{\mu}$, it follows that an equivalent formulation of the NEC is $T_{\mu\nu}k^{\mu}k^{\nu} \geqslant 0$ for all future-directed null vectors $k^{\mu}$ (i.e., $k_{\mu}k^{\mu} = 0$ with $k^0 > 0$).
    }
    For a stress-energy tensor of the form given in eq.\,(\ref{eq:AnisoTMUNU}), the NEC is satisfied if and only if
  \begin{align}
  \epsilon + P_r \geqslant 0 \,,~~
\epsilon + P_t \geqslant 0\,.   \label{eq:NEC}
  \end{align}  
\item[$\circ$] Weak Energy Condition (WEC).

The WEC asserts that $T_{\mu\nu}v^{\mu}v^{\nu} \geqslant 0$ for all timelike vectors $v^{\mu}$ (i.e., $v^{\mu}v_{\mu} < 0$).\footnote{
Since the truth value of the inequality $T_{\mu\nu}v^{\mu}v^{\nu} \geqslant 0$ is unaffected if the nonzero vector $v^{\mu}$ is replaced by any nonzero scalar multiple of $v^{\mu}$, it follows that the WEC is equivalent to the statement that $T_{\mu\nu}v^{\mu}v^{\nu} \geqslant 0$ for all normalized future-directed timelike vectors $v^{\mu}$ (i.e., vectors that satisfy $v^{\mu}v_{\mu} = -1$ and point in the direction of increasing proper time, $v^0>0$).
}
For a stress-energy tensor of the form given in eq.\,(\ref{eq:AnisoTMUNU}), the WEC is satisfied if and only if
  \begin{align}
  \epsilon \geqslant 0 \,,~~
  \epsilon + P_r \geqslant 0 \,,~~
\epsilon + P_t \geqslant 0\,.   \label{eq:WEC}
  \end{align}  
\item[$\circ$] Dominant Energy Condition (DEC).

For any future-directed timelike
vector $v^{\mu}$, 
the DEC requires that $T_{\mu\nu}v^{\mu}$ is neither past-directed nor spacelike.\footnote{
An equivalent formulation of the DEC is that $T_{\mu\nu}v^{\mu}$ is neither past-directed
nor spacelike for all normalized future-directed timelike vectors $v^{\mu}$.
} 
For a stress-energy tensor of the form given in eq.\,(\ref{eq:AnisoTMUNU}), the DEC is satisfied if and only if
 \begin{align}
\epsilon \geqslant |P_r|\,,~~
\epsilon \geqslant |P_t|\,.\label{eq:DEC}
 \end{align}
 \item[$\circ$] Strong Energy Condition (SEC).

The SEC asserts that
 $(T_{\mu\nu} - \frac{1}{2}Tg_{\mu\nu})v^{\mu}v^{\nu}\geqslant 0$ for any timelike
vector $v^{\mu}$.\footnote{
An equivalent formulation of the SEC is that $(T_{\mu\nu} - \frac{1}{2}Tg_{\mu\nu})v^{\mu}v^{\nu}\geqslant 0$ for all
normalized future-directed timelike vectors $v^{\mu}$.
}
For a stress-energy tensor of the form given in eq.\,(\ref{eq:AnisoTMUNU}), the SEC is satisfied if and only if
  \begin{align}
  \epsilon + P_r \geqslant 0 \,,~~
\epsilon + P_t \geqslant 0\,,~~
 \epsilon + P_r + 2P_t \geqslant 0\,,\label{eq:SEC}
 \end{align}
\end{itemize}
with the following chains of implications
\begin{align}
\textrm{DEC} \implies \textrm{WEC}  \implies
\textrm{NEC} \impliedby \textrm{SEC}\,.
\end{align}
It is important to emphasize that the energy conditions are not direct consequences of Einstein's field equations; rather, they represent physical assumptions based on reasonable expectations regarding the behavior of matter. These conditions are generally assumed to be valid for ordinary matter, such as radiation and standard forms of matter, although they may be violated in the presence of exotic forms of matter.

In the isotropic limit, the system formed by 
eqs.\,(\ref{eq:Anisotro1},\,\ref{eq:Anisotro2}) can be closed and solved once an EoS relating the energy density and the pressure is specified. In this work, we consider barotropic EoS, in which the pressure is a function of the energy density alone, i.e., $P = P(\epsilon)$, without any dependence on the specific volume or other thermodynamic variables. 
We can now impose an additional condition on the pressure and energy density.
The causality bound on a barotropic EoS is related to the requirement that the speed of sound in the fluid cannot exceed the speed of light. 
In relativistic hydrodynamics, the speed of sound is defined as the propagation speed of small adiabatic perturbations in a fluid, and is given by
\begin{align}
c_s^2 \equiv \frac{dP}{d\epsilon}\,.   
\end{align}
Intuitively, $c_s^2$ quantifies how efficiently pressure reacts to local changes in energy density, and therefore how fast small perturbations can propagate through the fluid.
To ensure causality, the speed of sound must satisfy the condition $c_s^2 \leqslant 1$. This condition ensures that disturbances (e.g., sound waves) propagate at speeds less than or equal to the speed of light, in accordance with the principles of relativity. 
We consider the requirement $c_s^2 \leqslant 1$ as an additional condition to be imposed alongside the energy conditions. However, we expect that the condition $c_s^2 \leqslant 1$ is not entirely independent of the latter. In fact, the DEC requires that the flux of energy-momentum measured by an observer is causal and directed along the observer's proper time. This is often interpreted as prohibiting superluminal 
propagation of energy.\footnote{
This statement is not entirely accurate. It is interesting to note that if energy-momentum could flow along spacelike vectors, it would not necessarily result in violations of causality\,\cite{Liberati:2001sd}.
}
Intuitively, it may therefore be plausible to expect that a violation of the causality condition could also manifest as a violation of the DEC.

Finally, in view of later applications, let us comment on the SEC. The physical content of the SEC can be understood in terms of the so-called timelike convergence condition 
(one can prove the complete equivalence between the two conditions in the case of a zero cosmological constant)\,\cite{Hawking:1973uf}. The timelike convergence condition asserts that the expansion of a congruence of timelike geodesics (with zero vorticity) monotonically decreases along a timelike geodesic, meaning the geodesics tend to converge or get closer together. 
In other words, the SEC essentially asserts that gravity is always attractive. 
Many classical matter configurations, at least from a mathematical standpoint, are known to violate the SEC. For example, a scalar field with a positive potential can lead to a violation of this condition. We will return to this issue in section\,\ref{sec:BS}.

\subsection{Equation of state: energy conditions and implications}

We consider the linear EoS with the relativistic energy density $\epsilon$ given, as function of the pressure $P$, by 
\begin{align}
\epsilon(P) = \epsilon_0 + P/\omega\,,\label{eq:LinEOS}
\end{align}
with 
\begin{align}
\omega = \frac{dP}{d\epsilon} \equiv c_s^2\,,\label{eq:Speed}
\end{align}
the square of the speed of sound
 $c_s$.  
 This EoS is particularly significant as it saturates the causality bound.
In fact, assuming constant speed of sound, causality is saturated by the value $\omega = 1$, and this is the value upon which we will focus for the remainder of the work. 
Values of $\omega > 1$ violate causality, while $\omega < 1$ satisfy the constraint.
 \begin{figure}[!t]
	\centering
\includegraphics[width=0.495\textwidth]{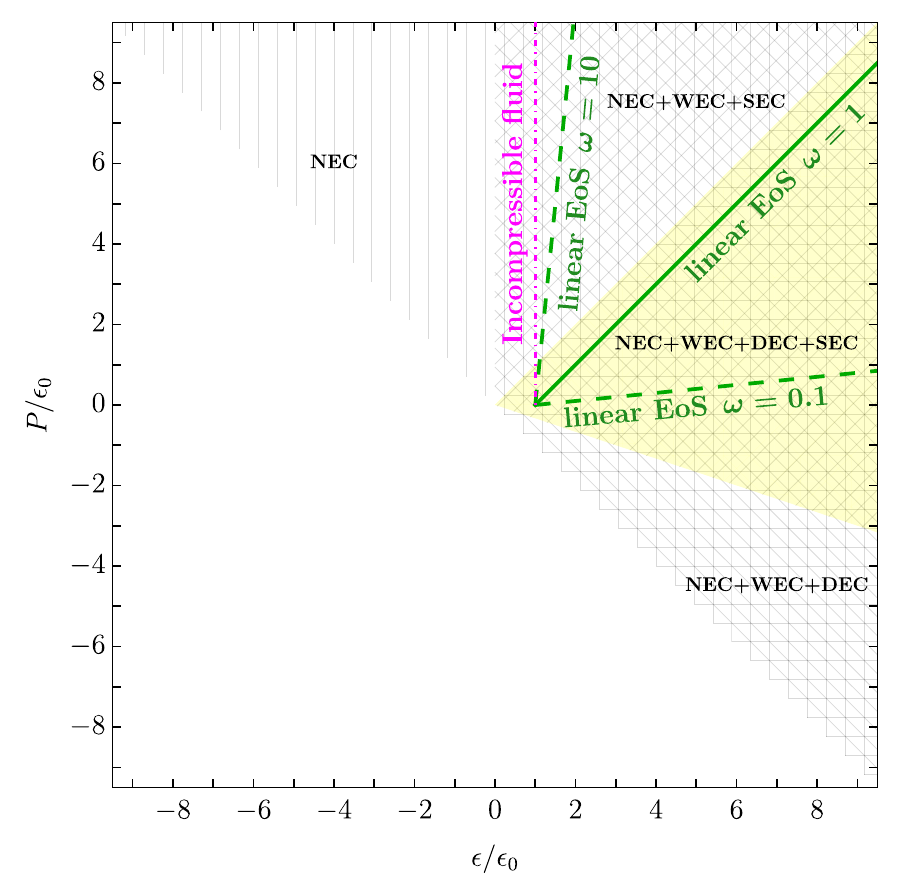}
	\caption{\it  
Relationship between the linear EoS in eq.\,(\ref{eq:LinEOS}) and the energy conditions discussed in eqs.\,(\ref{eq:NEC}-\ref{eq:SEC}) in the energy density/pressure plane.
We rewrite eq.\,(\ref{eq:LinEOS}) in the form 
$P/\epsilon_0 = \omega(\epsilon/\epsilon_0 -1)$, and consider both pressure and energy density in units of $\epsilon_0$.
In the hatched region: horizontal lines indicate where the DEC is satisfied, vertical lines indicate where the NEC is satisfied, lines rotated counterclockwise by 45$^\circ$ indicate where the SEC is satisfied, and lines rotated clockwise by 45$^\circ$ indicate where the WEC is satisfied. In the yellow region, all energy conditions are satisfied. 
The linear EoS that satisfy the causality condition occupy this region. We also show the curve corresponding to a case of causality violation with $\omega = 10$. We observe that this corresponds to a violation of the DEC, although the WEC (and thus NEC) and SEC remain satisfied. 
The case of constant-density stars (a.k.a. incompressible stars) is indicated by a vertical magenta dot-dashed line
 }
\label{fig:EnergyCondLinearEOS}
\end{figure}
In fig.\,\ref{fig:EnergyCondLinearEOS},
we attempt to interpret the linear EoS in terms of the energy conditions. 
The linear EoS with $\omega \leqslant 1$ satisfies all the energy conditions discussed in eqs.\,(\ref{eq:NEC}-\ref{eq:SEC}) (refer to the caption of fig.\,\ref{fig:EnergyCondLinearEOS} for details).
As previously noted, on the contrary, we find that the condition $\omega > 1$ leads to a violation of the DEC. 
The limit $\omega\to \infty$ reproduces the (unphysical) case of the so-called constant-density stars, in which the energy density is constant. This case is indicated by a vertical magenta dot-dashed line in fig.\,\ref{fig:EnergyCondLinearEOS}.

Parallel to the analysis based on the linear EoS, we consider a set of EoS commonly used to describe the properties of neutron star matter\,\cite{Antonopoulou:2022yot}.
 \begin{figure}[!t]
	\centering
\includegraphics[width=0.495\textwidth]{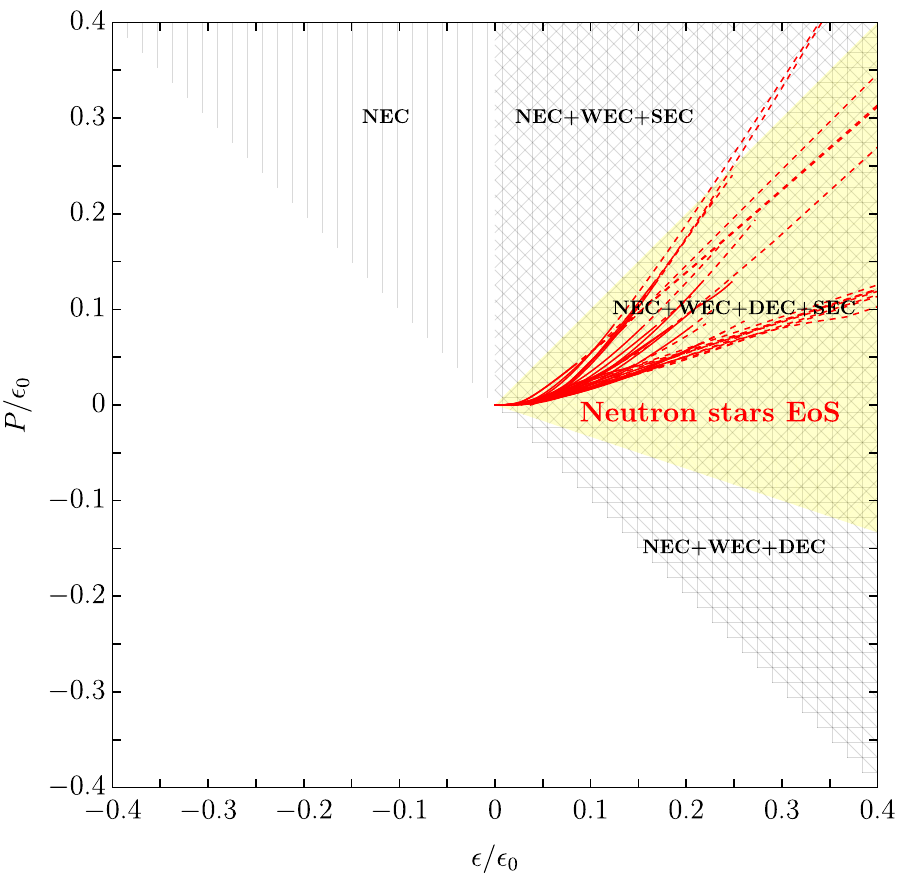}
	\caption{\it  
 Same as in fig.\,\ref{fig:EnergyCondLinearEOS} but considering realistic neutron star EoS. 
 Note that in this case pressure and energy density are normalized with respect to the value 
  $\epsilon_0 = m_n^4c^5/\pi^2\hbar^3  \simeq  
1.64\times 10^{37}$ erg/cm$^3$.
For each curve, the dashed portion indicates the values for which $P(\epsilon) > P(\epsilon_c^{\textrm{max}})$, see text for details.
}
\label{fig:EnergyCondNS}
\end{figure}
In fig.\,\ref{fig:EnergyCondNS}, we repeat the same analysis as before, comparing the EoS of neutron stars with the constraints imposed by the energy conditions. 
We find that in this case, all the EoS analyzed fall within the region where all energy conditions are satisfied, with the sole exception of some of them. If extrapolated to very large values of energy and pressure, they enter the region where the DEC---and thus the causality condition, according to our previous intuition---is violated.
We note that at this stage, we are simply plotting the equations of state without taking into account which values of pressure and energy density actually lead to stable solutions of the previously derived equilibrium equations. 
In order to obtain a more realistic physical picture, we proceed to solve eqs.\,(\ref{eq:Anisotro1},\,\ref{eq:Anisotro2}) in the isotropic limit, which correspond to the Tolman-Oppenheimer-Volkoff (TOV) equations 
\begin{align}
\frac{dm(r)}{dr} & = 4\pi r^2\epsilon(r)\,,\label{eq:TOV1}\\
\frac{dP(r)}{dr} & = 
-\frac{
[\epsilon(r) + P(r)][4\pi r^3 P(r) + m(r)]
}{
r^2[1-2m(r)/r]
}\,,\label{eq:TOV2}
\end{align}
with $m(r)$ 
the mass-energy enclosed within the radial distance $r$.
The TOV equation system can be expressed in a dimensionless form by defining rescaled quantities
\begin{align}
\tilde{\epsilon} \equiv 
\frac{\epsilon}{\epsilon_{0}}\,,~~~
\tilde{P} \equiv 
\frac{P}{\epsilon_{0}}\,,~~~
\tilde{r}\equiv r\sqrt{\epsilon_0}\,,~~~
\tilde{m}\equiv m
\sqrt{\epsilon_0}\,.\label{eq:Rescaling}
\end{align}
In units where $c=G_N=1$, $\epsilon_0$ has dimensions equivalent to those of an inverse square length (or, equivalently, an inverse square mass). 
Eqs.\,(\ref{eq:TOV1},\,\ref{eq:TOV2}) can be easely solved numerically from the center of the star outwards 
with boundary conditions $P(r=0)=P_c$ and $m(r=0)=0$. 
The radius $R$ is defined by the condition $P(R) = 0$ and the mass is given by $M=m(R)$. 
The rescaling in eq.\,(\ref{eq:Rescaling}) can be applied when solving the TOV equations 
both in the case of the linear EoS in eq.\,(\ref{eq:LinEOS}) and 
in the case of neutron stars with realistic EoS.
In the case of the linear EoS, $\epsilon_0$ is naturally identified with the parameter $\epsilon_0$ that  enters in the definition of 
the EoS; given that $P(R) = 0$, in the case of the linear EoS $\epsilon_0$ corresponds to the surface energy density of the star.
In the case of neutron stars, a convenient choice for the dimensionfull paramater $\epsilon_0$ is given by $\epsilon_0 = m_n^4c^5/\pi^2\hbar^3  \simeq  
1.64\times 10^{37}$ erg/cm$^3$. 
The mass-radius curves are found by integrating
the structure equations varying the central energy density up
to the one that corresponds to maximum mass.  
Once the maximum mass is reached, any further increase in central energy density leads to a decrease in the total mass, and the solutions become unstable due to radial perturbations.
In fig.\,\ref{fig:EnergyCondNS}, the dashed portion of the EoS gives rise to unstable solutions, and only the non-dashed part should be considered as realistic.
For each of the EoS we analyze, we calculate the maximum value of the central density beyond which stable solutions cannot be obtained, $\epsilon_c^{\textrm{max}}$. 
In fig.\,\ref{fig:EnergyCondNS}, for each EoS, we show as a dashed line the values for which $P(\epsilon) > P(\epsilon_c^{\textrm{max}})$.
We note that, if we restrict our analysis exclusively to the solid line portion of the EoS, i.e., the part for which stable solutions exist, we remain within the region where all energy conditions are satisfied.
 \begin{figure}[!t]
	\centering
\includegraphics[width=0.495\textwidth]{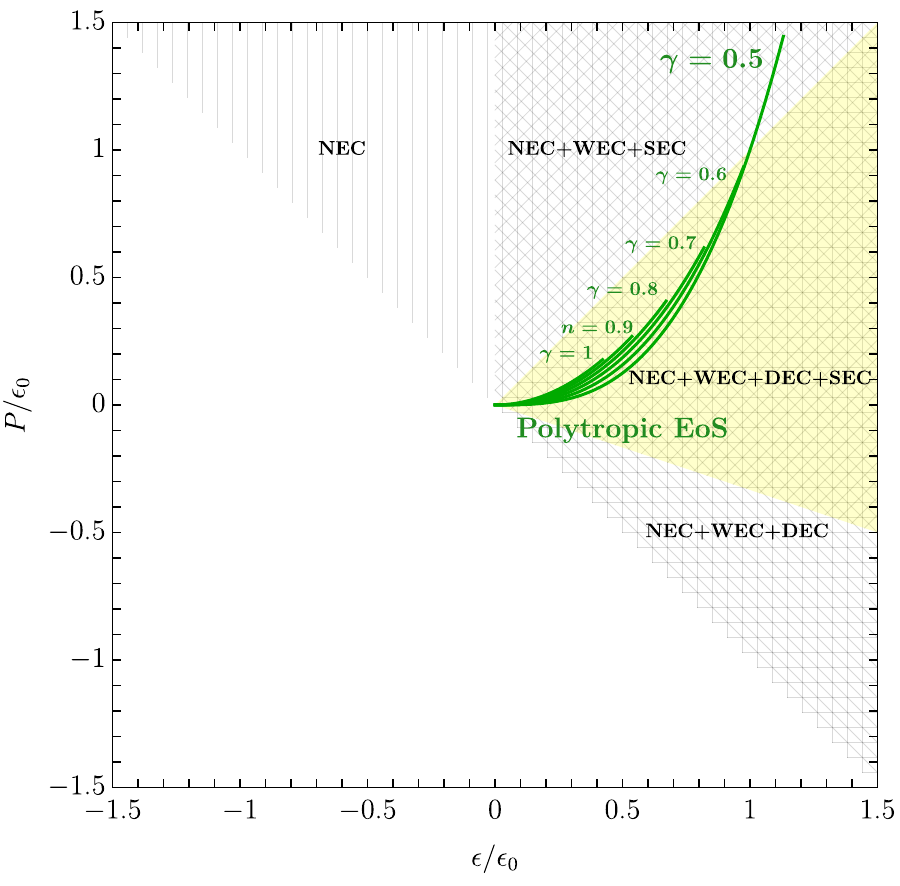}
	\caption{\it  
 Same as in fig.\,\ref{fig:EnergyCondLinearEOS} but considering polytropic EoS, cf. eq.\,(\ref{eq:PolyEoS}). 
 Pressure and energy density are normalized with respect to the quantity $\epsilon_0 \equiv 1/K^{\gamma}$.
 Each curve terminates at values of energy density and pressure beyond which no stable solutions exist.
}
\label{fig:EnergyCondPoly}
\end{figure}
As a final benchmark example, we consider stars supported by a polytropic EoS of the form
\begin{align}\label{eq:PolyEoS}
\epsilon(P) = \left(\frac{P}{K}\right)^{\frac{\gamma}{\gamma+1}}\,.
\end{align}
We focus on polytropic indices in the range $\gamma\in [0.5,1]$, where this EoS provides the closest approximation to the realistic neutron-star case. 
We note that, also in this case, the TOV equations can be written in a manifestly dimensionless form by introducing, following the notation of eq.\,(\ref{eq:Rescaling}), the quantity $\epsilon_0 \equiv 1/K^{\gamma}$. 
We present our results in fig.\,\ref{fig:EnergyCondPoly}. 
We observe that for 
$\gamma=0.5$ the EoS, at sufficiently large energy density and pressure, enters the region in which the DEC is violated. For larger values of 
$n$, by contrast, all stable solutions satisfy all the energy conditions.

\subsection{Equation of state: stiffness and implications}\label{sec:Stiffness}

An important property of an EoS is its stiffness.
 In words, stiffness quantifies the capacity of the EoS to accommodate matter within a specified volume with larger
stiffness that corresponds to more incompressible matter.

 When matter exhibits greater compressibility, it is possible for a greater quantity of matter to occupy the same volume leading to an elevated average density and thereby characterizing the EoS as {\it soft}. Conversely, EoS models featuring less compressible matter are associated with lower average densities and are termed {\it stiff}.
Examined from an alternative but complementary viewpoint, 
stiffness refers to the sensitivity of pressure to changes in energy density.
Specifically, stiffness quantifies how rapidly pressure changes with variations in energy density within the system. A stiffer EoS implies that small changes in energy density lead to larger changes in pressure, while a softer EoS indicates that pressure is less responsive to alterations in energy density.  
From this point of view, therefore, 
stiffness becomes related to the speed of sound since the latter, defined by $c_s = \sqrt{dP/d\epsilon}$, is directly sensitive to the rate of variation $dP/d\epsilon$.
Eq.\,(\ref{eq:Speed}), written for finite change of pressure and density, $\Delta P = c_s^2\Delta \epsilon$ with $c_s^2 \leqslant 1$, implies that the linear EoS provides the stiffer EoS. 

Even from these considerations alone, it is evident that stiffness plays a crucial role with respect to both compactness and tidal deformability. 
We consider first the case of compactness. 
The na\"{\i}ve expectations are the following. 
\begin{itemize}
\item[$\circ$] \textbf{Stiffer EoS}. A stiffer EoS implies that the pressure within the object increases more rapidly with density, and  that the material resists compression more strongly.  
Higher pressure within the object, as provided by a stiffer EoS, allows the object to support more mass against gravitational collapse within a given radius. This is because the pressure acts as an opposing force to gravity, preventing the object from collapsing further. Consequently, a stiffer EoS generally results in a higher compactness.
\item[$\circ$] \textbf{Softer EoS}. Conversely, a softer EoS means that the pressure increases more slowly with density, and the material is less resistant to compression,  leading to lower pressures at a given density compared to a stiffer EoS. 
Lower pressure within the object, as provided by a softer EoS, means that the object can support less mass against gravitational collapse within a given radius. This is because the weaker pressure is less effective at counteracting the gravitational force pulling the object inward.
Consequently, a softer EoS generally results in a smaller compactness.
\end{itemize}

\begin{figure}[!t]
	\centering
\includegraphics[width=0.495\textwidth]{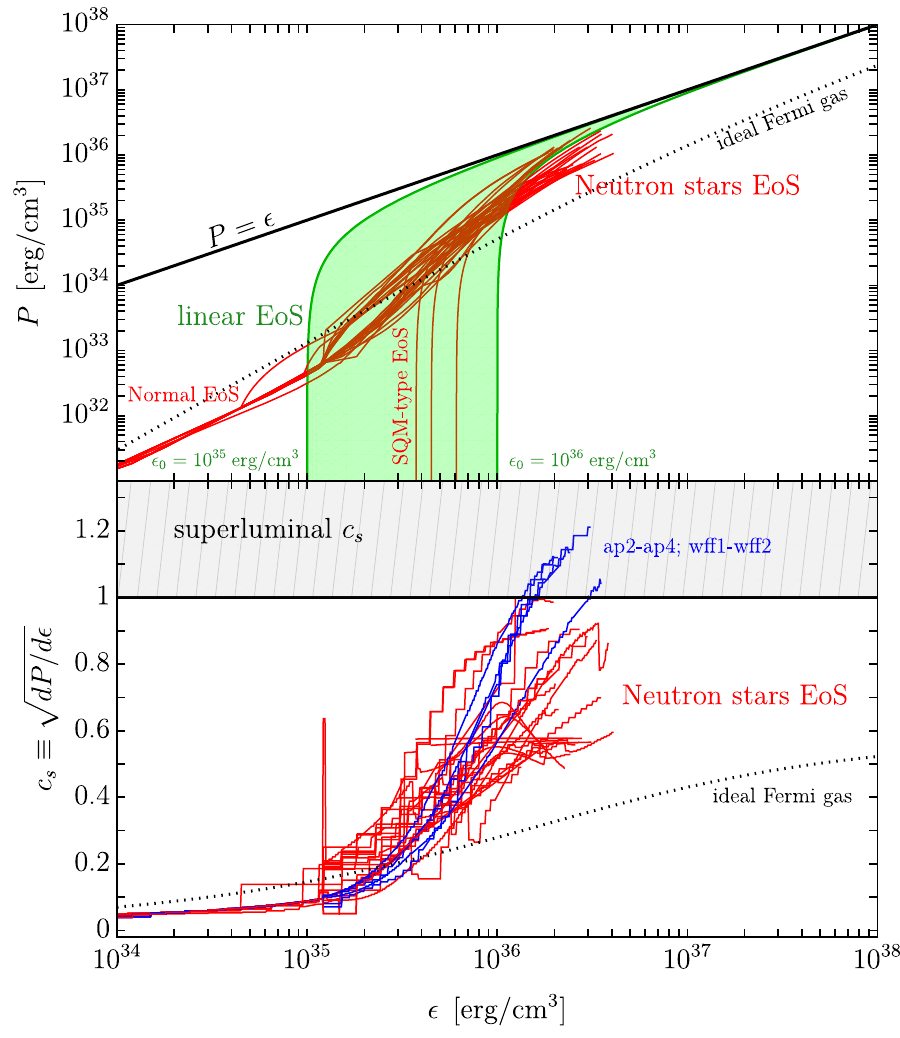}
	\caption{\it  
Top panel. 
Comparison between commonly employed equations of state for describing neutron star properties (red lines) and the linear EoS specified in  eq.\,(\ref{eq:LinEOS}) with $\omega = 1$.
 The region shaded in green 
 corresponds to $10^{35}
 \leqslant
 \epsilon_0\,\,[\textrm{erg}/\textrm{cm}^3]
 \leqslant 10^{36}$.
Bottom panel. Speed of sound $c_s$ (cf. eq.\,(\ref{eq:Speed})) as function of the energy density $\epsilon$ for the same neutron star EoS shown in the top panel. 
We highlight in blue the EoS that violate the causality condition $c_s < 1$. The labels correspond to the EoS: ap2-ap4\,\cite{Akmal:1998cf},
wff1-wff2\,\cite{Wiringa:1988tp}. 
 }
\label{fig:RealisticEoS}
\end{figure}
In fig.\,\ref{fig:RealisticEoS}, 
we compare the linear EoS that saturates the causality condition with the EoS describing neutron star matter (red lines). 
In the top panel of fig.\,\ref{fig:RealisticEoS} we plot the pressure $P$ as function of the energy density $\epsilon$ while in the bottom panel we show the speed of sound $c_s$ as function of the energy density $\epsilon$.
We remark that, in the case of neutron stars, both pressure and speed of sound have been drawn up to the value $\epsilon_{c}^{\textrm{max}}$, corresponding to the central density value associated with the maximum mass configuration for each of the considered EoS. 

As far as the neutron star EoS are concerned, we can broadly divide them into two categories. 
{\it i)} {\it Normal EoS}. These are EoS for which  the energy density goes to zero when the pressure vanishes.  
{\it ii)} 
{\it SQM-type EoS}. 
These are EoS for which  the energy density goes to a constant when the pressure vanishes. 
This physics-case corresponds to the so-called strange quark matter (SQM) stars.



We note that, among the Normal EoS, the ap2-ap4\,\cite{Akmal:1998cf} and 
wff1-wff2\,\cite{Wiringa:1988tp} EoS violate causality at high densities. 
We also find that the sly\,\cite{Douchin:2001sv} EoS also violates causality, but it does so for densities higher than those corresponding to the maximum mass configuration. Hence, it is not depicted in blue in fig.\,\ref{fig:RealisticEoS}.

It should be stressed that, in the case of neutron stars, this is not necessarily a problem. The reasons being that, rigorously stated, the quantity  $\sqrt{dP/d\epsilon}$ 
corresponds to the hydrodynamic phase velocity of sound waves in the neutron star
matter, and only  in the absence of dispersion and absorption it would be the velocity of signals in the medium, cf. ref.\,\cite{1978PhR201H}.
The bound $\sqrt{dP/d\epsilon} < 1$, therefore,  holds exactly only in the idealized case in which neutron star matter is neither dispersive nor
absorptive. 

Given the dependence on $\epsilon$ of the sound speed $c_s$, it may be clearer to adopt a modified ``global'' definition of the sound speed, calculated as an average over the range of energy densities present within the neutron star\,\cite{Saes:2021fzr,Saes:2024xmv}. 
Since, as discussed before, the speed of sound is a typical indicator of the
stiffness of an EoS, its average would represent the
mean stiffness. 
In the case of neutron star EoS for which  the energy density is zero when the pressure vanishes ({\it Normal EoS}), 
we write\,\cite{Saes:2021fzr,Saes:2024xmv}
\begin{align}
\langle
c_s^2 
\rangle & \equiv 
\frac{1}{\epsilon_c}
\int_0^{\epsilon_c}
c_s(\epsilon)^2 d\epsilon 
= 
\frac{1}{\epsilon_c}
\int_0^{\epsilon_c}\frac{dP}{d\epsilon}d\epsilon 
= \frac{P(\epsilon_c) - 
P(0)}{\epsilon_c}\nn\\
& = \frac{P_c}{\epsilon_c}\,,
\label{eq:AverageSoS}
\end{align}
so that we just need to compute the ratio $P_c/\epsilon_c$ and, to emphasize the relationship between stiffness and compactness, discuss it as a function of $\mathcal{C}$. 
We note that in the case in which the energy density goes to a constant when the pressure vanishes ({\it SQM-type EoS}) the above definition must be modified according to
\begin{align}
\langle
c_s^2 
\rangle_{\textrm{SQM}} & \equiv 
\frac{1}{\epsilon_c-\epsilon_R}
\int_{\epsilon_R}^{\epsilon_c}
c_s(\epsilon)^2 d\epsilon 
= \frac{P(\epsilon_c) - 
P(\epsilon_R)}{\epsilon_c-\epsilon_R}\nn\\
& = 
\frac{P_c}{\epsilon_c-\epsilon_R}\,,
\end{align}
with $\epsilon_R$ the energy density evaluated at the surface of the star  where the pressure vanishes, $P(\epsilon_R) = 0$.
\begin{figure}[!t]
	\centering
\includegraphics[width=0.495\textwidth]{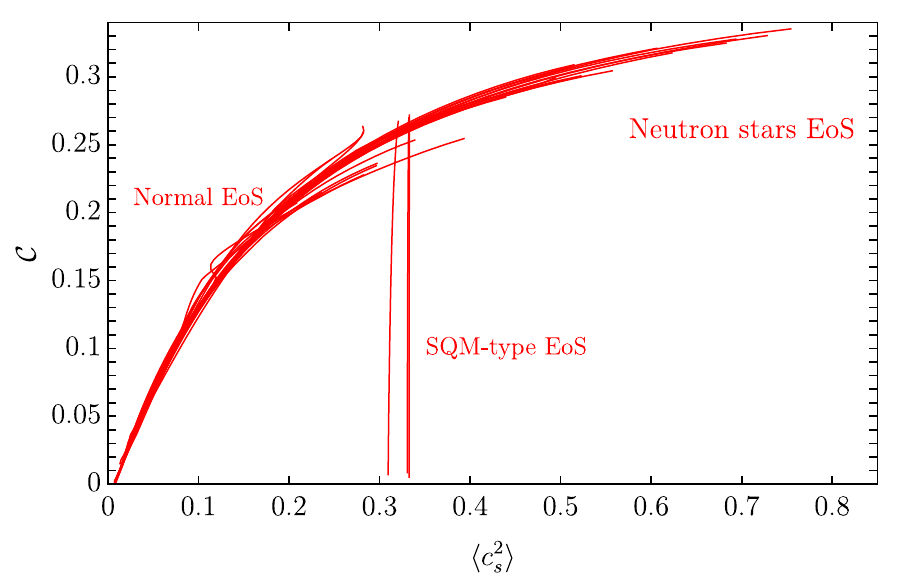}
	\caption{ \it 
 Relation between the average speed of sound squared (or, equivalently, the ratio of central pressure to central energy density, cf. eq.\,(\ref{eq:AverageSoS})) 
and the compactness $\mathcal{C}$ for the set of realistic
neutron star EoS analyzed in fig.\,\ref{fig:RealisticEoS}.
 }
\label{fig:AverageVelocity}
\end{figure}

In  fig.\,\ref{fig:AverageVelocity}, we show the compactness $\mathcal{C}$
as function of 
$\langle
c_s^2 
\rangle$ for the same neutron star EoS analyzed in fig.\,\ref{fig:RealisticEoS}.  
We highlight two points. 
{\it i)} For all the analyzed EoS, the average speed of sound is always sub-luminal and 
{\it ii)} in agreement with the qualitative discussion we put forward previously,
we see that, in the case of normal EoS, 
increasing stiffness (that is, increasing $\langle
c_s^2 
\rangle $) corresponds to more compact objects. In the case of SQM-type EoS, we note that the averaged speed of sound is constant.

We now consider the case of tidal deformability. 
The na\"{\i}ve expectations are the following. 
\begin{itemize}
\item[$\circ$] \textbf{Stiffer EoS}. In the presence of an external gravitational field, such as that from a companion in a binary system, a stiffer EoS will resist deformation more strongly. Therefore, for a given tidal force, a stiffer EoS typically leads to a lower tidal deformability because the object is less easily deformed.
\item[$\circ$] \textbf{Softer EoS}. 
On the other hand, a softer EoS can deform more readily under the influence of tidal forces. Hence, a softer EoS generally leads to a higher tidal deformability for a given tidal force.
\end{itemize}
To substantiate (or challenge) these insights, we proceed to compute the tidal Love number $k_2$ and both the tidal deformability $\lambda_{\textrm{Tidal}}$ and $\Lambda$. We closely follow the discussion in ref.\,\cite{Hinderer:2009ca}. 
For the sake of completeness, we present below the key steps of the discussion, some of which will be important in the subsequent analysis. 
We perturb both the metric in eq.\,(\ref{eq:SphMetric}) and the stress-energy tensor in eq.\,(\ref{eq:AnisoTMUNU}), considering the isotropic limit.
As far as the metric is concerned, 
we write
\begin{align}
g_{\mu\nu} = g_{\mu\nu}^{(0)} + h_{\mu\nu}\,,\label{eq:MetricPerturbations}
\end{align}
where now the unperturbed metric $g_{\mu\nu}^{(0)}$ is the one given in eq.\,(\ref{eq:SphMetric}).
We focus on static, even-parity, and quadrupolar ($l = 2$) metric perturbations, 
as these are the only modes sourced by a stationary external tidal field and therefore the only sector contributing to the (mass-type) tidal Love number.
The metric perturbation takes the form
\begin{align}
h_{\mu\nu}dx^{\mu}dx^{\nu}& = 
Y_{20}(\theta,\varphi)\bigg[
e^{\nu(r)}H_0(r)dt^2 + \nn\\
&
e^{\lambda(r)}H_2(r)dr^2 + 
r^2K(r)
(d\theta^2 + \sin^2\theta d\varphi^2)
\bigg]\,,
\end{align}
where 
$H_0(r)$, $H_2(r)$ and $K(r)$ describe the radial dependence of each perturbed metric component 
and $Y_{20}$ is the $(l,m) = (2,0)$ spherical harmonics.  
Given the metric in eq.\,(\ref{eq:MetricPerturbations}), we derive the first-order perturbative form of the Einstein tensor, which we denote by $\delta G_{\mu\nu}$.
The isotropic stress-energy tensor 
 is perturbed as follows. We write
\begin{align}
    T_{\mu\nu} = T^{(0)}_{\mu\nu} + \delta T_{\mu\nu}\,,
\end{align}
where the unperturbed value in the isotropic limit is given by $T^{(0)}_{\mu\nu} = (\epsilon + P)u^{(0)}_{\mu}u^{(0)}_{\nu} + Pg^{(0)}_{\mu\nu}$ with the unperturbed four-velocity given by 
$u^{(0)}_{\mu} = (-e^{\nu(r)/2},0,0,0)$. 
The perturbation $\delta T_{\mu\nu}$ takes the form
\begin{align}
\delta T_{\mu\nu} = 
&(\delta\epsilon + \delta P)
u^{(0)}_{\mu}u^{(0)}_{\nu} 
+
\delta P g^{(0)}_{\mu\nu} 
+ \nn\\
&
(\epsilon + P)\left[
u^{(0)}_{\mu}
\delta u_{\nu}
+
u^{(0)}_{\nu}
\delta u_{\mu}
\right] + P h_{\mu\nu}\,.
\end{align}
The perturbations $\delta\epsilon$ and $\delta P$ are related by the EoS $P=P(\epsilon)$. In fact, we write 
\begin{align}
P(\epsilon+\delta\epsilon) = 
P(\epsilon)+
\underbrace{\frac{dP}{d\epsilon}\delta\epsilon}_{\equiv\,\delta P}~~\Rightarrow~~
\delta\epsilon = \frac{\delta P}{
dP/d\epsilon
}\,.
\end{align}
The perturbed four-velocity 
takes the form $u_{\mu} = 
u_{\mu}^{(0)}+ \delta u_{\mu}$. The perturbation $\delta u_{\mu}$ can be computed from the condition $u_{\mu}u^{\mu} = 
u^{\mu}u^{\nu}g_{\mu\nu} = -1$, using the previous decomposition and the fact that
$u^{(0)}_{\mu}u^{(0)}_{\nu}
g^{(0)\mu\nu} = -1$. 
We find
$\delta u_{\mu} = 
(Y_{20}(\theta,\varphi)H_0(r)e^{\nu(r)/2}/2,0,0,0)$ and 
$\delta u^{\mu} = 
(-Y_{20}(\theta,\varphi)H_0(r)e^{-\nu(r)/2}/2,0,0,0)$. Note that only the temporal component is non-zero, as a consequence of the assumption of static perturbations. The perturbation $\delta T_{\mu\nu}$ takes a particularly simple form if we consider $\delta T^{\mu}_{~\nu}$, as we obtain $\delta T^{\mu}_{~\nu} =  
\textrm{diag}(-\delta\epsilon, 
\delta P,\delta P,\delta P)$. 
We now consider the the perturbed Einstein field equations, 
$\delta G^{\mu}_{~\nu} = 8\pi \delta T^{\mu}_{~\nu}$. 
We observe that\,\cite{Hinderer:2007mb}:
{\it i)} from $\delta G^{\theta}_{~\theta}-\delta G^{\varphi}_{~\varphi} =0$, we obtain 
$H_2(r) = H_0(r)$; {\it ii)} 
from $\delta G^{r}_{~\theta}= 0$ we derive 
$K^\prime(r) = H_0^{\prime}(r) + 
H_0(r)\nu^{\prime}(r)$; {\it iii)} 
from $\delta G^{\theta}_{~\theta} + \delta G^{\varphi}_{~\varphi} = 16\pi\delta P$, we extract an expression for $\delta P$ that reads 
$\delta P = Y_{20}(\theta,\varphi)e^{-\lambda(r)}H_0(r)[\lambda^{\prime}(r)+\nu^{\prime}(r)]/16\pi r$; 
{\it iv)}  
from $\delta G^{t}_{~t} + \delta G^{r}_{~r} = 8\pi(\delta T^{t}_{~t} - \delta T^{r}_{~r})$, we get a second-order differential equation for $H_0(r)$.
This latter, after applying the TOV equations to the unperturbed quantities, takes the form
\begin{align}
\frac{dH_0}{dr} & = \beta\,,\label{eq:EqH01}\\
\frac{d\beta}{dr} & =
2\left(
1-\frac{2m}{r}
\right)^{-1}H_0\bigg\{-2\pi
\left[
5\epsilon + 9P 
+ \frac{d\epsilon}{dP}(\epsilon + P)\right] \nn\\
& +\frac{3}{r^2} +
2\left(
1-\frac{2m}{r}
\right)^{-1}
\left(
\frac{m}{r^2} + 
4\pi r P
\right)^2\bigg\} \nn\\
& 
+\frac{2\beta}{r}
\left(
1-\frac{2m}{r}
\right)^{-1}
\left[
-1+\frac{m}{r} + 
2\pi r^2(\epsilon - P)
\right]\,.\label{eq:EqH02}
\end{align}
We impose the boundary conditions 
$H_0(r) = a_0r^2$ and 
$H^{\prime}_0(r) = 2a_0r$ as
$r\to 0$. 
These conditions arise from imposing regularity at the origin.
The constant $a_0$  can be chosen arbitrarily as it cancels in the expression for the Love number. 
It is crucial to analyze the equation for $H_0$ in the region external to the star (where we set $\epsilon=P=0$ and $m(r) = M$). 
Performing the change of variable  defined by $x\equiv \frac{r}{M}-1$, in the external region the equation 
for $H_0$ takes the form
\begin{align}
(x^2-1)\frac{d^2H_0}{dx^2} 
+2x\frac{dH_0}{dx}
-H_0\left(
6+\frac{4}{x^2-1}
\right) = 0\,,
\end{align}
with generic solution given, in terms of the radial variable $r$, by 
\begin{align}
&H_0(r) = \left(\frac{r}{M}\right)^2
\left(
1-\frac{2M}{r}
\right)\bigg\{
-3a + \nn\\
&
+b\bigg[
\frac{M(M-r)(2M^2+6Mr - 3r^2)}{r^2(2M-r)^2} -
\frac{3}{2}\log
\frac{r}{r-2M}
\bigg]\bigg\}\,.\label{eq:AsySolution}
\end{align}
The limit $r\gg M$ gives
\begin{align}
 H_0(r) =&
 \left[
 -3a\left(\frac{r}{M}\right)^2
 + O\left(\frac{r}{M}\right)
 \right] \nn\\
 &+ 
 \left[
-\frac{8}{5}b
\left(\frac{M}{r}\right)^3 + 
O\left(\frac{M^4}{r^4}\right)
 \right]\,.
\end{align}
We are now in a position to compare with the expansion given in eq.\,(\ref{eq:AsyMe}). 
This will allow us to obtain the coefficients $a$ and $b$ as functions of the tidal deformability. We decompose the tensor multipole moments in eq.\,(\ref{eq:AsyMe}) as 
\begin{align}
Q_{ij} = \sum_{m=-2}^{+2}
Q_{2m} \mathcal{Y}^{2m}_{ij}\,,~~~
\mathcal{E}_{ij} = \sum_{m=-2}^{+2}
\mathcal{E}_{2m} \mathcal{Y}^{2m}_{ij}\,,
\end{align}
where the symmetric, trace-free tensors 
$\mathcal{Y}^{lm}_{ij}$ are defined by 
$Y_{lm}(\theta,\varphi) = \mathcal{Y}^{lm}_{ij}n_i n_j$ 
where the angular dependence is contained in the unit vector  
$\boldsymbol{n} = (\sin\theta\cos\varphi,\sin\theta\sin\varphi,\cos\theta)$ with components $n_i$. We focus on the term with $m=0$, and we extract
\begin{align}
a = \frac{1}{3}M^2\mathcal{E}\,,~~~~~
b = \frac{15}{8}\frac{\lambda_{\textrm{Tidal}}\mathcal{E}}{M^3}\,,
\end{align}
where $\mathcal{E}\equiv \mathcal{E}_{20}$ is the magnitudes of the $l = 2$, $m = 0$
spherical harmonic coefficients of the tidal tensor.
Eq.\,(\ref{eq:AsySolution}), therefore, reads
\begin{align}
&H_0(r) = 
\mathcal{E}r^2\left(1-\frac{2M}{r}\right)
\bigg\{-1 + \frac{15\lambda_{\textrm{Tidal}}}{8M^5}\times \nn\\
&
\bigg[
\frac{M(M-r)(2M^2+6Mr - 3r^2)}{r^2(2M-r)^2} -
\frac{3}{2}\log
\frac{r}{r-2M}
\bigg]
\bigg\}\,.\label{eq:ExplicitH0}
\end{align}
We now introduce the Love number, see eq.\,(\ref{eq:LoveNumber}), and solve for $k_2$.
After defining the quantity
\begin{align}
Y \equiv \frac{R\beta(R)}{H_0(R)}\,,\label{eq:YTidl}    
\end{align}
we use in the right-hand side the solution in eq.\,(\ref{eq:ExplicitH0}) evaluated in $r=R$ and solve for $k_2$. 
The Love number is
\begin{align}
k_2 =   
&\frac{8\mathcal{C}^5}{5}
(1-2\mathcal{C})^2
\left[
2+2\mathcal{C}(Y-1)-Y
\right]\times \nn\\
&
\bigg\{
2\mathcal{C}\left[
6-3Y+3\mathcal{C}(5Y-8)
\right] \nn\\
& 
+4\mathcal{C}^3\left[
13-11Y+\mathcal{C}(3Y-2) 
+2\mathcal{C}^2(1+Y)
\right] \nn\\
& 
+3(1-2\mathcal{C})^2
\left[
2-Y+2\mathcal{C}(Y-1)
\right]\log(1-2\mathcal{C})
\bigg\}^{-1}\,.
\label{eq:k2formula}
\end{align}
Because $H_0$ and $H^{\prime}_0$
 are continuous between the interior and the vacuum at the surface, we can compute $k_2$ by evaluating $Y$ while integrating eqs.\,(\ref{eq:EqH01},\,\ref{eq:EqH02}) from the origin to $r=R$.
This completes the calculation of $k_2$ except for an important detail that we will now discuss. 

Let us rewrite eqs.\,(\ref{eq:EqH01},\,\ref{eq:EqH02}) in the equivalent form 
\begin{align}
\frac{dH_0}{dr^2} + C_1(r) \frac{dH_0}{dr} + C_0(r)H_0 = 0\,,\label{eq:HSchematic}
\end{align}
with 
\begin{align}
C_1 & \equiv -\frac{2}{r}
\left(
1-\frac{2m}{r}
\right)^{-1}
\left[
-1+\frac{m}{r} + 
2\pi r^2(\epsilon - P)
\right]\,,\\
C_0 & \equiv -
2\left(
1-\frac{2m}{r}
\right)^{-1}\bigg\{-2\pi
\left[
5\epsilon + 9P 
+ \frac{d\epsilon}{dP}(\epsilon + P)\right] \nn\\
& +\frac{3}{r^2} +
2\left(
1-\frac{2m}{r}
\right)^{-1}
\left(
\frac{m}{r^2} + 
4\pi r P
\right)^2\bigg\}\,.
\end{align}
As far as the coefficient $C_0$ is concerned, let us focus on the term that contains the derivative $d\epsilon/dP$. 
We write
\begin{align}
C_0 & \ni 
\frac{4\pi(\epsilon + P)}{1-2m/r}
\frac{d\epsilon}{dP} =
\frac{4\pi(\epsilon + P)}{1-2m/r}
\frac{d\epsilon}{dr}\frac{dr}{dP} \nn\\
& = -\frac{4\pi r^2}{(4\pi r^3 P + m)}
\frac{d\epsilon}{dr}\,,\label{eq:SingularTerm}
\end{align}
where in the last step we used eq.\,(\ref{eq:TOV2}).
For stars with a nonzero density at the surface, the radial profile of the energy density has a Heaviside theta discontinuity at $r=R$.
For example, in stars described by a linear EoS, the energy density is $\epsilon_0$ at the surface of the star (where the pressure is zero) and vanishes outside. More precisely, if we write the radial profile of the energy density as $\epsilon(r) = 
\epsilon(r)[1-\vartheta(r-R)]$, with 
$\vartheta(x) = 0 \text{ for } x < 0, \, \vartheta(x) = 1 \text{ for } x \geq 0$, then $d\epsilon/dr$ has a delta function singularity at $r=R$, 
$\left.d\epsilon/dr\right|_{r=R} = -\epsilon(R)\delta(r-R)$. 
Consequently, eq.\,(\ref{eq:SingularTerm}) contains, at $r=R$, a singular term given by 
\begin{align}
C_0^{\textrm{sing}} \equiv 
4\pi \epsilon(R)\delta(r-R)\frac{R^2}{M}\,.\label{eq:SingC0}
\end{align}
To account for this singular contribution, we can proceed as 
follows 
(see ref.\,\cite{Damour:2009vw} for the case of constant density stars, 
and ref.\,\cite{Hinderer:2009ca} for the case of neutron stars with SQM-type EoS).
We note that if we introduce the variable $y(r)\equiv rH_0^{\prime}(r)/H_0(r)$, eq.\,(\ref{eq:HSchematic}) takes the form
\begin{align}
y^{\prime}(r) = \frac{[1-y(r)]y(r)}{r} 
-C_1(r) y(r) - r C_0(r)\,.
\end{align}
Consequently, the singular term described in eq.\,(\ref{eq:SingC0}) can be incorporated by introducing a Heaviside theta discontinuity in the variable $y(r)$
\begin{align}
y^{\textrm{sing}} = -4\pi\epsilon(R)\vartheta(r-R)\frac{R^3}{M}\,.    
\end{align}
Eq.\,(\ref{eq:YTidl}), therefore, gets a correction given by 
\begin{align}
Y =
\frac{R\beta(R)}{H_0(R)} - 
\frac{4\pi R^3\epsilon(R)}{M}\,.\label{eq:CorrectedY}
\end{align}
This correction is important  in the case of neutron stars supported by an SQM-type EoS and in the more general case of compact objects supported by a linear EoS.
\begin{figure}[!t]
	\centering
\includegraphics[width=0.495\textwidth]{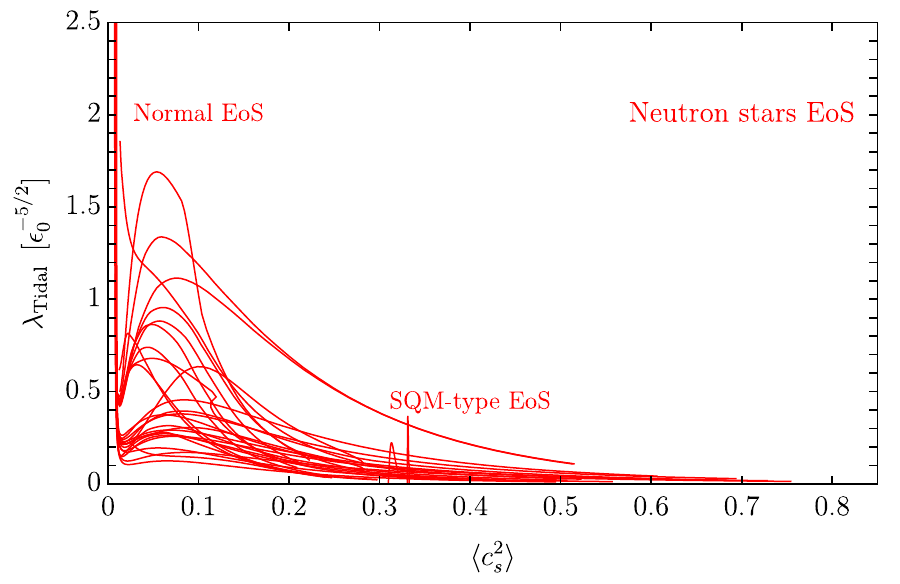}
	\caption{ \it 
 Relation between the average speed of sound squared (or, equivalently, the ratio of central pressure to central energy density, cf. eq.\,(\ref{eq:AverageSoS})) 
and the 
tidal deformability $\lambda_{\textrm{Tidal}}$ for the set of realistic
neutron star EoS analyzed in fig.\,\ref{fig:RealisticEoS}.  
SQM-type EoS are described by the (barely distinguishable) vertical red lines corresponding to their label.
 }
\label{fig:AverageVelocitylam}
\end{figure}
In fig.\,\ref{fig:AverageVelocitylam}, we show the tidal deformability $\lambda_{\textrm{Tidal}}$ as function of 
$\langle
c_s^2 
\rangle$ for the same neutron star EoS analyzed in fig.\,\ref{fig:RealisticEoS} and fig.\,\ref{fig:AverageVelocity}. 
This figure confirms our previous intuition, and shows that the tidal deformability $\lambda_{\textrm{Tidal}}$ decreases for increasing stiffness. 
\begin{figure}[!t]
	\centering
\includegraphics[width=0.495\textwidth]{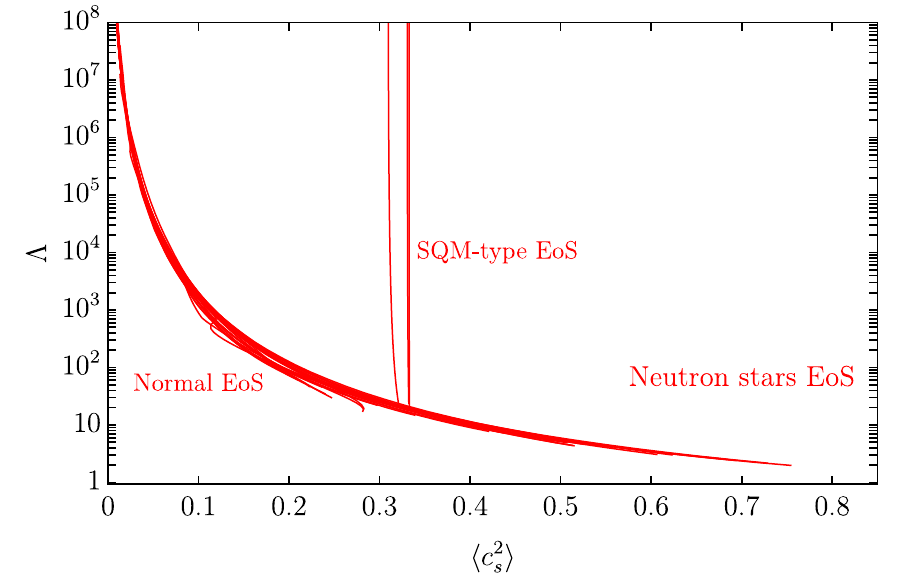}
	\caption{ \it 
 Relation between the average speed of sound squared (or, equivalently, the ratio of central pressure to central energy density, cf. eq.\,(\ref{eq:AverageSoS})) 
and the 
tidal deformability $\Lambda$ for the set of realistic
neutron star EoS analyzed in fig.\,\ref{fig:RealisticEoS}. 
 }
\label{fig:AverageVelocityLambda}
\end{figure}
Finally, in fig.\,\ref{fig:AverageVelocityLambda}, we show  the tidal deformability $\Lambda$ 
as function of 
$\langle
c_s^2 
\rangle$ 
for the same neutron star EoS analyzed in fig.\,\ref{fig:RealisticEoS} and fig.\,\ref{fig:AverageVelocity}.  
Increasing the stiffness of the EoS  corresponds to smaller values of the tidal deformability $\Lambda$. 
In particular, the plot shows that the tidal deformability $\Lambda$ follows the scaling $\Lambda \propto \mathcal{C}^{-5}$ as expected on the basis of its definition.
\begin{figure}[!t]
	\centering
\includegraphics[width=0.495\textwidth]{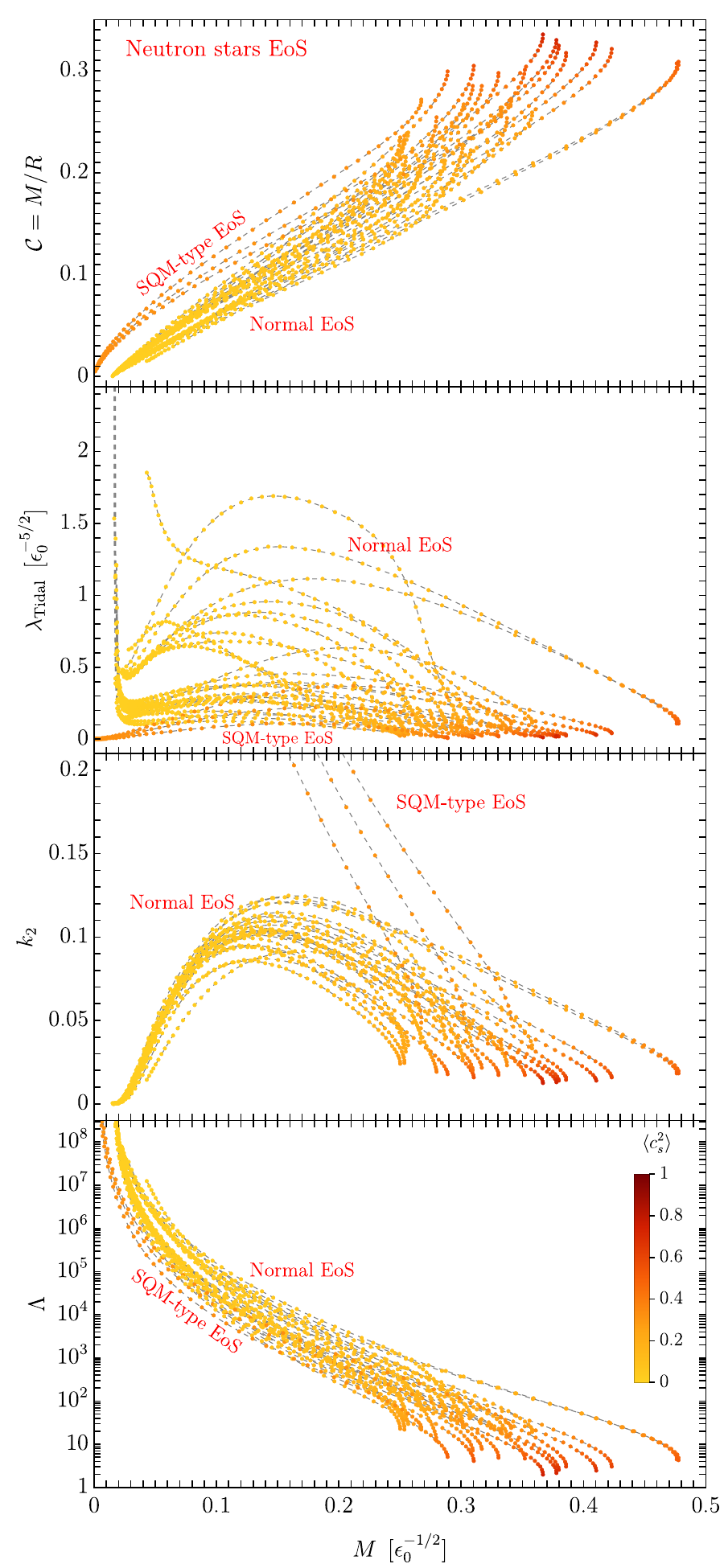}
	\caption{ \it 
From top to bottom: compactness, tidal deformability $\lambda_{\textrm{Tidal}}$, tidal Love number $k_2$ and tidal  deformability $\Lambda$ as function of the neutron star mass $M$.
All dimensionfull quantities are written in units of  
 $\epsilon_0 = m_n^4c^5/\pi^2\hbar^3$.
 The colored legend keeps track of the stiffness parameter $\langle c_s^2\rangle$.
 }
\label{fig:CompaLambdaNS}
\end{figure}

\begin{figure}[!t]
	\centering
\includegraphics[width=0.495\textwidth]{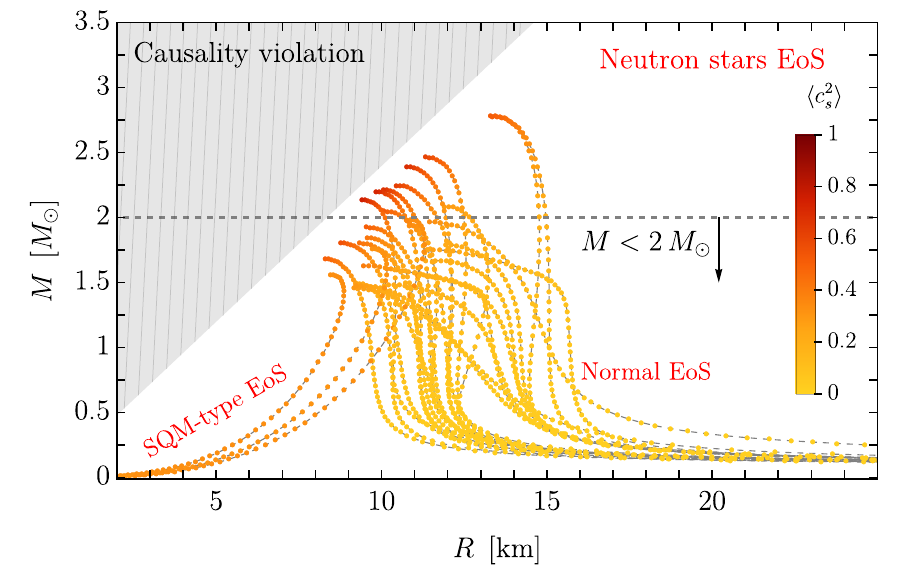}
	\caption{ \it 
 Mass-radius diagram for the set of realistic
neutron star EoS analyzed in fig.\,\ref{fig:RealisticEoS}. 
The region shaded in gray is excluded by violation of causality. 
The colored legend refers, for each of the analyzed EoS, to the average speed of sound $\langle c_s^2 \rangle$. 
The grey horizontal line corresponds to $M=2\,M_{\odot}$. Only the EoS characterized by a mass-radius curve above this value are capable of explaining the mass of the most massive neutron star known (PSR J0952–0607, with  approximately $M=2.35\,M_{\odot}$, cf. ref.\,\cite{Romani:2022jhd}).
 }
\label{fig:NeutronStarMassRadius}
\end{figure}

Given the definition of the tidal deformability $\Lambda$ in terms of tidal Love number and compactness given in 
eq.\,(\ref{eq:Tidal}), it is instructive to analyze separately the various contributions.
In fig.\,\ref{fig:CompaLambdaNS} we show 
a comparison between (from top to bottom) compactness,  
tidal deformability $\lambda_{\textrm{Tidal}}$, tidal Love number 
and tidal deformability parameter $\Lambda$
as function of the neutron star mass ($x$-axis in common). 
We choose to display these quantities as a function of mass because the latter is the most astrophysically relevant parameter, as mass is the measurable quantity during binary inspiral.
Importantly, the novelty in this analysis is that in all these plot we keep track of the value of the stiffness parameter $\langle c_s^2\rangle$ by means of the colored legend provided in the bottom panel. 
We highlight a number of important points. 
{\it i)} From the top panel, we observe that, for a fixed mass, a stiffer EoS corresponds to larger values of compactness. Keeping the mass constant, this implies that a stiffer equation of state corresponds to a smaller radius for a given mass.
{\it ii)} The plot of $\lambda_{\textrm{Tidal}}$ confirms our previous intuition: the tidal deformability tends to increase for softer EoS. 
As discussed in the previous point, a stiffer equation of state corresponds to a smaller radius for a given mass. 
Consequently, $\lambda_{\textrm{Tidal}}$ becomes smaller since proportional to $R^5$. 
The plot of $\lambda_{\textrm{Tidal}}$ also shows that, for fixed mass, 
the tidal deformability is generically smaller in the case of SQM-type EoS compared to normal EoS. 
This behavior can be understood by looking at the mass-radius diagram shown in 
fig.\,\ref{fig:NeutronStarMassRadius}.
For fixed mass on the vertical $y$-axis, 
SQM-type EoS are characterized by smaller values or $R$ and $\lambda_{\textrm{Tidal}}$ becomes 
smaller.
At very small masses (left end of the $x$-axis in fig.\,\ref{fig:CompaLambdaNS}), the tidal deformability 
$\lambda_{\textrm{Tidal}}$ approaches zero in the case of SQM-type EoS, whereas it tends toward large values in the case of normal EoS. 
This is a consequence of the fact that 
SQM-type EoS are essentially characterized by constant density in the bulk of the star and, therefore, 
$R\to 0$ as $M\to 0$ since we approximately  
have $R \propto M^{1/3}$.
As a consequence, $\lambda_{\textrm{Tidal}} \to 0$. 
From the plot depicted in fig.\,\ref{fig:NeutronStarMassRadius}, it is also evident that the mass-radius curve for a neutron star with a normal EoS approaches a constant value of $M$ as the radius increases. This behavior signifies the Newtonian limit, wherein the mass becomes essentially independent of the radius as the latter increases.
As a consequence, for very small values of $M$ in the case of normal EoS $\lambda_{\textrm{Tidal}}$ tends to large values since driven by $R^5$.
{\it iii)} The Love number $k_2$, defined in 
eq.\,(\ref{eq:LoveNumber}), factors out from the definition of $\lambda_{\textrm{Tidal}}$ the $R^5$-dependence.  
Consequently, we now observe an opposite behavior compared to what discussed in the case of $\lambda_{\textrm{Tidal}}$.
For decreasing values of $M$, the tidal Love number $k_2$ for SQM-type EoS becomes larger compared to the case of normal EoS (reaching a maximum value at around $k_2 \simeq 0.7$, not shown in the scale of the plot). 
{\it iv)} Finally, 
in the bottom panel of fig.\,\ref{fig:CompaLambdaNS}) we show the dimensionless tidal deformability $\Lambda$, cf. eq.\,(\ref{eq:Tidal}), as function of the mass. 
In this plot, the functional dependence  of $\Lambda$ qualitatively follows the scaling $\Lambda \sim \mathcal{C}^{-5}$. 
In this plot, we observe that neutron stars characterized by SQM-type EoS exhibit smaller tidal deformabilities $\Lambda$ compared to those with normal EoS. This phenomenon arises from the fact that, despite possessing larger tidal Love numbers, SQM-type EoS neutron stars are more compact than those with normal EoS. Consequently, the combined effect leads to a reduction in the value of $\Lambda$.

\subsection{Causality bound on compactness, tidal deformability and tidal Love number}\label{sec:Tidal}

In this section, we solve the TOV system for the linear EoS defined in eq.\,(\ref{eq:LinEOS}). We focus on the case with $\omega = 1$. 

\begin{figure}[!t]
	\centering
\includegraphics[width=0.495\textwidth]{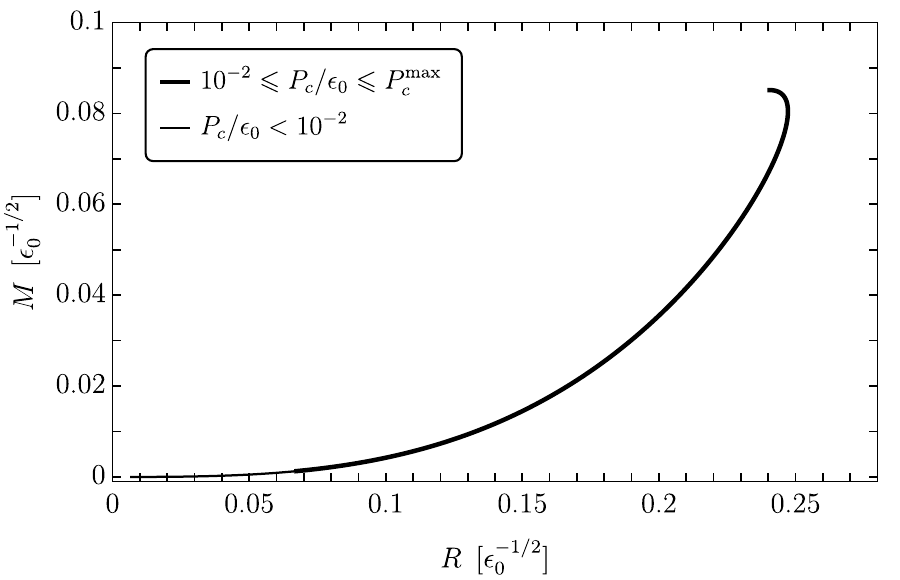}
	\caption{ \it 
Mass-radius diagram for the linear EoS defined in eq.\,(\ref{eq:LinEOS}) with $\omega = 1$. 
The thicker line corresponds to central pressure values within the interval $10^{-2} 
\leqslant P_c/\epsilon_0
\leqslant P_c^{\textrm{max}}
$ with $P_c^{\textrm{max}} = 2.018$.
 }
\label{fig:MassRadiusLinear}
\end{figure}

The mass-radius diagram is shown in fig.\,\ref{fig:MassRadiusLinear}.
In agreement with previous literature, the maximal compactness is given by 
$\mathcal{C}_{\textrm{max}} = 0.354$\,\cite{1984ApJL,Glendenning:1992dr,Lattimer:2006xb,Urbano:2018nrs} and corresponds to 
$(R_{\textrm{max}},M_{\textrm{max}}) = (0.241, 0.085)$ both in units of $\epsilon_0^{-1/2}$. 
After converting $R$ and $M$, respectively, into kilometers and solar masses, as $\epsilon_0$ varies, we find the relationship
\begin{align}
M_{\textrm{max}} 
= 0.239\left(
\frac{R_{\textrm{max}}}{\textrm{km}}
\right)\,M_{\odot}\,.\label{eq:CauslityMassRa}
\end{align}
This condition, taking now $R_{\textrm{max}}$ as variable, defines the lower boundary of the region shaded in gray in fig.\,\ref{fig:NeutronStarMassRadius}.

\begin{figure}[!t]
	\centering
\includegraphics[width=0.495\textwidth]{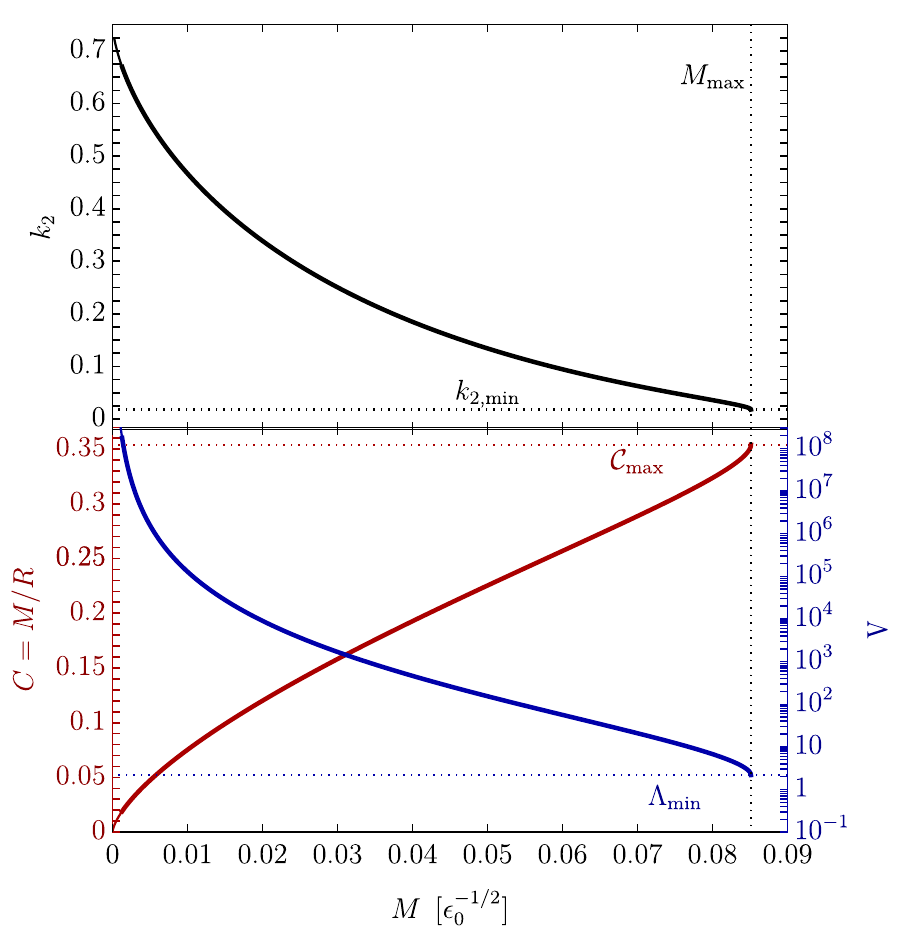}
	\caption{ \it 
Tidal Love number (top panel), compactness (bottom panel, left-side $y$-axis) and tidal deformability $\Lambda$ (bottom panel, right-side $y$-axis) as function of mass (in units of $\epsilon_0^{-1/2}$) for the linear EoS with $\omega = 1$. 
 }
\label{fig:TidalCompaLinearEoS}
\end{figure}

In fig.\,\ref{fig:TidalCompaLinearEoS} we show compactness, tidal deformability $\Lambda$ and tidal Love number $k_2$ for the linear EoS with $\omega = 1$ as function on mass.
The behavior of these curves follows 
what we have already discussed in the case of neutron star with a SQM-type EoS.
However, we are now pushing the speed of sound $c_s$ up to the maximal value allowed by causality, $c_s = 1$ (compared to the value $c_s \simeq 0.6$ that characterizes SQM-type EoS). 

At the maximum compactness and mass value, we find a minimum value for the tidal deformability and tidal Love number
\begin{align}
\Lambda_{\textrm{min}} = 2.186\,,~~~~
k_{2,\textrm{min}} = 0.018\,.\label{eq:CausalTidal}
\end{align}
This is shown in fig.\,\ref{fig:TidalCompaLinearEoS} as function of the mass with the two aforementioned values located at the far right as we move towards higher mass values.

At this stage, and in view of some considerations that will be made later, it is instructive to explicitly examine how the tidal deformability $\Lambda$ is extracted.
\begin{figure}[!t]
	\centering
\includegraphics[width=0.495\textwidth]{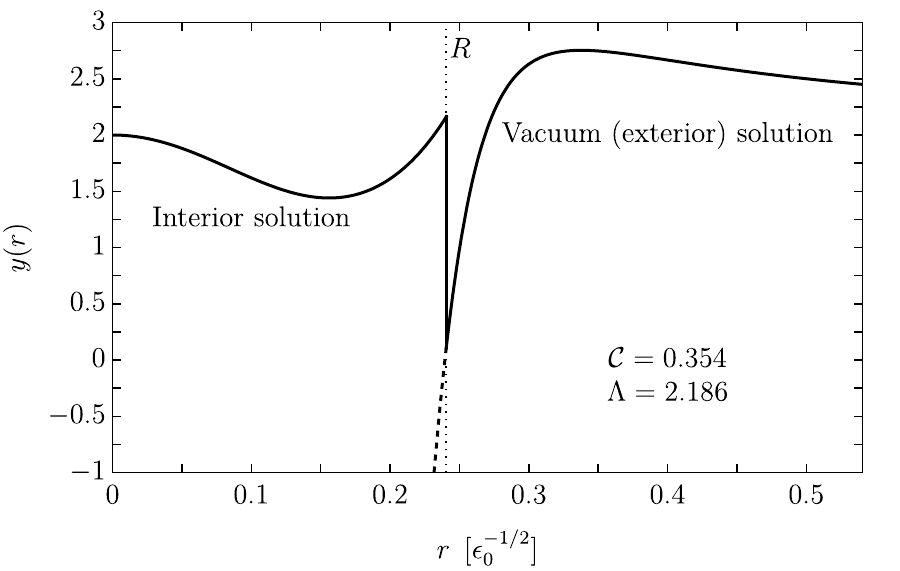}
	\caption{ \it 
Radial profile of the function $y(r)\equiv rH_0^{\prime}(r)/H_0(r)$, whose value at $R$ is used in the determination of the tidal deformability $\Lambda$. We present the specific case of an unperturbed solution with maximum compactness. 
We separate the spacetime into an interior region of the star ($r<R$, where $R$ is its radius), in which $y(r)$ is evaluated by integrating eqs.\,(\ref{eq:EqH01},\,\ref{eq:EqH02}), and an exterior vacuum region, where $y(r)$ is calculated analytically, see eqs.\,(\ref{eq:Analy1}-\ref{eq:Analy4}). 
Note the jump at $r=R$, due to the correction introduced in eq.\,(\ref{eq:CorrectedY}).
}
\label{fig:ySolLinEoS}
\end{figure}
In fig.\,\ref{fig:ySolLinEoS}, we focus on the solution with maximal compactness. 
In terms of the radial coordinate, we separate the interior region ($r<R$) from the exterior region ($r>R$). In the interior region, we solve eqs.\,(\ref{eq:EqH01},\,\ref{eq:EqH02}) and plot the radial profile of the quantity 
$y(r)\equiv rH_0^{\prime}(r)/H_0(r)$. 
Afterward, we calculate the value of $Y$, defined as in eq.\,(\ref{eq:CorrectedY}). 
Finally, we compute $\Lambda = 2k_2/3\mathcal{C}^5$ with $k_2$ as in 
eq.\,(\ref{eq:k2formula}). 
In the exterior region, we make use of eq.\,(\ref{eq:ExplicitH0}) to reconstruct the value of the variable $y(r)$, which in this region we denote as $y_{\textrm{vac}}(r)$. We find
\begin{align}
y_{\textrm{vac}}(x) & = \frac{2}{x-1}\bigg[
x+ \frac{120\Lambda}{
(x^2-1)^2 f(x,\Lambda)
+30x\Lambda g(x)
}\bigg]\,,\label{eq:Analy1}\\
f(x,\Lambda) & \equiv -16-45\Lambda\log\left(\frac{x+1}{x-1}\right)\,,\label{eq:Analy2}\\
g(x) & \equiv 3(x+1)^2 - 6(x+1) -2\,,\label{eq:Analy3}\\
x & \equiv \frac{r}{M} - 1 =
\frac{r/R}{\mathcal{C}} - 1\,.\label{eq:Analy4}
\end{align}
In fig.\,\ref{fig:ySolLinEoS}, the radial profile of $y_{\textrm{vac}}(r)$, with the value of $\Lambda$ calculated as previously described, is shown for $r>R$. 
We particularly emphasize the behavior at $r=R$, where the role of the correction introduced in eq.\,(\ref{eq:CorrectedY}) becomes evident  in connecting the interior solution with the exterior one. 
Treating the singular term present in eq.\,(\ref{eq:HSchematic}) as done in the text, which led to the introduction of the correction defined in eq.\,(\ref{eq:CorrectedY}), is certainly one way to account for the non-zero surface energy density that characterizes stars supported by a linear EoS. However, it remains a model, as we expect that in more realistic situations, the energy density will ultimately approach zero, even if perhaps through a rapid transition, as it crosses the surface of the star. From our analysis, we expect that the details of this transition may be relevant for an accurate determination of tidal deformability. 
This aspect will be addressed in section\,\ref{sec:BS}, where we will discuss the tidal deformability of solitonic boson stars.

For the moment, we focus on the possibility of using the information gathered thus far to delineate the parameter space of ECOs.
\begin{figure}[!t]
	\centering
\includegraphics[width=0.495\textwidth]{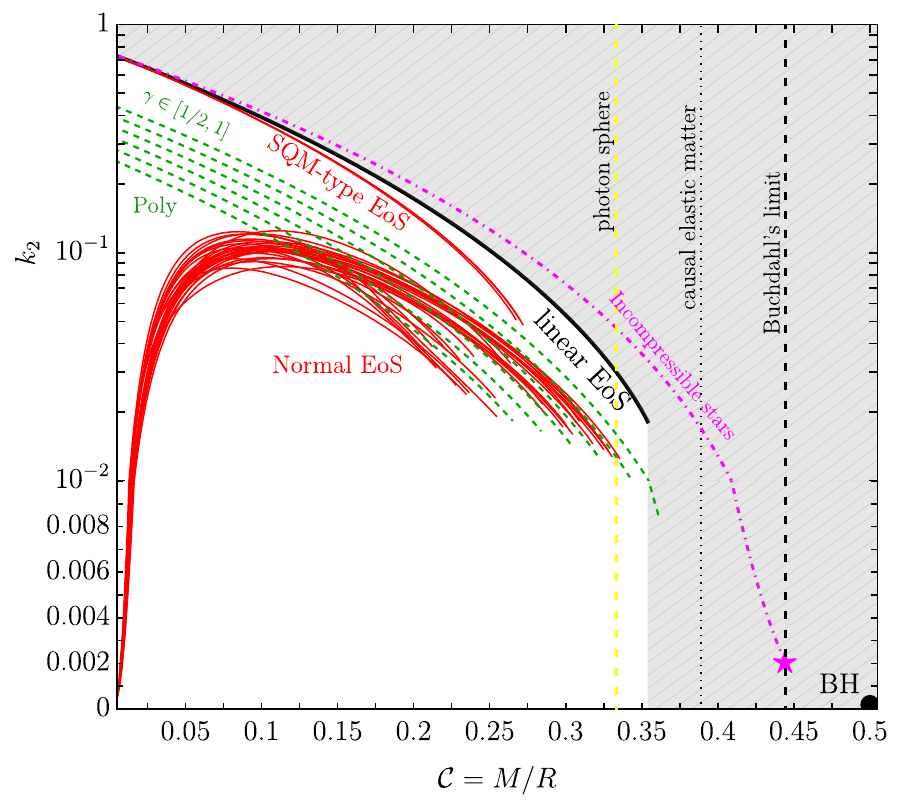}
	\caption{ \it 
Tidal Love number as function of compactness.  
The vertical lines demarcate the compactness value corresponding to the existence of a photon sphere ($\mathcal{C} = 1/3$), the maximal compactness obtainable with radially stable causal elastic matter ($\mathcal{C} \approx 0.389$) and the Buchdahl's limit  ($\mathcal{C} = 4/9$).
We show the tidal Love number of neutron stars (red lines), that corresponding to the linear EoS with $\omega = 1$ (solid black line), and that of constant-density stars (dot-dashed magenta line). The polytropic EoS cases are indicated by green dashed lines, corresponding to $\gamma \in [0.5,1]$.
 }
\label{fig:KeyPlotCk2}
\end{figure}
First of all, we focus on pure dimensionless numbers, and 
we consider tidal Love number and tidal deformability $\Lambda$ as function of compactness. 

In fig.\,\ref{fig:KeyPlotCk2}, we show the parameter space $(\mathcal{C},k_2)$. 
The solid black line represents the curve corresponding to the linear EoS with $\omega = 1$. 
The red lines corresponds to neutron stars with both normal and SQM-type EoS. 
Black holes occupy the bottom-right corner of this parameter space with 
$(\mathcal{C},k_2)_{\textrm{BH}} = (0.5,0)$. 
The line corresponding the the linear EoS stops at $\mathcal{C}_{\textrm{max}} = 0.354$. 
The only possibilities to exceed this value are to violate causality or invalidate some of the assumptions underlying our computation. 
We discuss an example for both cases. 

With regard to the violation of causality, a limiting case is represented by constant-density stars, in which $\epsilon$ is constant and, formally, the speed of sound is infinite. The tidal deformability and the tidal Love number of these idealized, incompressible stars can be computed by following the same steps used in the case of the linear EoS, simply by setting the energy density to a constant value (cf. also ref.\,\cite{Damour:2009vw}). 
In fig.\,\ref{fig:KeyPlotCk2}, we show, with a magenta dot-dashed line, the value of the tidal Love number $k_2$ as a function of compactness for incompressible stars. As expected, for a given value of compactness, the value of the tidal Love number exceeds the one defined by the linear EoS and saturates the causality violation, having an infinite speed of sound. 
As well known, the maximum compactness (reached in the limit of infinite central pressure) saturates the Buchdahl limit, and at this point we find 
\begin{align}
(\mathcal{C},k_2)_{\textrm{Buchdahl}} = (4/9, 
2\times 10^{-3})\,.
\end{align}
Concerning causality violation, another example is provided by a polytropic EoS with index 
$\gamma = 0.5$ (see fig.\,\ref{fig:EnergyCondPoly} and the corresponding discussion in the text). 
The case of polytropic EoS is shown in fig.\,\ref{fig:KeyPlotCk2} by green dashed lines. As expected from the violation of the DEC, the curve corresponding to $\gamma = 0.5$ is the one that enters the gray region, indicating a violation of causality.

With regard instead to the possibility of relaxing one of our assumptions, ref.\,\cite{Alho:2022bki} considers stars made of elastic anisotropic materials (cf. also ref.\,\cite{Karlovini:2004gq}).
In this setup, the role of anisotropy makes it possible to enhance the compactness of ECOs. 
However, it should be noted that even in this case, once the conditions of radial stability and causal propagation of perturbations are imposed, the maximum compactness allowed does not deviate significantly from our causality bound, settling at a maximum value of $\mathcal{C} \approx 0.389$.
Barring these cases, 
we argue that the region shaded in gray violates the requirement of relativistic causality. 

\begin{figure}[!t]
	\centering
\includegraphics[width=0.495\textwidth]{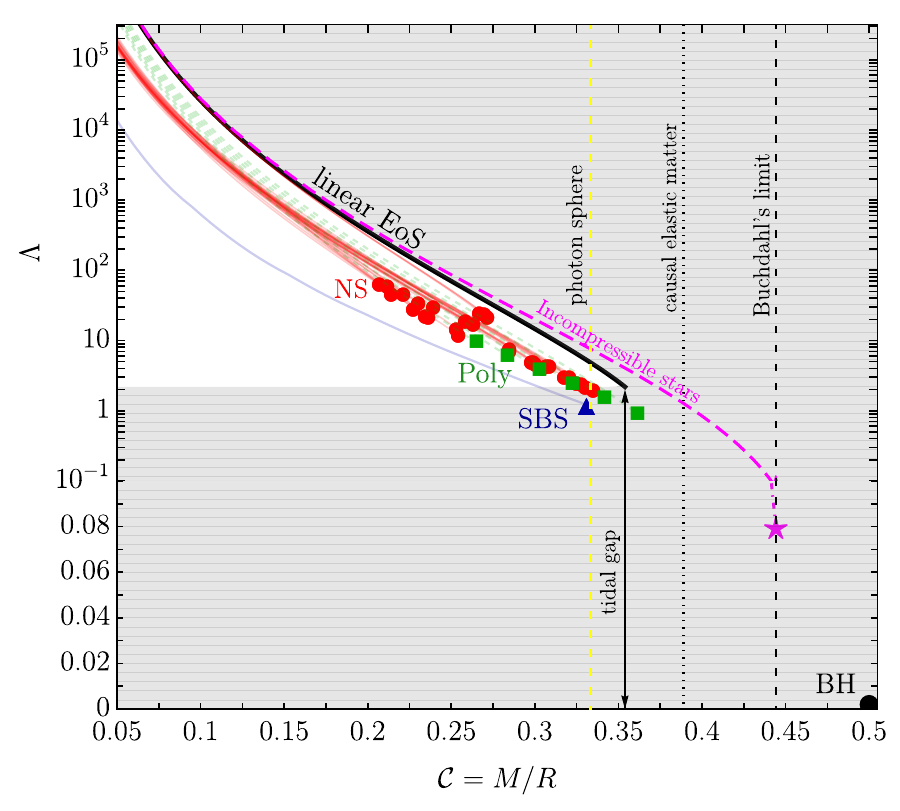}
	\caption{\it  
Tidal deformability $\Lambda$ as function of compactness. 
The vertical lines demarcate the compactness value corresponding to the existence of a photon sphere ($\mathcal{C} = 1/3$), the maximal compactness obtainable with radially stable causal elastic matter ($\mathcal{C} \approx 0.389$) and the Buchdahl's limit  ($\mathcal{C} = 4/9$).
We show the compactness/tidal deformability relation for neutron stars (red lines), ECOs characterized by linear EoS with $\omega = 1$ (solid black line), constant-density stars (dot-dashed magenta line), and 
solitonic boson stars (solid blue lines) with $\xi = 0.186$ (cf. section\,\ref{sec:BS}). The polytropic EoS cases are indicated by green dashed lines, corresponding to $\gamma \in [0.5,1]$.
 For each line, a symbol marks the stable configuration with maximal compactness.
 }
\label{fig:KeyPlotCLambda}
\end{figure}

Of even greater interest is examining what happens in the plane $(\mathcal{C},\Lambda)$, where we depict the tidal deformability $\Lambda$ as a function of compactness. We show our result in 
fig.\,\ref{fig:KeyPlotCLambda}.
Given the significance, both theoretical and phenomenological, of these parameters, we believe that this plane represents the most natural arena in which to understand where ECOs might be located relative to known astrophysical objects. 
Black holes occupy again the bottom-right corner of this parameter space with 
$(\mathcal{C},\Lambda)_{\textrm{BH}} = (0.5,0)$. 
As in the previous figure, we show the curve corresponding to the linear EoS with $\omega = 1$ (solid black line) while neutron stars, with both normal and SQM-type EoS, are described by the red lines.

There are two aspects of this plot that we wish to emphasize. 
\begin{itemize}
\item[{\it i)}] Along the $x$-axis, the presence of a {\it compactness gap}, imposed by the condition $\mathcal{C} < \mathcal{C}_{\textrm{max}}$ that we previously discussed.
\item[{\it ii)}] Along the $y$-axis, we constrain $\Lambda$ according to the condition
\begin{align}
\boxed{
 \Lambda_{\textrm{min}} \equiv 2.186
 < \Lambda < 
 \Lambda_{\textrm{lin\,EoS}}(\mathcal{C})}
 \label{eq:MainBound}
\end{align}
with $\Lambda_{\textrm{lin\,EoS}}(\mathcal{C})$ the tidal deformability given, for fixed compactness, by the linear EoS.
\end{itemize}
We remark that causality-based constraints on the tidal deformability have also been investigated in previous work. In particular, ref.\,\cite{VanOeveren:2017xkv} derived an upper bound on $\Lambda$ for neutron stars by considering the maximally stiff causal equation of state above nuclear density. Our upper bound, obtained from the requirement of relativistic causality together with the compactness limit, is fully consistent with their result and naturally encompasses it: in our approach, the same causal requirement yields a universal upper limit on $\Lambda$ that applies not only to neutron-star matter but also to any physically motivated compact object satisfying the standard energy conditions.

Taken together, {\it i)} and {\it ii)} define a triangular-shaped allowed region in the $(\mathcal{C},\Lambda)$ parameter space, cf. fig.\,\ref{fig:KeyPlotCLambda}. 
Specifically, we observe the presence of a tidal gap, corresponding to the condition $\Lambda <  \Lambda_{\textrm{min}}$, which separates ECOs from black holes. 
 It is interesting to test the validity of the constraint in eq.\,(\ref{eq:MainBound}). Firstly, we note that the condition $\Lambda < \Lambda_{\textrm{lin\,EoS}}(\mathcal{C})$ is violated when considering ECOs described by linear EoS with $\omega > 1$. As before, the violation of causality is saturated by constant-density stars with infinite speed of sound, and in the figure we show the value of their tidal deformability as a function of compactness (magenta dot-dashed line). At the maximum compactness, we find 
 \begin{align}
(\mathcal{C},\Lambda)_{\textrm{Buchdahl}} = (4/9, 8\times 10^{-2})\,.
 \end{align}
 The curves in the $(\mathcal{C}, \Lambda)$ plane representing neutron stars correctly lie within the region constrained by the condition in eq.\,(\ref{eq:MainBound}).\footnote{With the sole exception of the EoS wff1 and wff2, for which we obtain the values $\Lambda_{\textrm{min}} =1.98$ and $\Lambda_{\textrm{min}} =2.17$ respectively, corresponding to the maximum mass on the stable branch. 
 However, as discussed in 
 fig.\,\ref{fig:RealisticEoS}, this is actually consistent with the fact that the EoS wff1 and wff2 violate causality at high density values corresponding to the maximum mass.} 
The polytropic EoS cases are represented, as 
$\gamma$ varies, by green dashed lines. The 
 $\gamma = 0.5$ case violates the causality bound in the high-compactness limit. This is because configurations near the maximum mass violate the DEC, as illustrated in fig.\,\ref{fig:EnergyCondPoly}.

We now anticipate some results that will be derived in section\,\ref{sec:BS}. 
Solitons are solutions of classical field equations with particle-like properties. They are localized in space, have finite energy, and are stable against decay into radiation. 
In quantum field theory, a non-topological soliton refers to a specific type of soliton field configuration that possesses a conserved Noether charge. This Noether charge ensures the stability of the non-topological soliton against transformation into typical particles of the same field. The stability arises from a fundamental energy consideration: for a fixed charge $Q$, the total mass of $Q$ individual free particles exceeds the energy of the non-topological soliton configuration itself. This energy advantage makes the non-topological soliton energetically favorable. 
The simplest examples of non-topological solitons are boson stars, i.e., configurations composed of complex scalar fields $\Phi$ minimally coupled to gravity and possessing a $U(1)$ global  symmetry. 
In particular, we focus on the case of the so-called solitonic boson stars\,\cite{Friedberg:1986tq}. 
We refer the reader to section\,\ref{sec:BS} for a detailed discussion of the properties of solitonic boson stars. Here, we instead focus on their position in the $(\mathcal{C}, \Lambda)$ diagram.
The curve of the solitonic boson stars is depicted in blue  in fig.\,\ref{fig:KeyPlotCLambda}. 
We focus on solutions that achieve the maximum possible compactness without encountering instabilities.
We observe that the curve describing the tidal deformability as a function of compactness reaches, at the maximum compactness, a value of $\Lambda$ that slightly violates the causality bound (specifically, we find $\Lambda_{\textrm{SBS,min}} = 1.21$). 
Consequently, this implies that in some way, solitonic boson stars violate one or more of the assumptions underlying the validity of our bound. 
We will address this issue in detail in section\,\ref{sec:BS}. 
This will provide us with the opportunity to understand under which circumstances it is possible to improve our lower limit on $\Lambda$. For the moment, we will focus on some observational/phenomenological consequences of our study.

\noindent
\section{Exotic compact objects in the sub-solar and solar mass range}\label{sec:Pheno}

In this section, we aim to connect the theoretical considerations discussed in the previous section with potential observational consequences. Specifically, in section\,\ref{sec:graveyard}, we will begin by exclusively considering the mass distribution of ECOs in relation to that of neutron stars and black holes. Then, in section\,\ref{sec:Fisher}, we will move on to discuss tidal deformability.

\subsection{ECOs in the stellar graveyard}\label{sec:graveyard}

The plot \href{https://www.ligo.caltech.edu/news/ligo20250826}{``masses in the stellar graveyard''}  produced by the LIGO/Virgo/KAGRA collaboration represents the distribution of masses for various compact objects detected through gravitational wave merger events.
These merger events include black hole binary systems, neutron star binary systems, systems where one component is a black hole and the other is a neutron star, and events of uncertain/mixed category in which the nature of the detected objects (whether they are neutron stars or black holes) is not definitively determined.   

\begin{figure*}[!t]
	\centering
\includegraphics[width=0.995\textwidth]{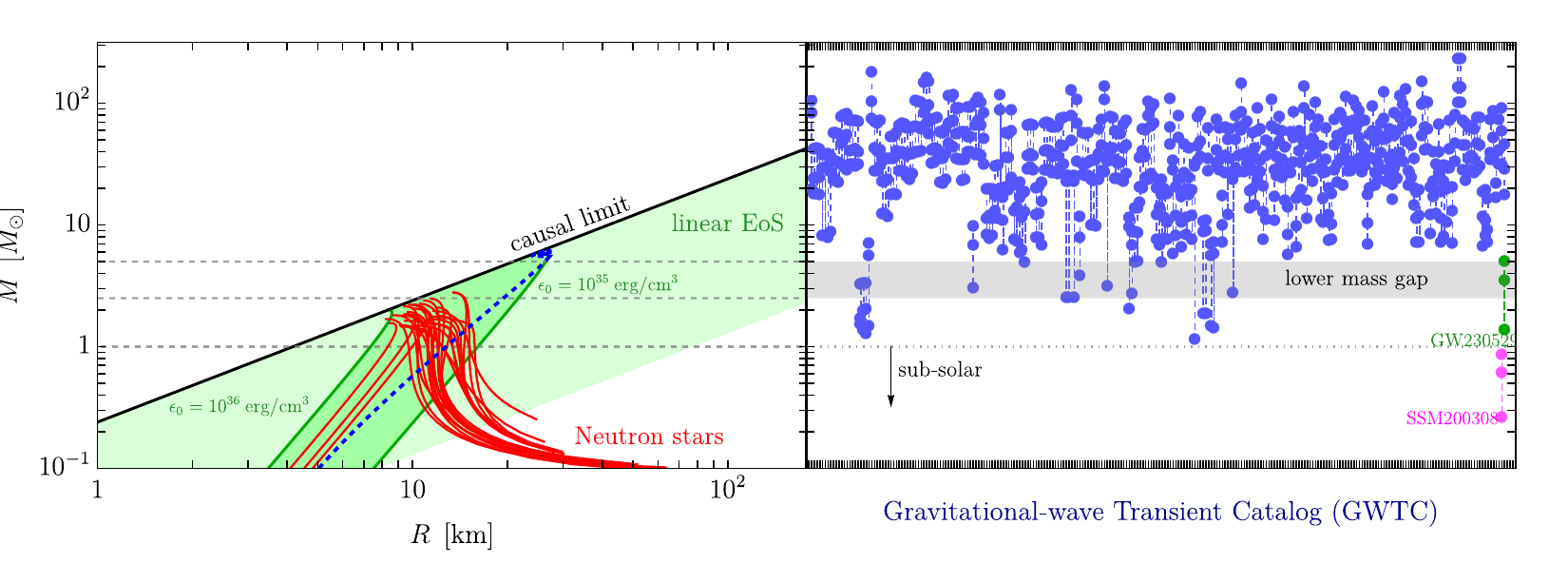}
	\caption{\it
Left panel. 
Mass–radius comparison. Red curves are neutron-stars. The green shaded band corresponds to the family of linear EoSs $\epsilon(P) = \epsilon_0 + P$, with the darker green band spanning 
$\epsilon_0 \in [10^{35},10^{36}]$ erg/cm$^3$. 
The black solid line marks the causal limit. The dashed blue line refers to solitonic boson stars with mass parameter $\mu = 3\times 10^{-10}$ eV and $f/\MPl = 0.186$.
Right panel.
\href{https://gwosc.org/eventapi/html/GWTC/}{Gravitational-wave Transient Catalog (GWTC)}. 
We also include: {\it i)} in magenta the sub-solar mass candidate
SSM200308, possibly composed by two sub-solar BHs with
masses $m_1 = 0.62^{+0.46}
_{-0.20}$ $M_{\odot}$ and 
$m_2 = 0.27^{+0.12}_{-0.10}$ $M_{\odot}$ and total mass $M = 0.88^{+0.35}_
{-0.08}$ $M_{\odot}$, cf. ref.\,\cite{Prunier:2023cyv}; 
{\it ii)} in green the recently reported observation GW230529, composed by 
 compact objects with mass $m_1 = 3.6^{+0.8}_{-1.2}$ $M_{\odot}$ and $m_2 = 1.4^{+0.6}_{-0.2}$ $M_{\odot}$ 
 and total mass $M = 5.1^{+0.6}_{-0.6}$ $M_{\odot}$, cf. ref.\,\cite{LIGOScientific:2024elc}.
}
\label{fig:MassComparison}
\end{figure*}

In the right panel of fig.\,\ref{fig:MassComparison}, we have reproduced the mass distributions (blue dots) of events reported in the Gravitational-wave Transient Catalog (GWTC), including those from the GWTC-4 release\,\cite{LIGOScientific:2025slb}.  
There are two specific mass ranges that are of particular interest.

The {\it lower mass gap} refers to the range of masses between the heaviest known neutron stars and the lightest known black holes, where few or no compact objects have been detected. 
This gap typically spans approximately from $2.5$ to $5$ $M_{\odot}$. 
The lower boundary of this region corresponds to the fact that neutron stars, remnants of supernova explosions, are tipically observed with masses up to about 2.5 $M_{\odot}$. This boundary of the lower mass gap has a firm theoretical understanding. 
When massive stars reach the end of their life cycle and exhaust their nuclear fuel, they undergo a catastrophic collapse, leading to the formation of a neutron star or a black hole, depending on the mass of the progenitor star. In the case of neutron stars, the collapse is halted by neutron degeneracy pressure, resulting in a compact object mainly composed  of neutrons. The maximum mass that a neutron star can attain is determined by the balance between the gravitational force trying to collapse the star further and the pressure exerted by neutron degeneracy. Although the exact value of this maximum mass depends on the specific EoS used to describe nuclear matter beyond the approximation of exact Fermi degeneracy, the analysis in fig.\,\ref{fig:NeutronStarMassRadius} demonstrates that indeed the maximum mass that a neutron star can attain is around $2$ to $2.5$ $M_{\odot}$.
On the other hand, the upper boundary of the lower mass gap---thus referring to a potential lower limit regarding the mass of black holes resulting from stellar collapse---is mostly based on theoretical considerations of stellar evolution and core-collapse physics that seems to suggest that astrophysical black holes with masses below a certain threshold might not form through conventional mechanism. The exact value of this threshold depends on factors such as the progenitor star's metallicity, rotation rate, and mass loss during its evolution and it is thought to be around 5 $M_{\odot}$. 
These theoretical considerations appear to be supported by measurements of black hole masses inferred from X-ray binary systems, although in this case
limitations in the observational methods 
make it difficult to detect objects at such low masses. Although, due to these uncertainties, it is not possible to firmly exclude the presence of astrophysical black holes in the lower mass gap, this remains a promising mass range that could allow us to distinguish neutron stars and black holes from a potential population of ECOs.

The {\it sub-solar region}. 
Although, as shown in fig.\,\ref{fig:NeutronStarMassRadius}, it is quite possible for neutron stars to have a sub-solar mass, it is also true that \href{https://stellarcollapse.org/index.php/nsmasses.html}{all neutron stars observed so far} have masses above 1 $M_{\odot}$\,\cite{Lattimer:2012nd,Ozel:2016oaf}.  
This observational evidence is supported by standard formation scenarios, which suggest that astrophysical compact objects typically have masses above $1$ $M_{\odot}$\,\cite{Suwa:2018uni}. 
Notably, the possible detection of a merger event in the sub-solar region is therefore considered one of the smoking-gun signature of a population of primordial black holes\,\cite{Green:2020jor}.

For these reasons, we have also included (last two data points) {\it i)} the sub-solar mass candidate
SSM200308\,\cite{Prunier:2023cyv}, possibly composed by two sub-solar BHs 
with
masses $m_1 = 0.62^{+0.46}
_{-0.20}$ $M_{\odot}$ and 
$m_2 = 0.27^{+0.12}_{-0.10}$ $M_{\odot}$
 and {\it ii)} the recently reported observation GW230529\,\cite{LIGOScientific:2024elc}, composed by 
 compact objects with mass $m_1 = 3.6^{+0.8}_{-1.2}$ $M_{\odot}$ and $m_2 = 1.4^{+0.6}_{-0.2}$ $M_{\odot}$. 
As claimed in ref.\,\cite{LIGOScientific:2024elc}, GW230529 provides further evidence of the existence of a population of compact objects with masses between the heaviest neutron stars and the lightest black holes observed in the Milky Way. Whether these are unexpectedly light astrophysical black holes, primordial black holes, or other types of ECOs remains one of the most pressing questions to be investigated.

In the left panel of fig.\,\ref{fig:MassComparison}, we confront the mass distribution of the observed events with theoretical models. 
The red lines correspond to the mass-radius relations for neutron stars (both with normal and SQM-type EoS, cf. fig.\,\ref{fig:NeutronStarMassRadius}). 
The light green band corresponds to the region of the mass-radius plane that can be populated by ECOs describable by a linear EoS. 
More specifically, this region is obtained by considering solutions of the TOV system, varying the central pressure (within the range $10^{-2} \leqslant P_c/\epsilon_0 \leqslant P_c^{\textrm{max}}$) and the parameter $\epsilon_0$. 
The darker green region is bounded by the values $10^{35}
\leqslant \epsilon_0\,[\textrm{erg}/\textrm{cm}^3]\leqslant 10^{36}$. 
The upper boundary of the green region (indicated by the solid black line) beyond which the central pressure would exceed the maximum value, defines the causality bound, cf. eq.\,(\ref{eq:CauslityMassRa}). 

The key message of this plot is as follows. As clearly shown in the figure, ECOs described by a linear EoS have the potential to populate a region of the mass-radius plane that encompasses both the lower mass gap and the sub-solar region, thus becoming potential candidates for the interpretation of any gravitational wave signal that lies within these mass ranges.

It is natural to ask, beyond the description made in terms of the linear EoS, whether there are explicit models of ECOs that indeed exhibit the same behavior. 
To answer this question, we preview here the results of a discussion that will be addressed in greater detail in section\,\ref{sec:BS}. 
Let us consider again, as discussed at the end of section\,\ref{sec:Tidal}, the case of solitonic boson stars. In the left panel of fig.\,\ref{fig:MassComparison}, we show the mass-radius relation for a specific realization of solitonic boson stars with a dashed blue line (see caption for details). 
As seen, this specific realization of the model falls precisely within the region previously identified by the linear EoS, again covering both the lower mass gap and the sub-solar region. 
The origin of this correspondence lies in the fact that, at least in terms of the background solution, solitonic boson stars, especially in the limit of maximum compactness, effectively behave like a fluid described by a linear EoS.

This discussion motivates the realistic possibility that the lower mass gap or the sub-solar region could contain a population of ECOs. This brings us back to the initial question.
Assuming we observe events within these mass ranges, how can we determine whether these events involve black holes or other ECOs?

\subsection{Tidal deformability and observational consequences}\label{sec:Fisher}

We begin with some general considerations. 
\begin{figure}[!t]
	\centering
\includegraphics[width=0.495\textwidth]{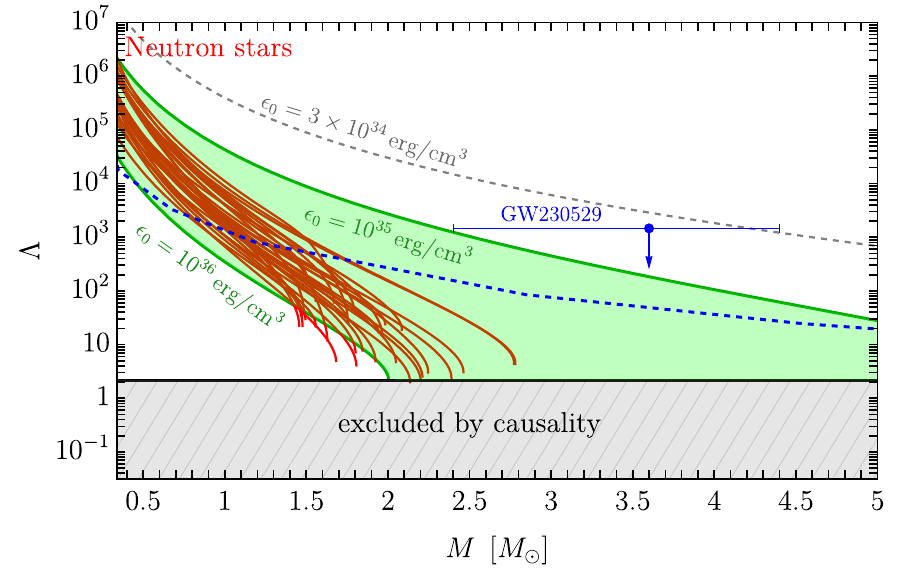}
	\caption{\it 
Tidal deformability parameter $\Lambda$ as function of mass in the sub-solar and lower mass gap range. 
The region shaded in gray corresponds to $\Lambda < \Lambda_{\textrm{min}}$, cf. eq.\,(\ref{eq:CausalTidal}).
The red lines correspond to the tidal deformability of neutron stars (both with normal EoS and SQM-type EoS) while the green band corresponds to ECOs with 
$10^{-2} \leqslant P_c/\epsilon_0 \leqslant P_c^{\textrm{max}}$ and  $10^{35}
\leqslant \epsilon_0\,[\textrm{erg}/\textrm{cm}^3]\leqslant 10^{36}$.
The blue dot with horizontal error bars corresponds to the upper limit 
$\Lambda \leqslant 1462$ quoted for the event GW230529, cf. ref.\,\cite{LIGOScientific:2024elc}. The dashed blue line corresponds to solitonic boson stars with mass parameter $\mu = 1.8\times 10^{-10}$ eV.
 }
\label{fig:TidalWithGW230529}
\end{figure}
In fig.\,\ref{fig:TidalWithGW230529},  we show the tidal deformability $\Lambda$ as a function of mass. We focus on mass values that encompass both the lower mass gap and the sub-solar region. The region shaded in gray is excluded 
by our causality bound. 
The red lines describe the tidal deformability of neutron stars, while the green shaded band corresponds to the linear equation of state and is obtained by varying $10^{-2} \leqslant P_c/\epsilon_0 \leqslant P_c^{\textrm{max}}$ and $10^{35}
\leqslant \epsilon_0\,[\textrm{erg}/\textrm{cm}^3]\leqslant 10^{36}$ (cf. fig.\,\ref{fig:MassComparison}). 
The main message of this figure is that ECOs described by a linear equation of state are characterized by a value of tidal deformability that could be large enough to be experimentally measured. This applies both to objects with sub-solar mass (where, as expected, the tidal deformability also becomes very large for neutron stars) and to masses within the lower mass gap. 

To quantify this claim, we begin---before subsequently discussing a more thorough statistical analysis---with some considerations prompted by the recent discovery of GW230529. 
Ref.\,\cite{LIGOScientific:2024elc} investigates the tidal constraints for both the primary and secondary components using waveform models that account for tidal effects. Regardless of whether GW230529 is analyzed with a neutron star/black hole model assuming only the tidal deformability of the primary compact object to be zero or a binary neutron star model that includes the tidal deformability of both objects, it is found that the tidal deformability of the secondary object remains unconstrained. On the other hand, 
the dimensionless tidal deformability of
the primary peaks at zero, consistent with a black hole. 
Nevertheless, an upper limit on the tidal deformability of the primary object is still extracted, with $\Lambda_1 \leqslant 1462$ at 90\% credibility.\footnote{In ref.\,\cite{LIGOScientific:2024elc},
the IMRPhenomPv2\_NRTidalv2 model is utilized, which permits tidal deformability on both merging objects but does not simulate tidal disruption. 
}
For illustrative purposes, we display this upper bound in fig.\,\ref{fig:TidalWithGW230529}. 
Although it is an upper bound rather than a true measurement, it is still noteworthy how it can provide relevant information for the case of ECOs. 
For example, ECOs described by a linear EoS with $\epsilon_0 \lesssim 3\times 10^{34}$ would be disfavoured, as they are characterized by a tidal deformability that could be ruled out (cf. the dashed gray line in fig.\,\ref{fig:TidalWithGW230529}). 
Motivated by these considerations and the hope that in the future an increasing number of events like GW230529 may be detected, we proceed to make some more quantitative considerations through a Fisher information matrix analysis.

\subsubsection{Details of the waveform model}

In the frequency domain, 
the gravitational
waveform can be obtained in the approximation
of the stationary phase\,\cite{Finn:1992xs}, and 
takes the form
\begin{align}
\tilde{h}(f) = \underbrace{\frac{M_c^{5/6}}{
\pi^{2/3}D_{\textrm{eff}}
}\sqrt{\frac{5}{24}}}_{\equiv\,\mathcal{A}}f^{-7/6}e^{i\Psi(f)} 
= \mathcal{A}f^{-7/6}e^{i\Psi(f)}\,.\label{eq:MainWaveform}
\end{align}
We define $\MT = M_1+M_2$ the total mass of the binary, $\eta = M_1 M_2/\MT^2$ the symmetric mass ratio and 
$M_c=\eta^{3/5}\MT$ the chirp mass.
$D_{\textrm{eff}}$ is the effective distance of the source.\footnote{
One tipycally defines 
$D_{\textrm{eff}}\equiv D_L/C_p$ where
$D_L$ is the luminosity distance and 
$C_p\equiv \frac{1}{2}\sqrt{(1+\cos^2\imath)F_+^2 + 
4\cos^2\imath F_{\cross}^2}$. 
The inclination angle is defined by $\imath = 
\cos^{-1}(\hat{\boldsymbol{k}}\cdot\hat{\boldsymbol{L})}$---being $\hat{\boldsymbol{k}}$ the propagation direction (that points from the source to the observer) and $\boldsymbol{L}$ the orbital angular momentum of the binary---and describes the orbital plane orientation.
The antenna pattern functions $F_+$ and 
$F_{\cross}$  describe the detector's  sensitivity to the $h_{+}$ and $h_{\cross}$ polarization
amplitudes for different source positions and orientations. 
Their explicit expression is given by 
\begin{align}
F_+ = \frac{1}{2}
(1+\cos^2\theta)
\cos(2\phi)\cos(2\psi) -
\cos\theta\sin(2\phi)\sin(2\psi)\,,\nn\\
F_{\cross} = \frac{1}{2}
(1+\cos^2\theta)
\cos(2\phi)\sin(2\psi) +
\cos\theta\sin(2\phi)\cos(2\psi)\,,
\end{align}
and, therefore, they depend on the source direction in the sky $(\theta,\phi)$  
(with the polar angle $\theta$ and azimuthal angle $\phi$ in the spherical coordinate system where the  $x$- and $y$-axis  coincide with the detector arms. If the source position is defined in the equatorial coordinate system, the polar and azimuthal angles are replaced by the right ascension $\alpha \in [-\pi,\pi]$ and declination $\delta \in [-\pi/2,\pi/2]$)
and the so-called polarization angle $\psi$.
}
For slow orbital velocity and weak gravitational
field, the wave phase is expanded as a power series in the post-Newtonian (PN)
orbital velocity parameter $v\equiv (\pi f \MT)^{1/3}$. 
A term proportional to $v^{2n}$ corresponds to the $n$-PN order of the 
approximation. 
We use the standard TaylorF2 waveform  augmented with the 5PN and 6PN tidal terms in the phase\,\cite{Wade:2014vqa,Lackey:2014fwa}. 
The phase is expressed as the
PN expansion
\begin{align}
\Psi(f) = &
2\pi ft_c -\phi_c -\frac{\pi}{4}+
 \nn\\
 &
 \frac{3}{128\eta v^{5}}
 \left(1 + \Psi^{\textrm{circ}}_{3.5\textrm{PN}} + 
 \Psi^{\textrm{spin}}_{2\textrm{PN}} 
 + 
 \Psi^{\textrm{Tidal}}_{6\textrm{PN}}\right)\,,
\end{align}
where $t_c$ and $\phi_c$ are the coalescence time and the coalescence phase.  
 In the bracket, $\Psi^{\textrm{circ}}_{3.5\textrm{pN}}$ 
 refers to the phase term associated with the circular approximation, computed to the $3.5$-PN order, 
 $\Psi^{\textrm{spin}}_{2\textrm{PN}}$ to spin effects up to $2$-PN order
 and $\Psi^{\textrm{Tidal}}_{6\textrm{PN}}$ represents the phase contribution due to the quadrupolar tidal interaction between the components of the binary system. 
The standard 3.5-PN circular contribution takes the form
\begin{align}
\Psi^{\textrm{circ}}_{3.5\textrm{PN}} = 
\sum_{n=2}^{7}c_n(\eta)v^n\,,
\end{align}
and we use the result of ref.\,\cite{Buonanno:2009zt} that we report, for completeness, in the following
\begin{align}
c_2& = \frac{20}{9}
\bigg(\frac{743}{336} + \frac{11}{4}\eta\bigg)\,,\\
c_3& = -16\pi\,,\\
c_4& = 
10
\bigg(\frac{3058673}{1016064} + \frac{5429}
{1008}\eta + \frac{617}{144}\eta^2\bigg)\,,
\\
c_5& =\pi
\bigg(\frac{38645}{756} - \frac{65}{9}\eta\bigg)\left[
1+3\log\left(\frac{v}{v_{\textrm{lso}}}\right)
\right]\,,\\
c_6& = 
\frac{11583231236531}
{4694215680} - \frac{640}{3}\pi^2 -
\frac{6848}{21}\gamma_{e} - \frac{6848}{
21}\log(4v)\nn\\
&
+\bigg(-\frac{15737765635}{3048192} + \frac{2255}{12}\pi^2
\bigg)\eta + \frac{76055}{1728}\eta^2 \nn\\
&- 
\frac{127825}{1296}\eta^3\,,
\\
c_7& =\pi\bigg(
\frac{77096675}{254016} + \frac{378515}{
1512}\eta -\frac{74045}{756}\eta^2
\bigg)\,,
\end{align}
with $\gamma_e \approx 0.577$ is the Euler constant. 
We use $v_{\textrm{lso}} = 1/\sqrt{6}$ (corresponding to the last stable orbit of the Schwarzschild metric).

The spins of the two compact objects are
indicated with $\boldsymbol{S}_{1,2}$. 
We indicate with
$\hat{\boldsymbol{L}}_N$ the unit vector in the direction of the binary's orbital angular momentum.  
We also define the dimensionless spin 
of the $i^{th}$ body as $\boldsymbol{\chi}_i \equiv 
\boldsymbol{S}_{i}/M_i^2$. 
We  have
\begin{align}
\Psi^{\textrm{spin}}_{2\textrm{PN}} & = 
4\beta_{1.5}v^3 - 10\sigma v^4\,,  
\end{align}
where $\beta_{1.5}$ is the 1.5-PN spin-orbit term\,\cite{Kidder:1995zr}
\begin{align}
\beta_{1.5} = 
\sum_{i=1,2}
\boldsymbol{\chi}_i\cdot 
\hat{\boldsymbol{L}}_N
\bigg(
\frac{113}{12}\frac{M_i^2}{\MT^2} +
\frac{25}{4}\eta
\bigg)\,.
\end{align}
The 2-PN term includes three contributions, $\sigma = \sigma_{s_1s_2} + \sigma_{\textrm{QM}} + \sigma_{\textrm{self spin}}$. 
The spin-spin interactions is\,\cite{Poisson:1997ha} 
\begin{align}
\sigma_{s_1s_2} = 
\frac{\eta}{48}
\big[721
(\boldsymbol{\chi}_1\cdot \hat{\boldsymbol{L}}_N)
(\boldsymbol{\chi}_2\cdot \hat{\boldsymbol{L}}_N)- 
247\boldsymbol{\chi}_1\cdot \boldsymbol{\chi}_2\big]\,.\label{eq:spinspin}
\end{align}
The quadrupole-monopole term is\,\cite{Poisson:1997ha}
\begin{align}
\sigma_{\textrm{QM}} = 
-\frac{5}{2}\sum_{i=1,2}
\frac{\bar{\mathcal{Q}}_i M_i^2}{\MT^2}
|\boldsymbol{\chi}_i|^2
\left[
3(\hat{\boldsymbol{L}}_N\cdot 
\hat{\boldsymbol{S}}_i)^2 - 1
\right]\,,\label{eq:spinQM}
\end{align}
where $\mathcal{Q}_i$ is the general-relativistic quadrupole-moment
scalar and we introduced the 
normalized quantity 
$\bar{\mathcal{Q}}_i 
\equiv \mathcal{Q}_i/M_i^3|\boldsymbol{\chi}_i|^2$.
The quadrupole moment describes the deviation of the mass distribution from spherical symmetry due to rotation.
In the case of a Kerr black hole, the quadrupole moment is given by 
$\mathcal{Q}_i = -|\boldsymbol{\chi}_i|^2 M_i^3$ (that is, $\bar{\mathcal{Q}}_i=-1$); the negative sign indicates  the oblate shape caused by the rotation of the black hole. 
Quadrupole moment of neutron stars were computed in ref.\,\cite{Laarakkers:1997hb} where it is shown that for a neutron star with mass
$1.4$ $M_{\odot}$ $\mathcal{Q}_i = -a|\boldsymbol{\chi}_i|^2 M_i^3$, 
with the parameter $a$ ranging from $4$ to $8$, depending on the neutron star EoS. 

\begin{figure}[!t]
	\centering
\includegraphics[width=0.495\textwidth]{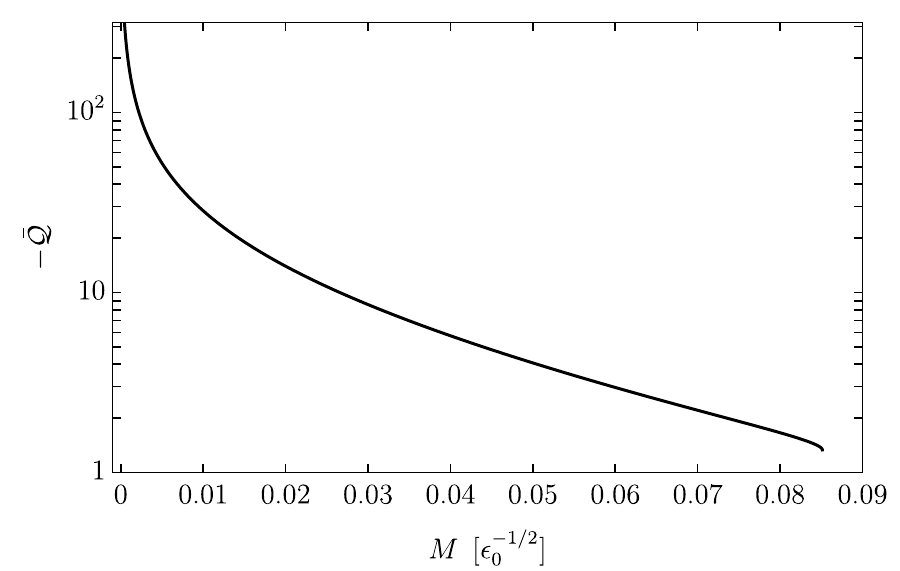}
	\caption{\it  
Dimensionless quadrupole moment computed for an ECO described by the linear EoS. 
 At the maximum mass, we find the causality limit 
 $-\bar{\mathcal{Q}}_{\textrm{min}} = 1.336$.
 }
\label{fig:QuadrupoleLinear}
\end{figure}

The possibility of investigating deviations from the black hole limit in the spin-induced quadrupole moment is an important test that may help distinguish black holes from neutron stars or ECOs, cf. ref.\,\cite{Krishnendu:2017shb,LIGOScientific:2020tif}. 
This perspective is particularly intriguing since the spin-induced quadrupole moment affects the waveform at a lower PN order compared to tidal effects. 
The accuracy of measuring spin-induced quadrupole moments is significantly 
influenced by the masses and spins of the binary system. Although there is a degeneracy, the inclusion of spin terms at various PN orders and non-spinning PN coefficients assists in mitigating correlations of $\bar{\mathcal{Q}}$ with spin and mass parameters, thereby enabling its measurement in spinning binary systems, cf. ref.\,\cite{Krishnendu:2019tjp,Lyu:2023zxv,Chia:2020psj}.
With these considerations in mind, we compute the quadrupole moment for the specific case of a linear EoS. We 
adopt the slow-rotation Hartle-Thorne approximation\,\cite{Hartle:1968si}, and show our result in fig.\,\ref{fig:QuadrupoleLinear}. 
In the limit of maximum compactness, $\bar{\mathcal{Q}}$ approaches the value of a black hole. For smaller masses, however, it may also deviate from the latter by an order of magnitude or more. 

The self-spin interaction 
\begin{align}
\sigma_{\textrm{self spin}}=
\frac{1}{96}
\sum_{i=1,2}
\left(\frac{M_i}{\MT}\right)^2
\bigg[7|\boldsymbol{\chi}_i|^2 - (
\boldsymbol{\chi}_i\cdot 
\hat{\boldsymbol{L}}_N
)^2\bigg]\,.\label{eq:selfspin}
\end{align}

The analysis with the TaylorF2 waveform is restricted to the case of spin aligned or spin anti-aligned. 
This means that we have 
$\boldsymbol{\chi}_i\cdot \hat{\boldsymbol{L}}_N = \pm |\boldsymbol{\chi}_i|$. 
In both cases, the spins do not have any components perpendicular to $\hat{\boldsymbol{L}}_N$ which ensures that the spins do not induce precession in the binary system.

In a binary neutron star system, the tidal deformability contribution
is given up to 6-PN order as\,\cite{Wade:2014vqa,Lackey:2014fwa}
\begin{align}
& \Psi^{\textrm{Tidal}}_{6\textrm{PN}}(f) = \nn\\
&-\left[
\frac{39\tilde{\Lambda}}{2}
v^{10} 
 +
 \left(
 \frac{3115}{64}\tilde{\Lambda}
 -
 \frac{6595}{364}\sqrt{1-4\eta}\delta\tilde{\Lambda}
 \right)v^{12}
\right]\,,
\end{align}
where the reduced tidal deformability 
$\tilde{\Lambda}$ and the asymmetric
tidal correction $\delta\tilde{\Lambda}$ are defined as
\begin{align}
\tilde{\Lambda} = 
\frac{8}{13}
\big[&\big(1+7\eta-31\eta^2\big)(\Lambda_1 + \Lambda_2)
\nn\\
&+
\sqrt{1-4\eta}\big(1+9\eta-11\eta^2\big)
(\Lambda_1 - \Lambda_2)\big]\,,
\end{align}
\begin{align}
\delta\tilde{\Lambda} = 
&\frac{1}{2}\bigg[
\sqrt{1-4\eta}\bigg(
1-\frac{13272}{
1319}\eta + 
\frac{8944}{1319}\eta^2\bigg)
(\Lambda_1 + \Lambda_2)\nn\\
&+
\bigg(
1-\frac{15910}
{1319}\eta + \frac{32850}{1319}\eta^2 + \frac{3380}{1319}\eta^3
\bigg)(\Lambda_1 - \Lambda_2)\bigg]\,,
\end{align}
with $M_1 \geqslant M_2$. 
The subscripts $1$ and $2$ indicate the individual neutron stars, and $\Lambda_{i}$ represents the dimensionless tidal deformabilities. 
Typically, $\delta\tilde{\Lambda}/\tilde{\Lambda} = O(10^{-2})$\,\cite{Favata:2013rwa}, and the contribution from  $\delta\tilde{\Lambda}$ is negligible. 

All in all, the waveform model described above depends on the following set of parameters 
\begin{align}
\boldsymbol{\vartheta} = \{
{\color{azure}{D_L,\imath,\theta,\phi,\psi,
t_c,\phi_c}},
{\color{harvardcrimson}{M_1,M_2,\chi_1,\chi_2,
\tilde{\Lambda},
\delta\tilde{\Lambda},
\bar{\mathcal{Q}}_1,
\bar{\mathcal{Q}}_2}}
\}\,.\label{eq:ParameterVector}
\end{align}
The parameters in azure are the so-called extrinsic parameters, while those in crimson red are intrinsic. Intrinsic parameters describe the physical properties of the gravitational wave source itself and are independent of the observer's location. On the other hand, extrinsic parameters are those that describe the orientation and location of the binary system with respect to the observer.  
The extrinsic parameters $\{
{\color{azure}{D_L,\imath,\theta,\phi,\psi}}
\}$ only enters in the amplitude of the waveform (cf. eq.\,(\ref{eq:MainWaveform})) and can be treated separately. 
More in detail, the correlations between 
these parameters and, in particular, the intrinsic  
parameters are negligible, so their inclusion will  not substantially change the measurement
errors of the intrinsic parameters. 
For this reason, we consider the waveform in the form  $\tilde{h}(f) = \mathcal{A}f^{-7/6}e^{i\Psi(f)}$, 
and fix the amplitude. 
This can be done in two ways. 
One possibility is to consider the luminosity distance $D_L$ of the detected event as fixed.
Assuming optimally oriented sources (that is, setting $(1+\cos^2\imath)F_+^2 + 
4\cos^2\imath F_{\cross}^2 = 4$), this gives $D_L = D_{\textrm{eff}}$. The amplitude, therefore, is fixed for a given value of the chirp mass characterizing the event. Alternatively, one can fix the so-called signal-to-noise ratio (SNR) of the event and extract the corresponding amplitude. 
The SNR of a gravitational wave event is a measure of how strong the gravitational wave signal is compared to the background noise in the detector. Intuitively, the SNR represents how clearly the gravitational wave signal can be distinguished from the random noise present in the detector data. Formally, it can be computed through the integral
\begin{align}
\textrm{SNR}^2 = 
4\int_{f_{\textrm{min}}}^{f_{\textrm{max}}}
\frac{|\tilde{h}(f)|^2}{S_n(f)}df \equiv \rho^2\,,
\end{align}
where $S_n(f)$ is the one-sided noise power spectral density of the detector, describing how the noise is distributed across different frequencies, and thus indicating the sensitivity of the instrument at each frequency. 
The integral is cut off by $f_{\textrm{min}}$ and $f_{\textrm{max}}$ to account for the frequency range in which the detector is sensitive and the signal is expected to lie. 
\begin{figure}[!t]
	\centering
\includegraphics[width=0.495\textwidth]{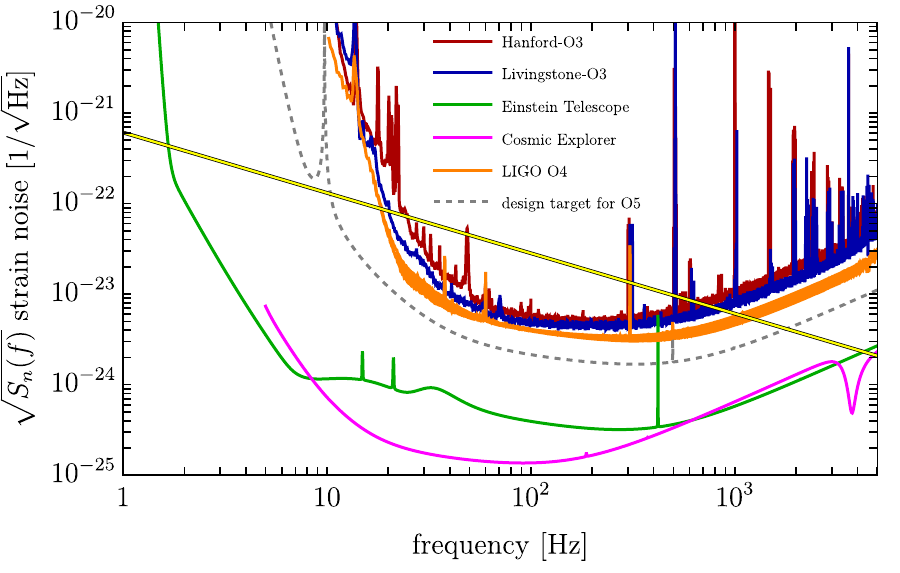}
	\caption{\it  
 Present and projected sensitivity curves for the Livingston and Hanford LIGO experiments during the O3, O4 and O5 observation runs, as well as for the next-generation Einstein Telescope and Cosmic Explorer.
 The yellow line corresponds to the value of $2|\tilde{h}(f)|\sqrt{f}$ for an event like GW230529 with $\textrm{SNR} \approx 10$.
 }
\label{fig:Noise}
\end{figure}
In fig.\,\ref{fig:Noise}, we show 
the square root of the noise power spectral density for different experiment (current and future) relevant to our analysis (see the caption for details). 
To provide an intuitive understanding of the signal strength relative to the noise in the detector,  
in fig.\,\ref{fig:Noise} we compare 
$\sqrt{S_n(f)}$ with 
$2|\tilde{h}(f)|\sqrt{f}$ considering a merger event with properties comparable with those of GW230529. 
In this case, the amplitude of the waveform is fixed so to get $\rho \simeq 10$ in the case of the sensitivity curve of the LIGO experiment during the O4 observation run. 
In the following, in order to encompass present and future prospects, we will focus on two possibilities. On the one hand, we consider the noise spectral density of Advanced LIGO (aLIGO) in the so-called Zero-Detuned High Power configuration\,\cite{LIGO:SensitivityCurves}. On the other hand, we consider the noise curve of the Einsten Telescope (ET),  a proposed third-generation gravitational-wave observatory\,\cite{Hild:2010id}.  
As far as lower frequency cut off is concerned, we consider $f_{\textrm{min}}^{\textrm{aLIGO}} = 10$ Hz and 
$f_{\textrm{min}}^{\textrm{ET}} = 1$ Hz. 
For the upper value, we choose in both cases the frequency at the innermost circular orbit, 
$f_{\textrm{max}}^{\textrm{aLIGO}} = f_{\textrm{max}}^{\textrm{ET}} = f_{\textrm{ISCO}}$ with 
\begin{align}
f_{\textrm{ISCO}} = 
(6^{3/2} \MT \pi)^{-1} = 
    4.4\left(\frac{M_{\odot}}{\MT}\right)\,\textrm{kHz}\,.\label{eq:fISCO}
\end{align}

However, it should be noted that binaries of stellar objects may have smaller maximal frequencies
because the less compact companion can be tidally disrupted during the inspiral. 
We can give an estimate of the tidal disruption radius (and the corresponding disruption frequency) as follows.
In a binary stellar system, the disruption radius 
is the distance from one star at which the tidal gravitational forces from its companion become strong enough to disrupt the star. This means that the gravitational force exerted by the companion star exceeds the self-gravitational force holding the disrupted star together, potentially leading to its disintegration. 
To derive the tidal disruption radius, we equate the tidal force exerted by the companion star to the self-gravitational force of the disrupted star\,\cite{1983bhwdbookS}. 
One finds, for the tidal disruption radius $R_{T,i}$ of the $i^{\text{th}}$ star in the binary system
\begin{align}
R_{T,i} = \left( \frac{2M_j}{M_i} \right)^{1/3}R_i\,,
\end{align}
where $M_j$ is the mass of the companion star, 
$R_i$ and $M_i$ are the radius and mass of the star being disrupted.
Consequently, we estimate the 
tidal disruption frequency to be\,\cite{Crescimbeni:2024cwh}
\begin{align}
f_T = \frac{1}{\pi}\sqrt{
\frac{\MT}{[\textrm{max}(
R_{T,1},R_{T,2}
)]^3}
}\label{eq:fT}
\end{align}
Assuming an equal-mass binary composed by ECOs, we find the relation
\begin{align}
f_T = 12\sqrt{6}\,\mathcal{C}^{3/2}f_{\textrm{ISCO}}
\approx 1.7
\left(\frac{\mathcal{C}}{0.15}
\right)^{3/2}
f_{\textrm{ISCO}}\,.
\end{align}
We observe how the tidal disruption frequency exceeds that of ISCO, unless we consider ECOs with compactness $\mathcal{C}\lesssim 0.1$.
\begin{figure}[!t]
	\centering
\includegraphics[width=0.495\textwidth]{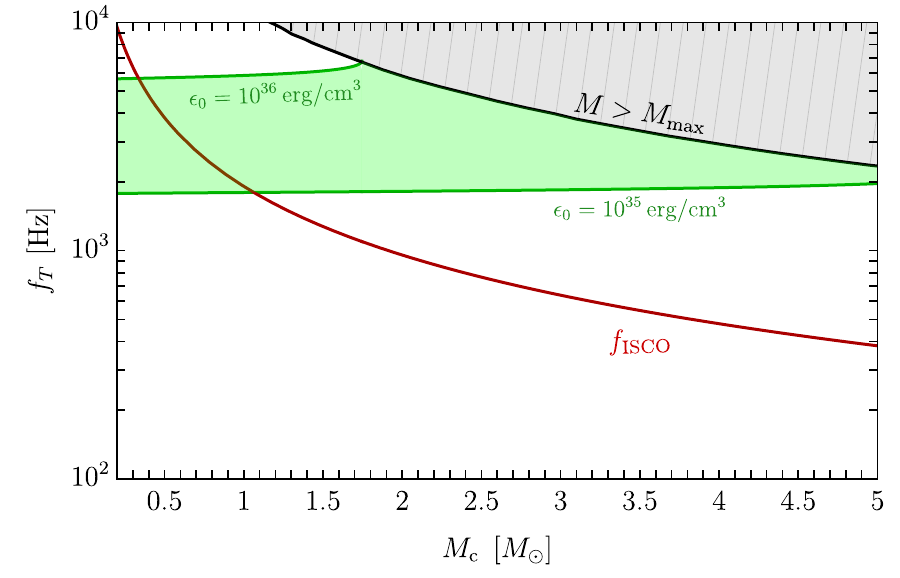}
	\caption{\it 
 Comparison between the ISCO frequency---red line, cf. eq.\,(\ref{eq:fISCO})---and the tidal disruption frequency---green band with $10^{35}
\leqslant \epsilon_0\,[\textrm{erg}/\textrm{cm}^3]\leqslant 10^{36}$, cf. eq.\,(\ref{eq:fT})---in the case of an equal-mass binary composed by two ECOs supported by the linear EoS.  
In this setup, $M_c = M/2^{1/5}$ with $M$ the common mass of the two ECOs.
 }
\label{fig:DisruptionFreq}
\end{figure}
In fig.\,\ref{fig:DisruptionFreq}, the region shaded in green corresponds to the disruption frequency 
$f_T$ computed in the case of ECOs described by the linear EoS with 
$10^{35}
\leqslant \epsilon_0\,[\textrm{erg}/\textrm{cm}^3]\leqslant 10^{36}$ (cf. fig.\,\ref{fig:MassComparison}). 
We consider the simplified case of an equal-mass binary and plot $f_T$ as function of $M_c = M/2^{1/5}$, where $M$ is the mass of the individual object in the binary system. 
We find that the disruption frequency 
$f_T$ is larger than $f_{\textrm{ISCO}}$ unless $M< 1$ $M_{\odot}$. 
In the sub-solar region, therefore, disruption effects may be significant, especially since frequencies of the order of 1 kHz can fall within the bandwidth of ground-based detectors.
For an observed signal, experiencing disruption before reaching the ISCO would amplify the discrepancy between the point-particle inspiral waveform and the true signal, thereby offering extra information. 
For instance, if a star is disrupted before the merger phase starts, the signal will be cut off at around the tidal disruption frequency.  
Modeling tidal disruption, consequently, could be relevant in the description of sub-solar events\,\cite{Cullen:2017oaz,Bandopadhyay:2022tbi}. 
Modeling tidal disruption effects goes beyond the scope of our analysis. 
We refer to ref.\,\cite{Crescimbeni:2024cwh} for a recent study focused on sub-solar searches, where tidal disruption effects are included by introducing a convolution of the waveform with a phenomenological tapering function that mimics the signal suppression due to disruption effects.

In conclusion, the signal parameters reduce to the set 
\begin{align}
\boldsymbol{\vartheta} = \{
{\color{azure}{t_c,\phi_c}},
{\color{harvardcrimson}{M_1,M_2,\chi_1,\chi_2,
\tilde{\Lambda},
\delta\tilde{\Lambda},
\bar{\mathcal{Q}}_1,
\bar{\mathcal{Q}}_2}}
\}\,,\label{eq:ParameterVectorSubset}
\end{align}
while the amplitude is kept fixed.
In the next section, instead of aiming for a comprehensive analysis that would exceed the scope of this work, we will seek to identify simplified combinations of these parameters that help emphasize the physics information we intend to convey.

\subsubsection{Fisher matrix analysis}

We begin with a brief formal discussion on the Fisher information matrix. This technique, well-established\,\cite{Helstrom1968}, is widely utilized in the theory of detection and measurement of gravitational-wave signals\,\cite{Finn:1992wt,Cutler:1994ys,Poisson:1995ef}, so we will restrict ourselves to outlining its key steps relevant to the type of study we are interested in.

Conceptually, the Fisher information matrix serves as a practical tool in gravitational-wave analysis, providing valuable insights into parameter uncertainties complementing the more exhaustive but computationally intensive Bayesian approach\,\cite{Thrane_Talbot_2020}. 
In a nutshell, the key aspect is the following. 
In the context of gravitational-wave data analysis, the main focus lies in determining the posterior distribution 
$p(\boldsymbol{\vartheta}|s)$ of a parameter set $\boldsymbol{\vartheta}$ given the observed total signal $s(t)$.
The latter, namely the detector output, is represented as 
 $s(t) = h(t,\boldsymbol{\vartheta}) + n(t)$, 
 where $h(t,\boldsymbol{\vartheta})$ is our model of the gravitational wave signal and $n(t)$ is the stationary noise component originating from the interferometer. 
Essentially, the posterior 
distribution $p(\boldsymbol{\vartheta}|s)$ gives the updated knowledge about $\boldsymbol{\vartheta}$ after taking the observed data $s(t)$ into account. It combines some prior assumptions about $\boldsymbol{\vartheta}$ with the likelihood of observing $s(t)$ given  $\boldsymbol{\vartheta}$, providing a probabilistic framework to infer the most plausible values of $\boldsymbol{\vartheta}$ in the context of a gravitational-wave detection. 
The posterior distribution can be approximated with\,\cite{Finn:1992wt}
\begin{align}
p(\boldsymbol{\vartheta}|s) 
\propto 
p^{(0)}(\boldsymbol{\vartheta})
\exp\left\{-\frac{1}{2}
\left(
h(\boldsymbol{\vartheta}) - s
\middle| 
h(\boldsymbol{\vartheta}) - s
\right)
\right\}\,,\label{eq:MainFinn}
\end{align}
where $p^{(0)}(\boldsymbol{\vartheta})$ 
is the aforementioned prior distribution, that is 
the {\it a priori} probability that the signal is characterized by $\boldsymbol{\vartheta}$. 
Priors are typically chosen based on theoretical models, previous observational data, and physical constraints relevant to the 
parameters $\boldsymbol{\vartheta}$ under investigation. 
The inner product $(\cdot|\cdot)$ appearing in eq.\,(\ref{eq:MainFinn}) is defined by
\begin{align}
\left(g \middle|h \right) \equiv 
2\int_{f_{\textrm{min}}}^{f_{\textrm{max}}}
\frac{
\tilde{g}^*(f)\tilde{h}(f) + 
\tilde{h}^*(f)\tilde{g}(f) 
}{S_n(f)}df\,,\label{eq:InnerFirst}
\end{align}
where $\tilde{g}(f)$ is the Fourier transform of $g(t)$. 
In a specific measurement scenario defined by the detector output $s(t)$, we estimate the true values of the source parameters by identifying the parameters set $\hat{\boldsymbol{\vartheta}}$ where the probability distribution function in 
eq.\,(\ref{eq:MainFinn}) reaches its maximum. This approach is known as the maximum-likelihood estimator. 
As we focus on the high SNR limit, the posterior distribution 
$p(\boldsymbol{\vartheta}|s) $ becomes sharply concentrated around this estimator. This is where the Fisher information matrix comes into play. 
Taylor expanding, assuming a nearly uniform prior distribution around $\hat{\boldsymbol{\vartheta}}$, and neglecting higher-order terms in the limit of large SNR, one arrives at
\begin{align}
  p(\boldsymbol{\vartheta}|s) \propto 
  p^{(0)}(\boldsymbol{\vartheta})
  \exp\left(
  -\frac{1}{2}
  \Gamma_{ij}\Delta\vartheta_i
  \Delta\vartheta_j
  \right)\,,
\end{align}
with $\Delta\vartheta_i \equiv 
\vartheta_i-\hat{\vartheta}_i$.
The key point, therefore, is that 
in the limit of large SNR the posterior distribution takes a Gaussian form.
 
The Fisher information matrix $\Gamma_{ij}$ is precisely defined through the previous equation as 
\begin{align}
\Gamma_{ij} \equiv
\left.\left(
\frac{\partial h}{\partial\vartheta_i}
\middle|
\frac{\partial h}{\partial\vartheta_j}
\right)\right|_{
\boldsymbol{\vartheta} = 
\hat{\boldsymbol{\vartheta}}
}\,,\label{eq:Fisher}
\end{align}
where the inner product $(\cdot|\cdot)$ is given by eq.\,(\ref{eq:InnerFirst}).
In eq.\,(\ref{eq:Fisher}), we indicate with $\vartheta_i$ the components of the vector $\boldsymbol{\vartheta}$ defined in 
eq.\,(\ref{eq:ParameterVector}). 
Therefore, formally, the computation of the Fisher information matrix proceeds by first taking the derivatives of the signal with respect to each parameter and then computing the inner product of their Fourier transforms.\footnote{Taking the Fourier transform of the partial derivative of the signal with respect to the parameters is equivalent to taking the derivatives of the Fourier transform of the signal with respect the those parameters. This equivalence simplifies the analysis and is widely used in practice.} Finally, 
the inner products must be evaluated at $\hat{\boldsymbol{\vartheta}}$. 

Following Gaussian statistics, 
the inverse of the Fisher information matrix provides an estimate of the covariance matrix $\Sigma_{ij} \equiv (\Gamma^{-1})_{ij}$. 
The square root of the diagonal elements of the inverse Fisher information matrix gives the standard deviation (uncertainty) of the parameter estimates
\begin{align}
\sigma_{\vartheta_i} \equiv \sqrt{\Sigma_{ii}}\,.\label{eq:FisherError}
\end{align}
On the other hand, non-zero off-diagonal elements indicate correlations between parameters. 
More in detail, the correlation coefficient $\rho_{ij}$ between parameters  $\vartheta_i$ and $\vartheta_j$ is given by 
$\rho_{ij} = \Sigma_{ij}/\sqrt{\Sigma_{ii}\Sigma_{jj}}$. 
We compute the elements of the Fisher matrix according to 
\begin{align}
\left(
\frac{\partial h}{\partial\vartheta_i}
\middle|
\frac{\partial h}{\partial\vartheta_j}
\right) = 
2\mathcal{A}^2 
\int_{f_{\textrm{min}}}^{f_{\textrm{max}}}
\frac{
[(\frac{\partial\Psi}{\partial\vartheta_i})^*
\frac{\partial\Psi}{\partial\vartheta_j} + 
\frac{\partial\Psi}{\partial\vartheta_j}
(\frac{\partial\Psi}{\partial\vartheta_i})^*
]
}{f^{7/3}S_n(f)}
df\,.
\end{align}
At this point, we should have defined all the relevant conceptual tools necessary to discuss the results of our analysis. 

We start from a simplified setup. 
We consider the set of parameters 
\begin{align}
\boldsymbol{\vartheta} = \{
{\color{azure}{t_c,\phi_c}},
{\color{harvardcrimson}{M_c,\eta,
\tilde{\Lambda}}}
\}\,.
\end{align}
We trade the two individual masses $M_{1,2}$ for the chirp mass $M_c$ and the symmetric mass ratio $\eta$, and we do not include the presence of spin.
Furthermore, the dependence on tidal deformability only enters at the leading 5-PN order via $\tilde{\Lambda}$, 
$\Psi^{\textrm{Tidal}}_{5\textrm{PN}}(f) = -
39\tilde{\Lambda}
v^{10}/2$. 
We fix the luminosity distance to be $D_{\textrm{eff}} = 400$ Mpc. 
We discuss two situations.

First, we consider the case of equal-mass black hole binaries, $\eta = 1/4$, $M_c = M/2^{1/5}$ and $\tilde{\Lambda} = 0$. 
Therefore, we run the Fisher information matrix analysis with 
$\hat{\boldsymbol{\vartheta}} = \{
t_c,\phi_c,M_c,{1}/{4},0
\}$. 
The central values of $t_c$ and $\phi_c$
are not important since their dependence drops from the computation of the Fisher matrix. 
By varying $M_c$, we compute the error 
$\sigma_{\tilde{\Lambda}}$, cf. eq.\,(\ref{eq:FisherError}).
\begin{figure}[!t]
	\centering
\includegraphics[width=0.495\textwidth]{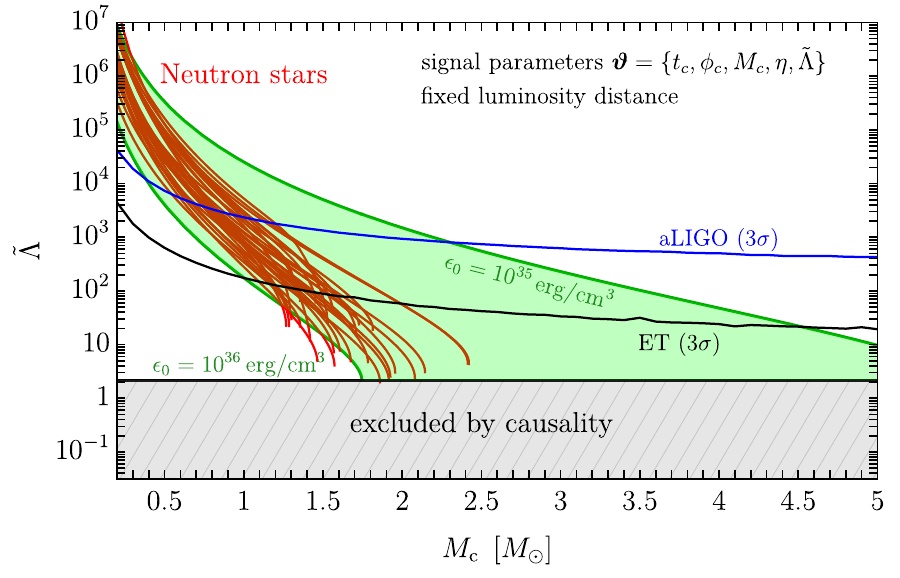}
	\caption{\it 
  Reduced tidal deformability parameter $\tilde{\Lambda}$ as function of chirp mass $M_c$ (assuming equal-mass binaries) in the sub-solar and lower mass gap range. 
The region shaded in gray corresponds to 
$\tilde{\Lambda} < \Lambda_{\textrm{min}}$, cf. eq.\,(\ref{eq:CausalTidal}).
The red lines correspond to neutron stars (both with normal EoS and SQM-type EoS) while the green band corresponds to ECOs with 
$10^{-2} \leqslant P_c/\epsilon_0 \leqslant P_c^{\textrm{max}}$ and  $10^{35}
\leqslant \epsilon_0\,[\textrm{erg}/\textrm{cm}^3]\leqslant 10^{36}$.   
The blue and black lines represent, respectively, the upper bound (at $3\sigma$) corresponding to aLIGO and ET, as obtained from the Fisher matrix analysis. The 
(effective) luminosity distance is fixed at $D_{\textrm{eff}} = 400$ Mpc.
}
\label{fig:TidalPlotFisher}
\end{figure}
We show our result in fig.\,\ref{fig:TidalPlotFisher}. 
The blue and black lines represent, respectively, the upper bound at $3\sigma$ corresponding to aLIGO and ET, compared with the theoretical models of neutron stars (red lines) and ECOs with a linear equation of state (green region, where $\epsilon_0$ is allowed to vary as indicated in the figure). The key message of this plot is that, in particular with the sensitivity of the future Einstein Telescope, it is plausible to measure values of $\tilde{\Lambda}$ that can distinguish objects in the mass gap from black holes.

Second, we focus on a specific merger event. We inject a signal with individual masses $M_1 = 3.5$ $M_{\odot}$ and $M_2 = 1.5$ $M_{\odot}$ (that is, $M_c \simeq 1.96$ $M_{\odot}$ and $\eta \simeq 0.21$). We assume that the signal strongly peaks at $\tilde{\Lambda} = 0$, and run the Fisher information matrix analysis with 
$\hat{\boldsymbol{\vartheta}} = \{
t_c,\phi_c,M_c \simeq 1.96,\eta \simeq 0.21,0
\}$. 
In this case, rather than considering $\sigma_{\tilde{\Lambda}}$, we try to extract the errors on the individual 
tidal deformabilities $\Lambda_{1,2}$.
\begin{figure}[!t]
	\centering
\includegraphics[width=0.495\textwidth]{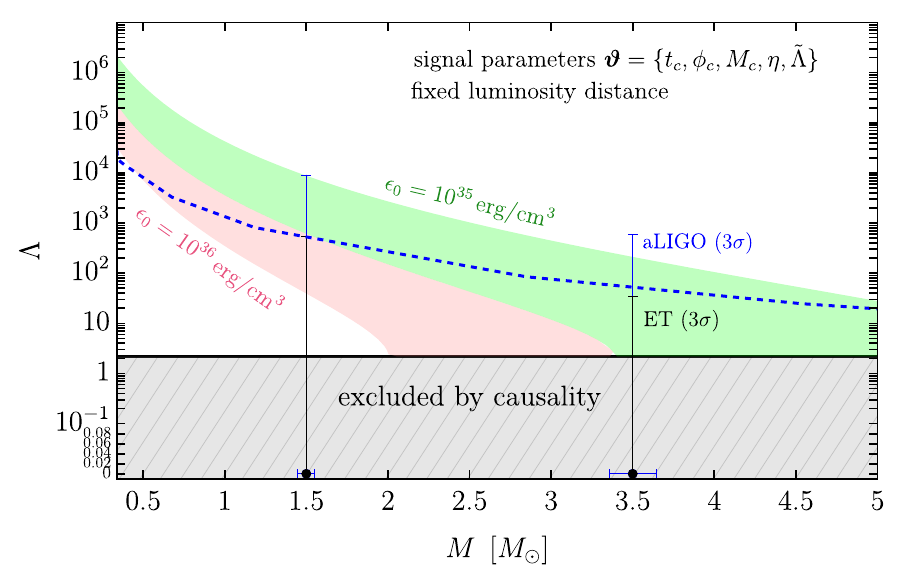}
	\caption{\it 
Tidal deformability $\Lambda$ as a function of mass. Hypothetical measurements for a $(3.5 + 1.5)\,M_{\odot}$ binary black hole system, with error bars estimated for an observation at an (effective) luminosity distance of $D_{\textrm{eff}} = 400$ Mpc by aLIGO and ET, shown in blue and black, respectively. The shaded regions denote all possible ECOs with a linear EoS that are consistent with the measurements of the smaller compact object by each detector (pink for ET and green for aLIGO). 
The dashed blue line corresponds to solitonic boson stars, with the mass parameter $\mu = 1.8\times 10^{-10}$ eV fixed so as to yield solutions marginally compatible with the tidal deformability measurement of the lighter object as observed by ET.
 }
\label{fig:FisherTest}
\end{figure}
We follow the analysis discussed in ref.\,\cite{Sennett:2017etc}. 
It is possible to extract information on the individual tidal deformabilities as follows.  
The Fisher matrix analysis, as previously explained, gives the uncertainties 
$\sigma_{\tilde{\Lambda}}$ and 
$\sigma_{\eta}$. 
These values provide an estimate of the precision on the individual tidal deformabilities according to\,\cite{Sennett:2017etc}
\begin{align}
\sigma_{\Lambda_1} & \leqslant 
\left\{
\left[
g_1(\eta)\sigma_{\tilde{\Lambda}}
\right]^2 + 
\left[
g_1^{\prime}(\eta)
\tilde{\Lambda}
\sigma_{\eta}
\right]^2
\right\}^{1/2}\,,\\
\sigma_{\Lambda_2} & \leqslant 
\left\{
\left[
g_2(\eta)\sigma_{\tilde{\Lambda}}
\right]^2 + 
\left[
g_2^{\prime}(\eta)
\tilde{\Lambda}
\sigma_{\eta}
\right]^2
\right\}^{1/2}\,,
\end{align}
where the functions $g_i(\eta)$ are given by 
\begin{align}
g_1(\eta) & = \frac{13}{16(1+7\eta - 31\eta^2)}\,,\\
g_2(\eta) & = 
\frac{13}{
8[1+7\eta-31\eta^2-
\sqrt{1-4\eta}(1+9\eta-11\eta^2)]
}\,.
\end{align}
We show our result in fig.\,\ref{fig:FisherTest}. 
 In this figure, 
 all possible ECOs supported by a linear EoS and consistent with the measurement of the lighter component are shown in green and pink for aLIGO and the ET, respectively. While the deformability measurements of each black hole, considered individually, are compatible with an ECO interpretation, they cannot both be ECOs simultaneously. We can therefore conclude, with significantly greater confidence, that third-generation detectors will be able to distinguish binary black hole systems from binary systems of ECOs. 
 As a specific example, the dashed blue line represents the case of solitonic boson stars, for which the value of the mass parameter $\mu$ is chosen to be compatible with the ET upper limit on the tidal deformability of the lighter mass in the binary system. In this case, we see that the heavier mass cannot also be a solitonic boson star, since it would exhibit a tidal deformability larger than that allowed by the uncertainty in the corresponding ET measurement.

\noindent
\section{The case of solitonic boson stars as black hole mimickers}\label{sec:BS}

Boson stars are hypothetical objects composed of self-gravitating scalar fields, as opposed to the fermionic matter that makes up most stars. These scalar fields could be composed of particles like hypothetical ultralight bosons. 

\subsection{Overview of boson star models: mass and compactness}\label{sec:BackBosonStar}

In the literature, three models of boson stars are typically considered, referred to as Minimal\,\cite{Kaup:1968zz,Ruffini:1969qy}, Massive\,\cite{Colpi:1986ye}, and Solitonic\,\cite{Friedberg:1986tq}, respectively (cf. ref.\,\cite{Schunck:2003kk} for a more comprehensive list of boson star models and ref.\,\cite{Liebling:2012fv} for an updated review encompassing various aspects of boson star physics).
Ref.\,\cite{Sennett:2017etc} (see also ref.\,\cite{Cardoso:2017cfl}) found that for solitonic boson stars the tidal deformability can
reach $\Lambda_{\textrm{min}} \approx 1.3$ (while in the case of massive boson stars $\Lambda_{\textrm{min}} \approx 280$). 
The lower bound for $\Lambda$ in the case of solitonic boson stars is {\it not} compatible with the causality bound extracted in this paper. It is, therefore, interesting to  investigate the connection between 
solitonic boson stars and the linear EoS discussed in section\,\ref{section:LinearEoS}. 
The discussion of Minimal, Massive, and Solitonic boson stars can be approached as follows. 

We consider a scalar field theory minimally coupled to Einstein gravity that enjoys a global $U(1)$ symmetry. 
The action takes the form
\begin{align}
S=\int d^4x \sqrt{-g}\left(\frac{1}{2}\MPl^2 R + 
\mathcal{L}_{\Phi}\right)\,,\label{eq:StartingAction}
\end{align}
where the Lagrangian density of the scalar field is given by $\mathcal{L}_{\Phi} = -g^{\mu\nu}
(\partial_{\mu}\Phi^{*})
(\partial_{\nu}\Phi) 
- V(|\Phi|^2)$. 
The Ricci scalar $R$ and the metric tensor $g_{\mu\nu}$ refer to the metric defined in eq.\,(\ref{eq:SphMetric}). 
$\MPl^2 \equiv 1/8\pi G_N$ is the so-called reduced Planck mass.

Minimal boson stars corresponds to the case in which the potential only features the mass term while the massive case  corresponds to a renormalizable quartic potential that includes self-interactions
\begin{align}
V_{\textrm{min}} & = \mu^2|\Phi|^2\,,  \\
V_{\textrm{mass}} & = \mu^2|\Phi|^2
 + g^2|\Phi|^4\,,
\end{align}
where $\Phi$ is a complex scalar field while $g$ and $\mu$ represent some fundamental coupling and mass scale, respectively.

The case of solitonic boson stars corresponds to a sixth-order non-renormalizable potential that takes the form
\begin{align}
V_{\textrm{sol}} & = 
\mu^2|\Phi|^2
 - \frac{4\mu^2}{f^2}|\Phi|^4 + \frac{4\mu^2}{f^4}
 |\Phi|^6 \label{eq:SolPote1}\\ 
 & = 
 \mu^2|\Phi|^2\left( 
 \frac{2}{f^2}
 |\Phi|^2 - 1
 \right)^2\,.\label{eq:SolPote2}
\end{align}
This potential features the presence of two degenerate minima in $\Phi = 0$ and $|\Phi| = f/\sqrt{2}$ separated by a potential barrier. 
As evident in eq.\,(\ref{eq:SolPote1}),
 the potential has a negative quartic coupling, but its boundedness from below is restored by higher-dimensional operators. 

 A recent comprehensive study on the properties of boson stars has been presented in ref.\,\cite{Boskovic:2021nfs}. 
 In particular, it was shown that in the high-compactness limit, solitonic boson stars exhibit an effectively linear EoS, thereby saturating the causality constraint\,\cite{Urbano:2018nrs}.

We decompose the scalar field in Fourier modes according to 
$\Phi(\boldsymbol{x},t) = \phi(r)e^{-i\omega t}$. The action in eq.\,(\ref{eq:StartingAction}) leads to the equations of motion
\begin{align}
\lambda^{\prime} & = 
\frac{1-e^{\lambda}}{r} + 
\frac{re^{\lambda}}{
\MPl^2
}\left(
V + \omega^2\phi^2e^{-\nu}+
e^{-\lambda}\phi^{\prime\,2}
\right)\,,\label{eq:Lambda}\\
\nu^{\prime} & =
\frac{e^{\lambda}-1}{r} + 
\frac{re^{\lambda}}{
\MPl^2
}
\left(
-V + \omega^2\phi^2e^{-\nu}+
e^{-\lambda}\phi^{\prime\,2}
\right)\,,\label{eq:Nu}\\
\phi^{\prime\prime} & = 
\left(
-\frac{2}{r} + \frac{\lambda^{\prime} - \nu^{\prime}}{2}
\right)\phi^{\prime}
+ e^{\lambda}\phi\left(
\frac{dV}{d\phi^2} - e^{-\nu}\omega^2
\right)\,,\label{eq:KG}
\end{align}
with $V= V(\phi^2)$.
These equations provide
a closed system and imply the continuity equation $\nabla_{\nu}T^{\mu\nu} = 0$, where $T_{\mu\nu}$ is
the scalar energy-momentum tensor 
$T_{\mu\nu} = \partial_{\mu}\Phi^*\partial_{\nu}\Phi + 
\partial_{\mu}\Phi\partial_{\nu}\Phi^* - g_{\mu\nu}(
g^{\rho\sigma}
\partial_{\rho}\Phi^*\partial_{\sigma}\Phi + V
)$. Energy density and pressure are given by 
\begin{align}
\rho & = \omega^2\phi^2e^{-\nu}+
e^{-\lambda}\phi^{\prime\,2} +V  = -T^{0}_{~0}\,,\\
P_r & = 
\omega^2\phi^2e^{-\nu}+
e^{-\lambda}\phi^{\prime\,2} -V  = + T^{r}_{~r}\,,\label{eq:RadialPressure}\\
P_{t} & = 
 \omega^2\phi^2e^{-\nu} - 
e^{-\lambda}\phi^{\prime\,2} -V = + T^{\theta}_{~\theta}\,,
\end{align}
with the expressions for $\rho$ and $P_r$ that enter on the right-hand side of eq.\,(\ref{eq:Lambda}) and eq.\,(\ref{eq:Nu}), respectively. 
As already noted in the pioneering work by Kaup, cf. ref.\,\cite{Kaup:1968zz}, 
boson stars are also characterized by a tangential pressure $P_{t}$ that is generally different from the radial one, $P_r$, 
indicating they are an example of anisotropic stars. 
It is indeed possible to recast the previos system of equations as follows. We introduce the  pressure anisotropy 
\begin{align}
\Delta \equiv P_r - P_{t} = 
2e^{-\lambda}\phi^{\prime\,2}\,.
\end{align}
The continuity equation $\nabla_{\nu}T^{\mu\nu} = 0$ thus takes the simple form 
\begin{align}
\frac{dP_r}{dr} = 
-\frac{2\Delta}{r}
- \frac{\nu^{\prime}}{2}(\rho + P_r)\,.\label{eq:PressureEq}
\end{align}
Using $e^{-\lambda(r)} = 1-2G_Nm(r)/r = 
1-m(r)/4\pi\MPl^2 r$, 
with $m(r)$ that can be interpreted as the mass-energy enclosed within the radius $r$, 
eq.\,(\ref{eq:Lambda}) becomes $m^{\prime} = 4\pi r^2\rho$ while 
for eq.\,(\ref{eq:Nu}) 
we find 
\begin{align}
\nu^{\prime} = 
\frac{4\pi r^3 P_r + m}{
r(4\pi\MPl^2 r - m)
}\,.
\end{align}
Eq.\,(\ref{eq:PressureEq}) thus gives 
\begin{align}
\frac{dP_r}{dr} = 
-\frac{
G_N(\rho + P_r)(4\pi r^3 P_r + m)
}{
r^2(1-2G_N m/r)
}-\frac{2\Delta}{r}\,,
\end{align}
which is indeed the TOV equation in the presence of pressure anisotropy. 

We numerically solve the system in 
eqs.\,(\ref{eq:Lambda}-\ref{eq:KG}). 
To set the boundary
conditions (cf., e.g., ref.\,\cite{Macedo:2013jja}), one asks the metric to be Minkowski and $\phi(r)$ to vanish at large distances. 
Specifically, this translates into the conditions 
$\nu(r \to \infty) = 0$ and 
$\phi(r \to \infty) = 0$. 
On the other hand, regularity at the origin requires 
$\phi(0) = \phi_c$,
$\phi^{\prime}(0) = 0$,
$\lambda(0) = 0$. 
The last two conditions are necessary in order for the terms $(1-e^{\lambda})/r$ and $\phi^{\prime}/r$, appearing in eqs.\,(\ref{eq:Lambda}-\ref{eq:KG}), not to produce singularities in the limit $r \to 0$. 

The boundary condition $\nu(r \to \infty) = 0$ can be replaced with a boundary condition at the origin.
Consider the metric in 
eq.\,(\ref{eq:SphMetric}) and write (with $d\Omega^2 = d\theta^2 + \sin^2\theta d\varphi^2$ the metric on the two-sphere) 
\begin{align}
ds^2 = -e^{\nu(r)-\nu(0)}e^{\nu(0)}dt^2 + 
e^{\lambda(r)}dr^2 + 
r^2 d\Omega^2\,.\label{eq:metrt}
\end{align}
We now change the time variable according to $\tilde{t} \equiv 
e^{\nu(0)/2}t$. Consequently, we write 
$ds^2 = -e^{\tilde{\nu}(r)}d\tilde{t}^2 + 
e^{\lambda(r)}dr^2 + 
r^2 d\Omega^2$ where we defined 
$\tilde{\nu}(r) \equiv 
\nu(r)-\nu(0)$. 
By construction, we now have 
$\tilde{\nu}(0) = 0$. 
At the same time, 
the scalar field profile also changes and can be brougth to its original form by means of 
a redefinition of the
scalar frequency
\begin{align}
\phi(r)e^{-i\omega t} 
= 
\phi(r)e^{-i\omega e^{-\nu(0)/2}\tilde{t}} & = 
\phi(r)e^{-i\tilde{\omega}\tilde{t}}\,,\nn\\
\tilde{\omega} &\equiv 
\omega e^{-\nu(0)/2}\,.
\end{align}
The upshot of this argument is that it is possible to numerically solve the system in eqs.\,(\ref{eq:Lambda}-\ref{eq:KG}) exactly as it stands, but by substituting $\nu\to \tilde{\nu}$ and $\omega\to \tilde{\omega}$, and now having the boundary condition $\tilde{\nu}(0) = 0$. The resulting system will be solved, for each value of $\phi_c$, only by a discrete set of eigenfrequencies $\tilde{\omega}$ (to be sought through a shooting method). We focus here on boson star solutions in the ground state, which correspond to the scalar profile having no nodes and to the lowest eigenfrequency $\tilde{\omega}$. Once the value of $\tilde{\omega}$ is found, it will be sufficient to find $\nu(0)$ such that $\nu(r\to \infty) = \tilde{\nu}(r\to \infty) + \nu(0) = 0$ and rescale $\omega = \tilde{\omega}e^{\nu(0)/2}$.

Lastly, we introduce the following dimensionless quantities
\begin{align}
x\equiv \mu r\,,~~
w \equiv \frac{\omega}{\mu}\,,~~
\varphi \equiv \frac{\sqrt{2}\phi}{f}\,,~~
\mathcal{V} \equiv \frac{2V}{\mu^2 f^2}\,.\label{eq:ADimVariaBack}
\end{align}
As a consequence, eqs.\,(\ref{eq:Lambda}-\ref{eq:KG})  become dimensionless. 
We also introduce the dimensionless ratio $\xi \equiv f/\MPl$. 
For completeness, we write 
\begin{align}
\lambda^{\prime} & = 
\frac{1-e^{\lambda}}{x} + 
\frac{xe^{\lambda}\xi^2}{
2
}\left(
\mathcal{V} + w^2\varphi^2e^{-\nu}+
e^{-\lambda}\varphi^{\prime\,2}
\right)\,,\label{eq:LambdaAdim}\\
\nu^{\prime} & =
\frac{e^{\lambda}-1}{x} + 
\frac{xe^{\lambda}\xi^2}{
2
}
\left(
-\mathcal{V} + w^2\varphi^2e^{-\nu}+
e^{-\lambda}\varphi^{\prime\,2}
\right)\,,\label{eq:NuAdim}\\
\varphi^{\prime\prime} & = 
\left(
-\frac{2}{x} + \frac{\lambda^{\prime} - \nu^{\prime}}{2}
\right)\varphi^{\prime}
+ e^{\lambda}\varphi\left(
\frac{d\mathcal{V}}{d\varphi^2} - e^{-\nu}w^2
\right)\,,\label{eq:KGAdim}    
\end{align}
where now $^{\prime}\equiv d/dx$.
It is also possible to introduce dimensionless energy density $\bar{\rho}$ and pressure $\bar{P}$ defined according to 
$\rho\equiv (\mu^2f^2)\bar{\rho}$ and $P_{i=r,t} \equiv (\mu^2f^2)\bar{P}_{i=r,t}$ so that the anisotropic TOV equation reads  
\begin{align}
\frac{d\bar{P}_r}{dx} 
= -\frac{\xi^2(\bar{\rho}+\bar{P}_r)(4\pi x^3\bar{P}_r + \bar{m}/\xi^2)}{x^2(1-\bar{m}/4\pi  x)} - \frac{2\bar{\Delta}}{x}\,,
\end{align}
where 
\begin{align}
\bar{m} &\equiv \frac{\mu m}{\MPl^2} = 4\pi x(1-e^{-\lambda})\,,\label{eq:RescaledMass}\\
\bar{\Delta} &\equiv 
\frac{\Delta}{\mu^2 f^2} = 
\frac{P_r-P_t}{\mu^2 f^2} = 
e^{-\lambda}\varphi^{\prime\,2}\,.\label{eq:PreAni}
\end{align}
Note that this equation has the same 
structure as eq.\,(\ref{eq:Anisotro2}). 
We also observe that, in this case, the anisotropy parameter is always positive.
\begin{figure}[!ht!]
	\centering
\includegraphics[width=0.495\textwidth]{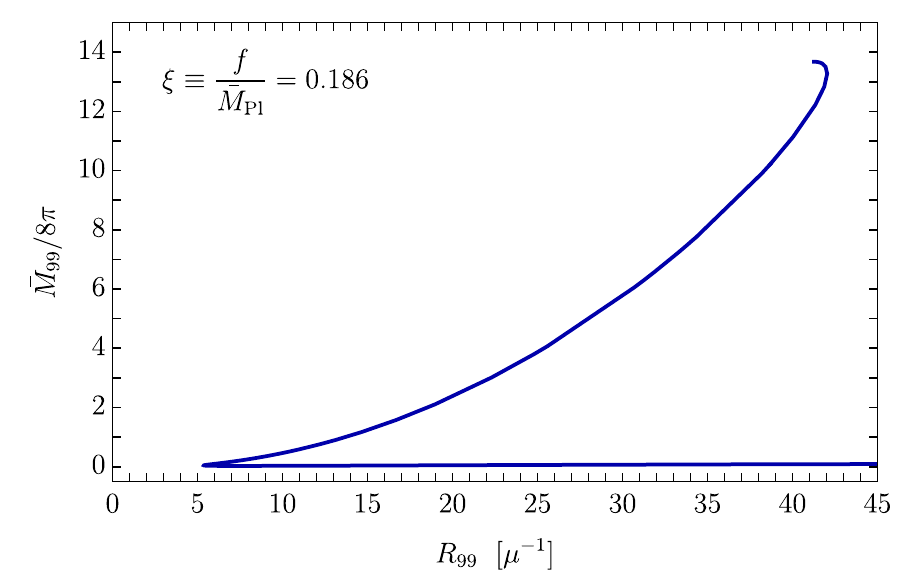}
	\caption{ \it 
Mass-radius diagram for solitonic boson stars with $\xi = 0.186$. 
The solutions shown correspond to the stable branch $(2)$ in fig.\,\ref{fig:BSStability}.
 }
\label{fig:BSMassRadius}
\end{figure}
\begin{figure}[!ht!]
	\centering
\includegraphics[width=0.495\textwidth]{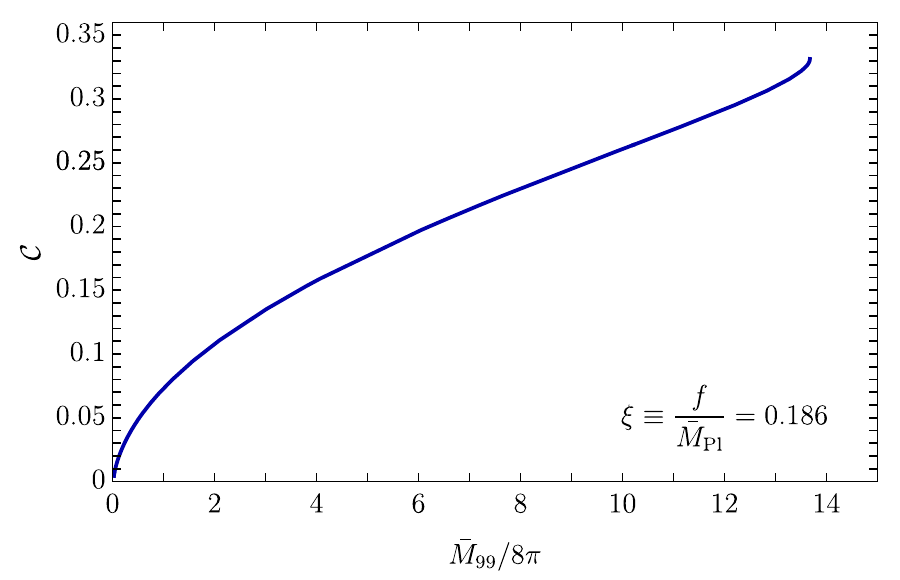}
	\caption{ \it 
Compactness as a function of mass for solitonic boson stars with $\xi = 0.186$. 
The solutions shown correspond to the stable branch $(2)$ in fig.\,\ref{fig:BSStability}.
 }
\label{fig:BSMassCompa}
\end{figure}

Consider the system in 
eqs.\,(\ref{eq:LambdaAdim}-\ref{eq:KGAdim}).
A local expansion
of $\lambda(x)$, $\nu(x)$ and $\varphi(x)$ around $x = 0$ gives
\begin{align}
\lambda(x) & \approx 
\frac{x^2\xi^2 \varphi_c^2}{6}
\bigg[
w^{2}e^{-\nu(0)} 
+ 
(-1+\varphi_c^2)^{2}
\bigg]\,,\\
\nu(x) & \approx 
\nu(0) + \frac{x^2\xi^2\varphi_c^2}{6}\bigg[
2w^{2}e^{-\nu(0)} - (-1+\varphi_c^2)^{2}
\bigg]\,,\\
\varphi(x) & \approx \varphi_c 
+\frac{x^2\varphi_c}{6}
\bigg[
-w^{2}e^{-\nu(0)} 
- 
(1 - \varphi_c^2)(-1 + 3\varphi_c^2)
\bigg]\,,
\end{align}
up to terms of order $O(x^4)$.  
As already discussed in the main body of the text, it is possible to shift to the function $\tilde{\nu}(x) \equiv \nu(x) - \nu(0)$ and to the frequency $\tilde{\omega}^2 = \omega^2e^{-\nu(0)}$ at the cost of having $\tilde{\nu}(\infty) \equiv \tilde{v}_{\infty} \neq 0$ (but with $\tilde{\nu}(0)=0$). 
The previous system reads 
\begin{align}
\lambda(x) & \approx 
\frac{x^2\xi^2 \varphi_c^2}{6}
\bigg[
\tilde{w}^{2}
+ 
(-1+\varphi_c^2)^{2}
\bigg]\,,\\
\tilde{\nu}(x) & \approx 
 \frac{x^2\xi^2\varphi_c^2}{6}\bigg[
2\tilde{w}^{2} - 
(-1+\varphi_c^2)^{2}
\bigg]\,,\\
\varphi(x) & \approx \varphi_c 
-\frac{x^2\varphi_c}{6}
\bigg[
\tilde{w}^{2} - 
(-1 + \varphi_c^2)(-1 + 3\varphi_c^2 )
\bigg]\,.
\end{align}
These are the boundary conditions (with, in addition, $\varphi^{\prime}(x)$ calculated accordingly) that we impose in the limit as $x\to 0$.

Unlike traditional stars, boson stars do not have a well-defined surface or sharp boundary where the star ends and the vacuum begins. 
We define the radius of the boson star as the radial distance within which 99\% of the total mass of the boson star is enclosed.  
Formally, using eq.\,(\ref{eq:RescaledMass}) we define the quantity $X_{99}$ by imposing the condition 
\begin{align}
X_{99}\left[
1-e^{-\lambda(X_{99})}
\right] = 0.99\lim_{x\to \infty}
x\left[
1-e^{-\lambda(x)}
\right]\,.\label{eq:SBSRadius99}
\end{align}
Consequently, the mass of the boson star is defined as
\begin{align}
 \bar{M}_{99} \equiv 4\pi X_{99}\left[
1-e^{-\lambda(X_{99})}
\right]\,.\label{eq:Mass99}
\end{align}
Consequently, the compactness is given by 
\begin{align}
\mathcal{C} = \frac{\bar{M}_{99}}{8\pi X_{99}}\,,\label{eq:C99}
\end{align}
and we also define $R_{99}\equiv X_{99}/\mu$.

Regarding the background solutions, we have validated our numerical results by comparing them with those discussed in ref.\,\cite{Boskovic:2021nfs}. 
In the following, we will focus on the solutions that achieve maximum compactness. 
Consistent with ref.\,\cite{Boskovic:2021nfs}, we find that the maximum compactness is achieved by considering $\xi = 0.186$, and it is on this value that we will focus for the remainder of our analysis.
In fig.\,\ref{fig:BSMassRadius}, we show the mass-radius diagram for solitonic boson stars, while in fig.\,\ref{fig:BSMassCompa}, 
we show the compactness as a function of mass. We remark that our solutions consistently terminate at the maximum mass, beyond which the solutions become unstable. 
For non-rotating  boson stars, radial stability involves examining the locations of turning points on the 
$\bar{M}_{99}$-$\varphi_c$ curve, with $\xi$ held constant. 
\begin{figure}[!t]
	\centering
\includegraphics[width=0.495\textwidth]{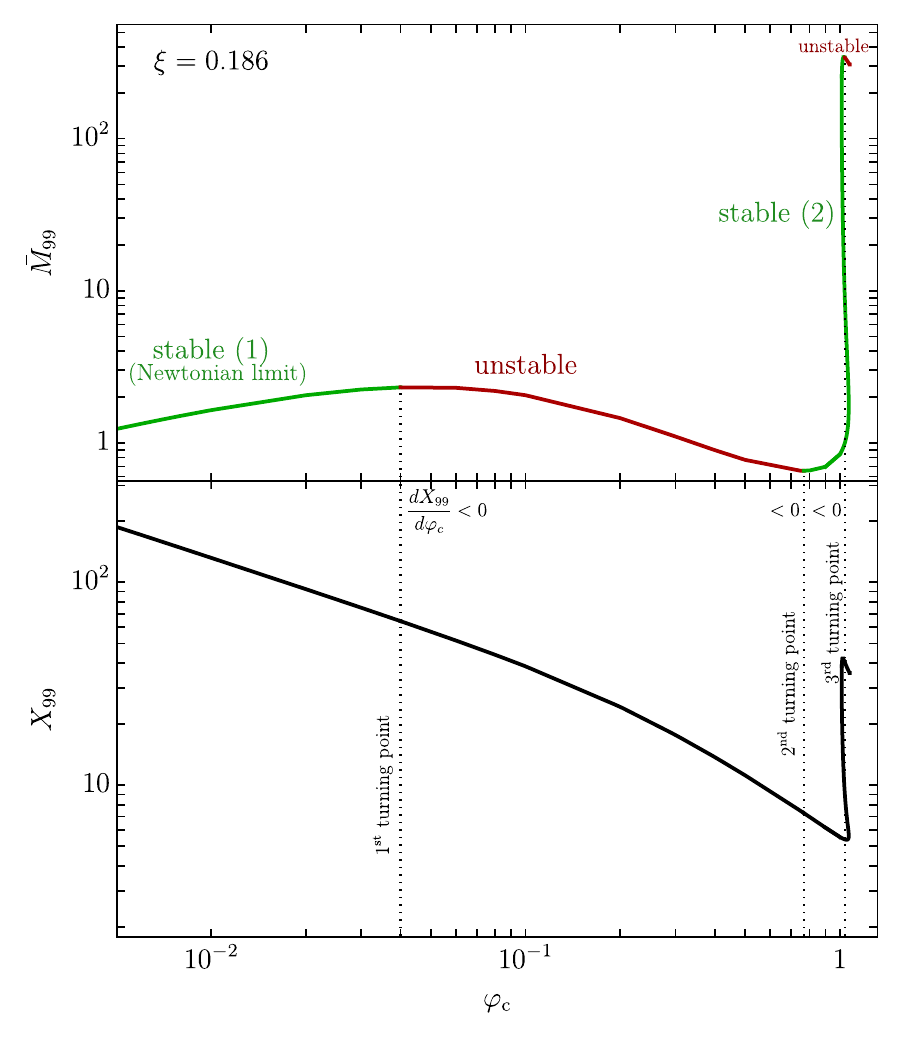}
	\caption{ \it
 Top panel.
 Soliton boson star mass $\bar{M}_{99}$ as function of the central field value $\varphi_c$. 
 Green (red) portions of the curve indicate stable (unstable) solutions.
 Bottom panel. 
 Soliton boson star radius $X_{99}$ as function of the central field value $\varphi_c$.
 The location of the three critical points $d\bar{M}_{99}/d\varphi_c = 0$ is marked with vertical dotted lines. 
 $\xi = 0.186$ is kept fixed.
}
\label{fig:BSStability}
\end{figure}
More in detail, stability theorems indicate that transitions between stable and
unstable configurations occur only at the critical points in the parameter space such that\,\cite{1983bhwdbookS,Liebling:2012fv,Siemonsen:2020hcg}
\begin{align}
\frac{d\bar{M}_{99}}{d\varphi_c} = 0\,,
\end{align}
and this requirement must be accompanied by the condition $dX_{99}/d\varphi_c < 0$ at the critical point. The rationale behind this condition is as follows. The linear perturbation equations pose a Sturm-Liouville boundary value problem on the finite interval $[0, R]$, where $R$ is the star's radius, for the perturbation frequency squared, $\Omega^2$. According to spectral theory, there are infinitely many real ordered eigenvalues, 
$\Omega_0^2 < \Omega_1^2 < \Omega_2^2 < \dots$, and the eigenfunction corresponding to the eigenvalue $\Omega_n^2$ (with $n\in \mathbb{N}_0$) has exactly $n$ nodes in the open interval $(0, R)$.
An eigenfunction with an odd number of nodes is  referred to as an odd mode, while one with an even number of nodes is called an even mode.
The lowest-frequency mode is called the fundamental mode, whereas modes with $n > 0$ are excited modes. 
The fundamental mode is stable if $\Omega_0^{2} >0$. When the fundamental mode is stable, all higher modes are also stable due to the ordering $\Omega_0^2 < \Omega_1^2 < \Omega_2^2 < \dots$. If the fundamental mode is unstable, the first excited mode can be either stable or unstable, given $\Omega_0^2 < \Omega_1^2$. If the first excited mode is stable, then all subsequent higher modes are stable as well. If the first excited mode is unstable, the stability of the second excited mode depends on $\Omega_1^2 < \Omega_2^2$. This reasoning can be extended to the $n$-th mode. Therefore, the fundamental mode is crucial in determining linear stability. 
In this respect, the key point is that the stability can be analyzed by noting that $dR/d\varphi_c > 0$ ($dR/d\varphi_c < 0$) at a critical
point corresponds to a change of sign of an odd (even) mode\,\cite{Santos:2024vdm}. 
It is, therefore, crucial to identify the critical points ${d\bar{M}_{99}}/{d\varphi_c} = 0$ and monitor the value of the derivative $dX_{99}/d\varphi_c$.
In particular, $dX_{99}/d\varphi_c < 0$ at a critical point indicates a change in the sign of the fundamental mode frequency, and thus the transition from a stable branch to an unstable one or vice versa.

To better illustrate this point, we show in the top panel of fig.\,\ref{fig:BSStability} the curve $\bar{M}_{99}(\varphi_{c})$ obtained by solving the structure equations as previously described. 
Let us follow the evolution of the curve $\bar{M}_{99}(\varphi_{c})$ starting from the limit $\varphi_{c}\to 0$.
\begin{figure}[!t]
	\centering
\includegraphics[width=0.495\textwidth]{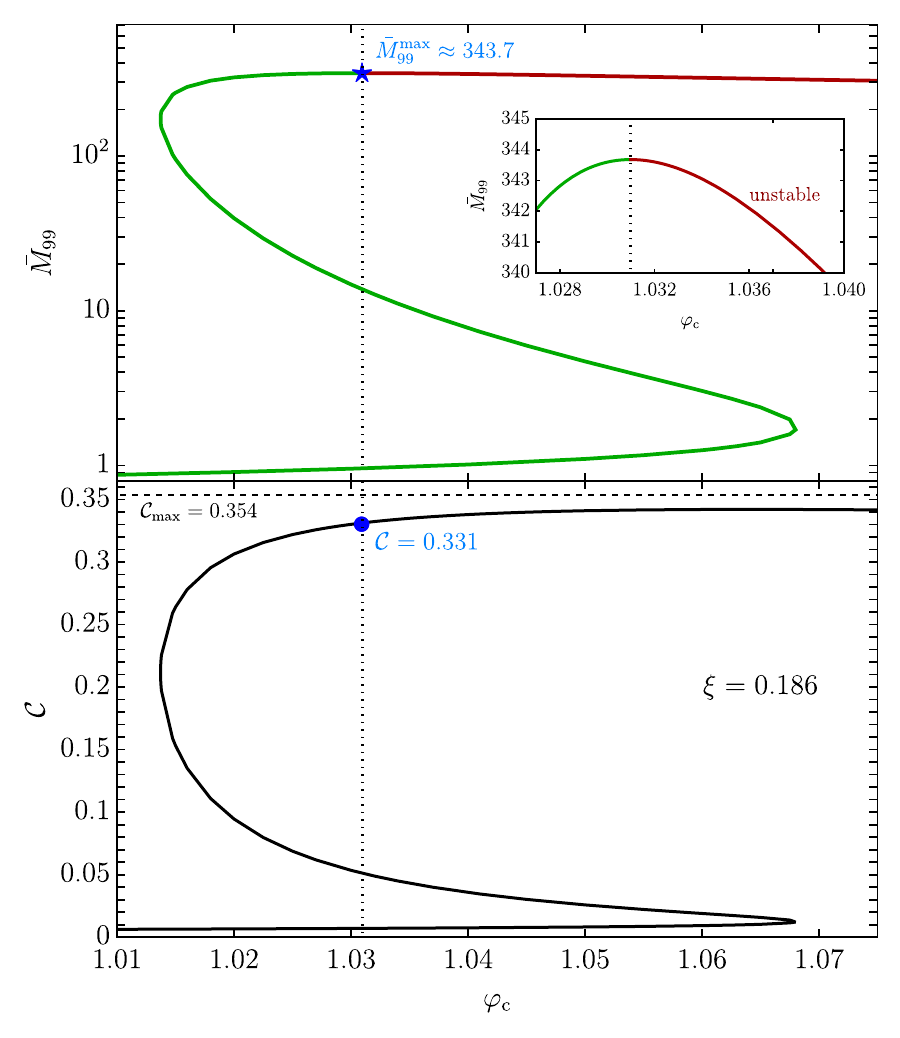}
	\caption{ \it 
 Top panel. Soliton boson star mass $\bar{M}_{99}$, cf. eq.\,(\ref{eq:Mass99}), as function of the central field value $\varphi_c$. 
 This plot is a zoomed-in version of fig.\,\ref{fig:BSStability}, focusing on central field values of order 1.
 Bottom panel. Soliton boson star compacteness, cf. eq.\,(\ref{eq:C99}), as function of the central field value $\varphi_c$. 
 The horizontal dashed line corresponds to 
 the causal limit $\mathcal{C}_{\textrm{max}}= 0.354$.
 In both panels we fix $\xi = 0.186$. 
The vertical dotted line corresponds to the central field value that gives the maximal mass, $\bar{M}_{99}^{\textrm{max}} \approx 343.7$ (indicated by the blue star in the upper panel).  
In the compactness plot, the maximum mass configuration is characterized by $\mathcal{C} \approx 0.331$ (indicated by the blue dot in the lower panel).
 }
\label{fig:BosonStarBack}
\end{figure}
In the limit $\varphi_c \to 0$, 
we have a first stable branch corresponding to the Newtonian limit of boson stars\,\cite{Liebling:2012fv}.
Subsequently, we encounter the first turning point which leads to an unstable branch. 
In the bottom panel of fig.\,\ref{fig:BSStability}, we show the evolution of the radius $X_{99}$ as a function of the central value of the field. As can be seen, at the first turning point (identified for clarity by vertical dotted lines), we have $dX_{99}/d\phi < 0$, and consequently, the frequency of the fundamental mode changes sign, indicating the transition from stability to instability.

Thereafter, a second turning point with $dX_{99}/d\phi < 0$ guides the solutions to a second stable branch characterized by $\varphi_c = O(1)$. 
Finally, a third turning point with $dX_{99}/d\phi < 0$, barely visible in fig.\,\ref{fig:BSStability}, leads to a final unstable branch. 
In fig.\,\ref{fig:BosonStarBack}, we zoom in on the final part of the $\bar{M}_{99}$-$\varphi_c$ curve.
 The presence of a third turning point becomes clearly visible only in the inset plot, where we have further zoomed in on the final section of the $\bar{M}_{99}$-$\varphi_c$ curve. 
 Numerically, we find the maximum mass 
 $\bar{M}_{99}^{\textrm{max}}\approx 343.7$ at $\varphi_{c} \approx 1.031$ (blue star in the top panel of fig.\,\ref{fig:BosonStarBack}). 
 In the bottom panel of fig.\,\ref{fig:BosonStarBack} we show the compactness $\mathcal{C}$ in 
 eq.\,(\ref{eq:C99}) as function of $\varphi_c$. 
It is interesting to note that the maximum mass configuration does not correspond to a maximum in compactness, 
as the curve $\mathcal{C}(\varphi_c)$ continues to increase (albeit very slowly) even when entering the final unstable branch.
Numerically, the value of compactness corresponding to the maximum mass is
 $\mathcal{C}\approx 0.331$ (blue dot in the bottom panel of fig.\,\ref{fig:BosonStarBack}) while proceeding along the unstable branch, even higher compactness values can be reached, up to $0.342$ in the case under consideration.

We remark that the curves in fig.\,\ref{fig:BSMassRadius} and fig.\,\ref{fig:BSMassCompa} correspond to the second stable branch in fig.\,\ref{fig:BosonStarBack}.
\begin{figure}[!t]
	\centering
\includegraphics[width=0.495\textwidth]{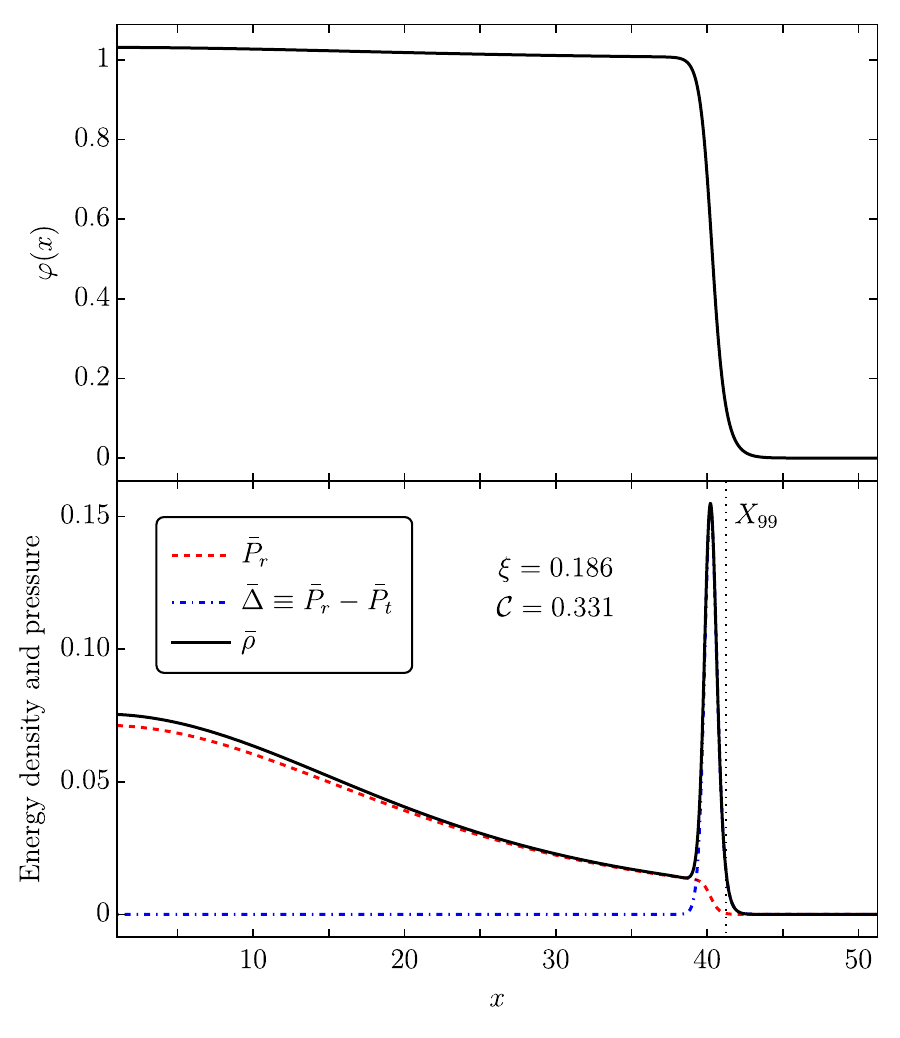}
	\caption{ \it 
 Top panel. Profile of the scalar field $\varphi(x)$ as a function of radial distance (normalized as in eq.\,(\ref{eq:ADimVariaBack})). Bottom panel. Radial profiles of energy density and pressure (in units of $\mu^2 f^2$). The latter is decomposed into the radial term and the anisotropic component. 
 In both panels, we fix $\xi = 0.186$ and display a configuration of a solitonic boson star approaching maximum compactness, $\mathcal{C}\approx 0.331$.
 }
\label{fig:BackgroundExample}
\end{figure}

In fig.\,\ref{fig:BackgroundExample}, we present an explicit solution corresponding to the maximum compactness configuration (specifically, while keeping $\xi = 0.186$, we consider $\mathcal{C} \approx 0.331$). 
The vertical black dotted line corresponds to the radius $X_{99}$ of the star, calculated as described in eq.\,(\ref{eq:SBSRadius99}). 
From this plot, we observe that the scalar field $\varphi$ (which, as a reminder, is normalized in units of $f/\sqrt{2}$, see eq.\,(\ref{eq:ADimVariaBack})) is approximately $\varphi \approx 1$ throughout most of the star (i.e., it is in the second degenerate minimum, which has a non-zero field value), except in the transition region near its surface, where it transitions exponentially quickly to the null value characteristic of the vacuum of the potential at the origin.

It is also interesting to comment on the behavior of the pressure and energy density. In the bulk of the star, we observe that the relationship $\bar{\rho} \approx \bar{P}_r$ holds (as already noted in ref.\,\cite{Boskovic:2021nfs}).
This relationship explains why, in situations of extreme compactness, solitonic boson stars can be considered a concrete example of stars supported by a linear EoS. 
Note, however, that the relationship $\bar{\rho} \approx \bar{P}_r$ fails in the transition region near the surface of the star. In this region, the energy density exhibits a pronounced spike corresponding to the transition of the field, dominated by the derivative term $\varphi^{\prime}$.  
In this situation, the anisotropic pressure term $\bar{\Delta}$ becomes dominant, see eq.\,(\ref{eq:PreAni}). 
However, the field transition between the two minima occurs in an extremely small boundary region (of order $1$ in units of $\mu^{-1}$) when compared to the dimensions of the bulk of the star. For this reason, solitonic boson stars continue to satisfy the compactness bound derived from the linear EoS (although the presence of an anisotropic pressure term may appear to preclude a direct comparison).
\begin{figure}[!t]
	\centering
\includegraphics[width=0.495\textwidth]{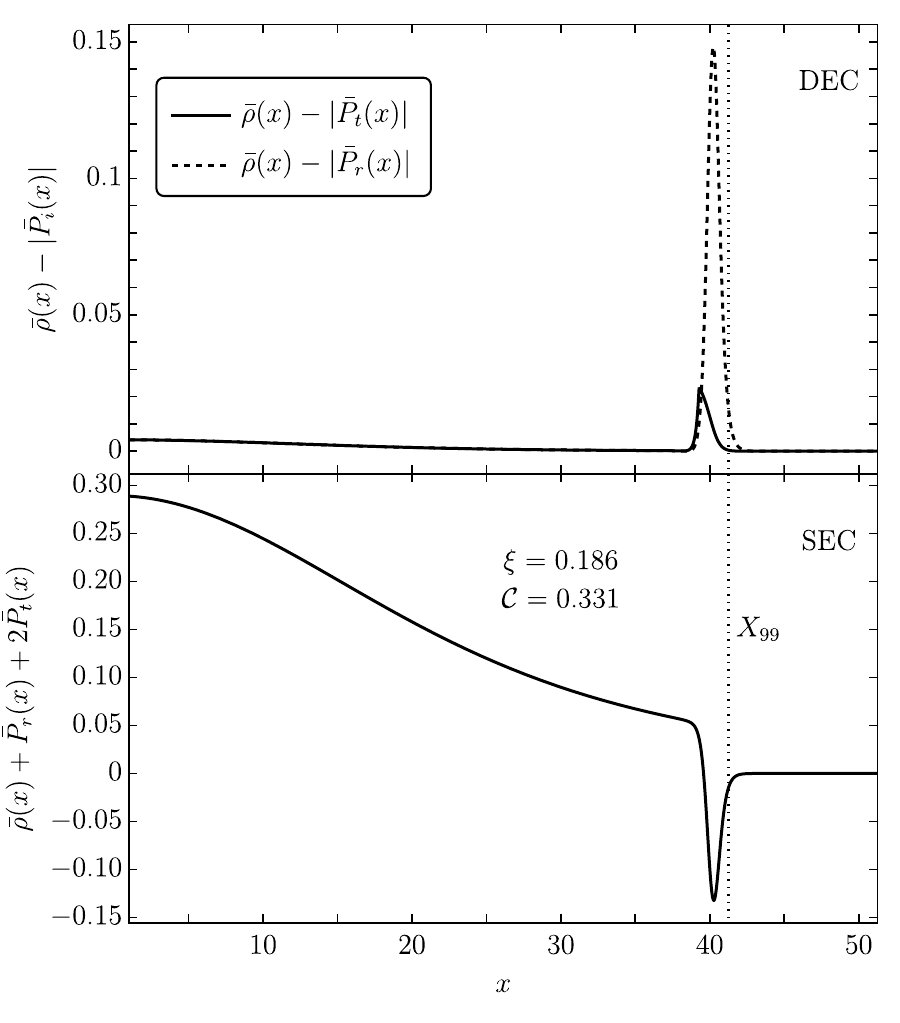}
	\caption{ \it 
 Energy conditions in the case of solitonic boson stars. 
 Top panel. We show, along the entire radial profile, the
 behavior of the two quantities $\bar{\rho} - |\bar{P}_r|$ and $\bar{\rho} - |\bar{P}_t|$, which appear in the definition of the DEC, see eq.\,(\ref{eq:DEC}).
 Bottom panel. We show, along the entire radial profile, 
 the
 behavior of the  quantity $\bar{\rho} + \bar{P}_r + 2\bar{P}_t$, which appears in 
 the last inequality that defines the SEC, see eq.\,(\ref{eq:SEC}).
 In both panels, we fix $\xi = 0.186$ and display a configuration of a solitonic boson star approaching maximum compactness, $\mathcal{C}\approx 0.331$.
 }
\label{fig:EnergyConditions}
\end{figure}
It is also instructive to consider what happens to the energy conditions, which we introduced and discussed in 
section\,\ref{section:LinearEoS}.
We focus again on the solution corresponding to the maximum compactness configuration ($\xi = 0.186$, $\mathcal{C} \approx 0.331$). 
In the top panel of fig.\,\ref{fig:EnergyConditions} we show, as function of the radial distance 
$x$ (see eq.\,(\ref{eq:ADimVariaBack})),
the behavior of the quantities 
$\bar{\rho} - |\bar{P}_r|$ and $\bar{\rho} - |\bar{P}_t|$. 
As discussed in eq.\,(\ref{eq:DEC}), 
when $\bar{\rho} - |\bar{P}_r| > 0$ and 
$\bar{\rho} - |\bar{P}_t| > 0$, the DEC is satisfied. The figure illustrates how this indeed occurs throughout the entire radial profile of the star.
In the bottom panel of fig.\,\ref{fig:EnergyConditions} we show, as function of the radial distance, 
the behavior of the quantity $\bar{\rho} + \bar{P}_r + 2\bar{P}_t$.
As discussed in eq.\,(\ref{eq:SEC}), 
when $\bar{\rho} + \bar{P}_r + 2\bar{P}_t < 0$, the SEC is violated.
The figure shows how $\bar{\rho} + \bar{P}_r + 2\bar{P}_t$ indeed becomes negative at the star's surface when the energy density peaks and then rapidly decreases to zero.
We thus find that solitonic boson stars violate the SEC in a radial region of order $\mu^{-1}$ near the surface. This result is consistent with the discussion in ref.\,\cite{Collodel:2022jly}.

\subsection{Tidal deformability of solitonic boson stars and comparison with the causality bound}

We now consider small perturbations to the metric and
scalar field. 
We write
\begin{align}
g_{\mu\nu} = g_{\mu\nu}^{(0)} + h_{\mu\nu}\,,~~~\Phi = \Phi_0 + \delta\Phi\,.
\end{align}
We focus on static, even-parity, and quadrupolar ($l = 2$) metric perturbations. The latter are described below eq.\,(\ref{eq:MetricPerturbations}).
The scalar field perturbation is described by 
\begin{align}
\delta\Phi(t,r,\theta,\varphi) =  
\phi_1(r)\frac{e^{-i\omega t}}{\mu r}
Y_{20}(\theta,\varphi)\,.
\end{align}
Note that we are using the same time dependence in the perturbation as in the description of the unperturbed field. 
We write the latter in the form 
$\Phi_0(t,r,\theta,\varphi) = \phi_0(r)e^{-i\omega t}$, where we now use the subscript $_0$ to indicate the unperturbed background scalar field. 

We consider the Einstein field equations 
 $R_{\mu\nu} - \frac{1}{2}g_{\mu\nu}R = T_{\mu\nu}/\MPl^2$ written in terms of the perturbed quantities (keeping only terms linear in the perturbations). 
 The key aspects of this analysis are the following. 
 First, the $r$-$r$ and $r$-$\theta$ components of the Einstein field equations can be used to algebraically eliminate, respectively, $K^{\prime}(r)$ and $K(r)$.      
Second, the $\theta$-$\theta$ and $\varphi$-$\varphi$ components of the Einstein field equations can be solved by setting
\begin{align}
   H_2(r) = H_0(r)\,. 
\end{align}
This is immediately evident by substituting into the perturbed 
equations both the expressions for 
$K(r)$, $K^{\prime}(r)$ and $K^{\prime\prime}(r)$ previously obtained, as well as the 
equations describing the background (both metric and scalar field, cf. eqs.\,(\ref{eq:Lambda}-\ref{eq:KG})) obtained in the preceding section.

Finally, the $t$-$t$ component of the Einstein field equations gives a second-order differential equation for $H_0(r)$. 
We find (by eliminating, as previously discussed, $K(r)$, $K^{\prime}(r)$ and $K^{\prime\prime}(r)$ and after using the background equations)
\begin{widetext}
\begin{align}
H_0^{\prime\prime} 
&+ \frac{H_0^{\prime}e^{\lambda}}{r}
\left[
1 + e^{-\lambda} - \frac{r^2V(\phi_0)}{\MPl^2}
\right] 
+
\frac{4e^{\lambda}\phi_1}{\mu r^2\MPl^2}
\left[
\phi_0^{\prime}
\left(
1-e^{-\lambda}
+\frac{r^2 P_r}{\MPl^2}
\right) 
+ r\phi_0\left(
2e^{-\nu}\omega^2  - 
U_0
\right)
\right] \nn\\
&+\frac{H_0e^{\lambda}}{r^2}
\left[
-\frac{2r^2V(\phi_0)}{\MPl^2} 
- 6 -e^{\lambda}\left(
1-e^{-\lambda}+\frac{r^2P_r}{\MPl^2}
\right)^2 + \frac{8r^2\phi_0^2}{\MPl^2}
e^{-\nu}\omega^2
\right] = 0\,,
\end{align}
\end{widetext}
where $U_0 \equiv U(\phi_0)$ with 
$U(\phi) \equiv dV/d|\Phi|^2$. 
The radial pressure is defined in terms of background quantities in eq.\,(\ref{eq:RadialPressure}). 

As a final step, it remains to write the equation describing the dynamics of the perturbed field $\phi_1$. 
We consider the Klein-Gordon equation
\begin{align}
\frac{1}{\sqrt{-g}}
\partial_{\mu}
\left[
\sqrt{-g}g^{\mu\nu}(\partial_{\nu}\Phi)
\right] - \frac{dV}{d|\Phi|^2}\Phi = 0\,,
\end{align}
and expand it to linear order in the perturbations, both in the metric and in the scalar field. We find
\begin{widetext}
\begin{align}
\phi_1^{\prime\prime} &+ 
\frac{\phi_1^{\prime}e^{\lambda}}{r}
\left[
1-e^{-\lambda} - \frac{r^2V(\phi_0)}{\MPl^2}
\right] 
- \mu H_0e^{\lambda}
\left[
\phi_0^{\prime}
\left(
-1+e^{-\lambda}-\frac{r^2P_r}{\MPl^2}
\right) 
+
r\phi_0\left(U_0 - 2e^{-\nu}\omega^2\right)
\right]\nn\\
&+
\frac{e^{\lambda}\phi_1}{r^2}
\left[
e^{-\lambda} - 7 
+\frac{r^2 V(\phi_0)}{\MPl^2} 
 +r^2e^{-\nu}\omega^2 
 - \frac{4e^{-\lambda}r^2(\phi_0^{\prime})^2}{\MPl^2}
 -r^2(U_0 + 2W_0\phi_0^2)
\right] = 0\,,
\end{align}
\end{widetext}
where $W_0 \equiv W(\phi_0)$ with 
$W(\phi) \equiv dU/d|\Phi|^2$.
Our results agree with those in ref.\,\cite{Sennett:2017etc}.
We can also, in this case, switch to dimensionless variables.
The equivalent of eq.\,(\ref{eq:ADimVariaBack}) now reads
\begin{align}
x\equiv \mu r\,,~~
w \equiv \frac{\omega}{\mu}\,,~~
\varphi_0 \equiv \frac{\sqrt{2}\phi_0}{f}\,,
~~
\varphi_1 \equiv \frac{\sqrt{2}\phi_1}{f}\,,
\end{align}
and, as far as the potential and its derivatives are concerned, we write 
\begin{align}
V = \frac{\mu^2f^2\mathcal{V}}{2}\,,~~
U = \frac{dV}{d|\Phi|^2} 
 \equiv \mu^2\mathcal{U}\,,~~
W =\frac{dU}{d|\Phi|^2} 
\equiv
\frac{2\mu^2}{f^2}\mathcal{W}\,,
\end{align}
with
\begin{align}
\mathcal{V}_0 & = 
\varphi_0^2(\varphi_0^2-1)^2
\,,\\
\mathcal{U}_0 & =  
(-1+\varphi_0^2)(-1+3\varphi_0^2)\,,\\
\mathcal{W}_0 & = -4+6\varphi_0^2\,.
\end{align}
All in all, we arrive at the differential equations 
\begin{widetext}
\begin{align}
H_0^{\prime\prime} 
&+ \frac{H_0^{\prime}e^{\lambda}}{x}
\left[
1 + e^{-\lambda} - \frac{\xi^2 x^2 \mathcal{V}_0}{2}
\right] 
+
\frac{2e^{\lambda}\xi^2\varphi_1}{x^2}
\left[
\varphi_0^{\prime}
\left(
1-e^{-\lambda}
+ x^2\xi^2 \bar{P}_r
\right) 
+ x\varphi_0\left(
2e^{-\tilde{\nu}}\tilde{w}^2  - 
\mathcal{U}_0
\right)
\right] \nn\\
&+\frac{H_0e^{\lambda}}{x^2}
\left[
-x^2\xi^2 \mathcal{V}_0 
- 6 -e^{\lambda}\left(
1-e^{-\lambda}+ x^2\xi^2\bar{P}_r
\right)^2 + 4x^2\xi^2\varphi_0^2
e^{-\tilde{\nu}}\tilde{w}^2
\right] = 0\,,\label{eq:H0Fin}\\
\varphi_1^{\prime\prime} &+ 
\frac{\varphi_1^{\prime}e^{\lambda}}{x}
\left(
1-e^{-\lambda} - \frac{x^2 \xi^2 \mathcal{V}_0}{2}
\right) 
- H_0e^{\lambda}
\left[
\varphi_0^{\prime}
\left(
-1+e^{-\lambda}- x^2\xi^2\bar{P}_r
\right) 
+
x\varphi_0\left(\mathcal{U}_0 - 2e^{-\tilde{\nu}}\tilde{w}^2\right)
\right]\nn\\
&+
\frac{e^{\lambda}\varphi_1}{x^2}
\left[
e^{-\lambda} - 7 
-\frac{3}{2}x^2\xi^2 \mathcal{V}_0 
 +x^2e^{-\tilde{\nu}}\tilde{w}^2(1+2\xi^2\varphi_0^2) 
 -4x^2\xi^2\bar{P}_r
 -x^2(\mathcal{U}_0 + 2\mathcal{W}_0\varphi_0^2)
\right] = 0\,,\label{eq:TidalPert2}
\end{align}
\end{widetext}
where now $^{\prime}\equiv d/dx$. Note also how we have transitioned to the variables $\tilde{\nu}$ and $\tilde{w}$, as described below eq.\,(\ref{eq:metrt}). 
Expanding around the origin reveals that 
\begin{align}
H_0(x) = \frac{1}{2}x^2H_0^{\prime\prime}(0) + O(x^3)\,,\\
\varphi_1(x) = 
\frac{1}{6}x^3
\varphi_1^{\prime\prime\prime}(0)
+ 
O(x^4)\,.
\end{align}
Furthermore, we note that eq.\,(\ref{eq:H0Fin}) is invariant under a simultaneous rescaling of $H_0$ and $\varphi_1$. We can use this property to normalize the initial condition on $H_0$. We thus write
\begin{align}
H_0(x) \approx x^2\,,~~~
\varphi_1(x) \approx x^3\varphi_{1,3}\,.
\end{align}
Furthermore, to stabilize the behavior at small values of $x$, it is possible to rescale the variables according to
\begin{align}
\tilde{H}_0(x) \equiv \frac{1}{x^2}H_0(x)\,,~~~~
\tilde{\varphi}_1(x) \equiv 
\frac{1}{x^3}\varphi_1(x)\,.
\end{align}
Consequently, the boundary conditions for these rescaled variables become
\begin{align}
 \tilde{H}_0(0) = 1\,,~~~
 \tilde{\varphi}_1(0) = \varphi_{1,3}\,,~~~
 \tilde{\varphi}_1^{\prime}(0) = 0\,.
\end{align}
The value of $\varphi_{1,3}$ can be obtained through a numerical shooting method by imposing the asymptotic condition $\tilde{\varphi}_{1}(\infty) = 0$.

The resolution strategy, therefore, is divided into two steps. First, we solve the background dynamics as described in 
section\,\ref{sec:BackBosonStar}; subsequently, we solve the equations for the perturbations, obtaining the radial profiles of $H_0(x)$ and $\varphi_1(x)$.
Once the solution for $H_0(x)$ is known, we compute the logarithmic derivative 
\begin{align}
y(x) \equiv  \frac{xH_0^{\prime}(x)}
{H_0(x)}\,.\label{eq:ySol}
\end{align}
We then compute the tidal deformability $\Lambda = 
2k_2/3\mathcal{C}^5$ with $k_2$ given by eq.\,(\ref{eq:k2formula}) 
and $Y\equiv y(x_{\textrm{ext}})$. 
The quantity $x_{\textrm{ext}}$ defines the radial distance at which the tidal deformability is calculated. 
\begin{figure}[!t]
	\centering
\includegraphics[width=0.495\textwidth]{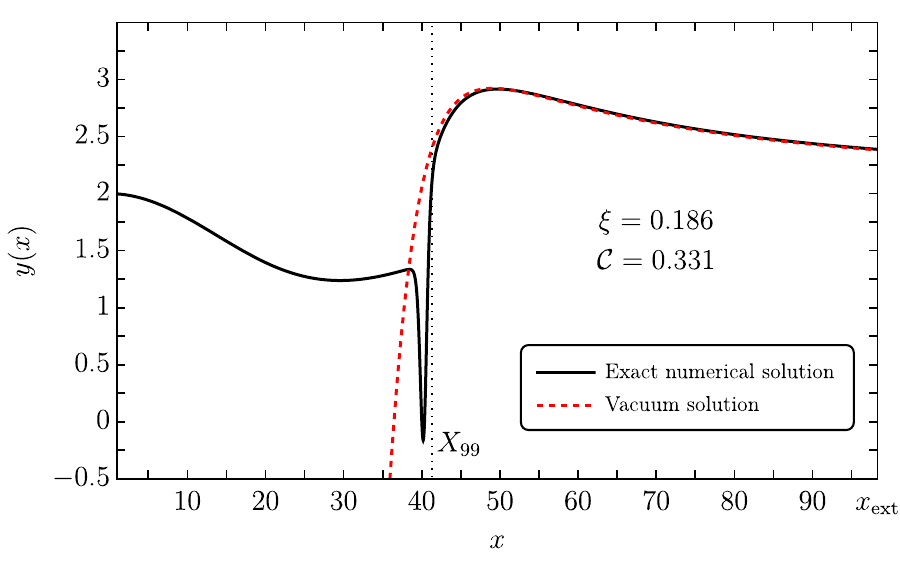}
	\caption{ \it 
 Radial profile of the function 
 $y(x)$ defined in eq.\,(\ref{eq:ySol}). 
 The solid black line corresponds to the 
 exact numerical solution of the perturbed equations, cf. eqs.\,(\ref{eq:H0Fin},\,\ref{eq:TidalPert2}). 
 The dashed red line corresponds to the 
 vacuum solution discussed in 
 eq.\,(\ref{eq:Analy1bis}).
 }
\label{fig:TidalExample}
\end{figure}
We illustrate our numerical procedure with the help of a specific example. 
We fix $\xi = 0.186$ and consider a background solution with maximal compactness $\mathcal{C} = 0.331$ and radius 
$X_{99} = 41.3$. 
In fig.\,\ref{fig:TidalExample}, we show in solid black the profile of the function $y(x)$ in eq.\,(\ref{eq:ySol}) obtained numerically by solving the equations for the perturbations as previously discussed. 
Note that we extend our solution well beyond the radius of the boson star. 
Far from the value $X_{99}$, we evaluate $Y\equiv y(x_{\textrm{ext}})$. 
Specifically, we show the typical value of $x_{\textrm{ext}}$ that we use in our analysis as the rightmost end of the $x$-axis in fig.\,\ref{fig:TidalExample}. 
We then compute the tidal deformability $\Lambda$. 
To validate our result, one can use the fact that, far from the center of the boson star, the scalar field decays exponentially and the system approaches the vacuum solution. 
Neglecting the vanishingly small contributions from the scalar field, $y(x)$ takes the form already discussed in eqs.\,(\ref{eq:Analy1}-\ref{eq:Analy4}), which we present here again to facilitate readability, as we are now employing slightly different notation
\begin{align}
y_{\textrm{vac}}(z) & = \frac{2}{z-1}\bigg[
z+ \frac{120\Lambda}{
(z^2-1)^2 f(z,\Lambda)
+30z\Lambda g(z)
}\bigg]\,,\label{eq:Analy1bis}\\
f(z,\Lambda) & \equiv -16-45\Lambda\log\left(\frac{z+1}{z-1}\right)\,,\label{eq:Analy2}\\
g(z) & \equiv 3(z+1)^2 - 6(z+1) -2\,,\label{eq:Analy3}\\
z & \equiv \frac{8\pi x}{\bar{M}}-1\,,\label{eq:Analy4bis}
\end{align}
with $x$ as in eq.\,(\ref{eq:ADimVariaBack}).
The vacuum solution in eq.\,(\ref{eq:Analy1bis}), if we plug in the value of $\Lambda$ previously computed, must coincide, in the vacuum region far from the boson star, with the full numerical solution. 
In fig.\,\ref{fig:TidalExample}, 
we show the vacuum solution in dashed red. 
As shown in the plot, we achieve a perfect agreement with the numerical solution in the region outside the surface of the boson star, particularly at the extraction value corresponding to $x_{\textrm{ext}}$.

It is also instructive to make a comparison with the analogous analysis presented in fig.\,\ref{fig:ySolLinEoS} for the linear EoS. 
We specifically note how the discontinuity present at the star's surface in the case described by the linear EoS is replaced by a continuous transition, which follows the same behavior across the surface but in a smooth manner.
\begin{figure}[!t]
	\centering
\includegraphics[width=0.495\textwidth]{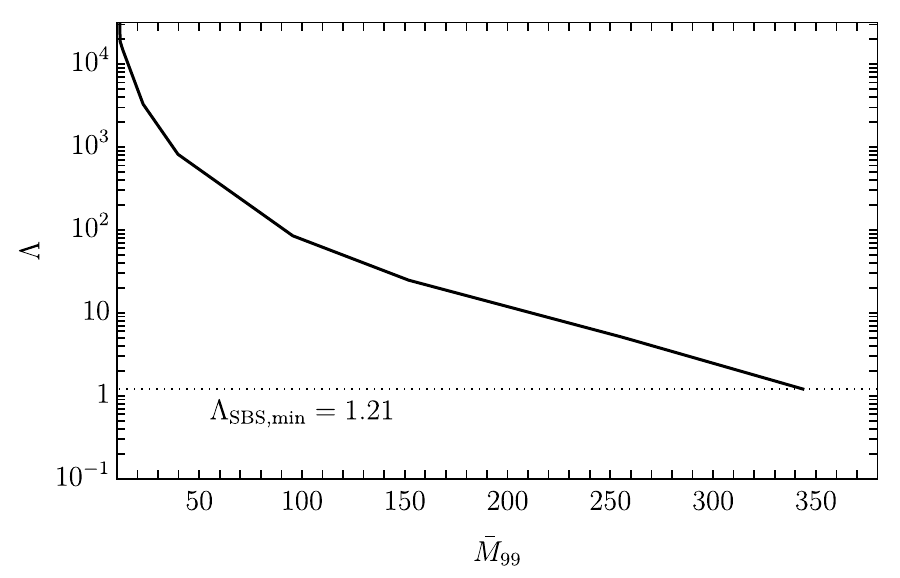}
	\caption{ \it 
 Tidal deformability as function of the mass parameter $\bar{M}_{99}$ in the case of solitonic boson stars with 
 $\xi = 0.186$.
 }
\label{fig:TidalSBS}
\end{figure}
In fig.\,\ref{fig:TidalSBS}, we show the behavior of the tidal deformability as a function of mass, considering, as usual, the value $\xi = 0.186$.
This is the curve we presented, after appropriately rescaling the mass, in
 fig.\,\ref{fig:TidalWithGW230529} and in fig.\,\ref{fig:FisherTest}, while in fig.\,\ref{fig:KeyPlotCLambda} the same tidal deformability is shown as a function of compactness.
 
At the maximum value of stable mass, see fig.\,\ref{fig:BosonStarBack}, we obtain the corresponding value of tidal deformability
\begin{align}
    \Lambda_{\textrm{SBS,min}} = 1.21\,,\label{eq:LimitLambdaSBS}
\end{align}
which is slightly smaller than the lower limit of causality obtained from the study of the linear EoS.

Our interpretation of this result is as follows. As discussed in section\,\ref{sec:Tidal}, the value of tidal deformability is sensitive to how the energy density of the star behaves as it crosses the surface. In the case of a linear EoS, the sharp Heaviside theta transition forced us to account for this discontinuity in an artificial manner (see eq.\,(\ref{eq:CorrectedY}) and the preceding discussion). In contrast, solitonic boson stars can continuously describe the behavior of the function $y(x)$ across the surface. The price to pay for this is a violation of the SEC.
However, solitonic boson stars continue to satisfy the DEC and thus maintain relativistic causality (to which the DEC is related; see the discussion in section\,\ref{section:LinearEoS}). 
Consequently, we can state that the value in eq.\,(\ref{eq:LimitLambdaSBS}) represents, in a certain sense, an improvement upon the bound in eq.\,(\ref{eq:CausalTidal}), 
obtained without violating our fundamental requirement of relativistic causality, while modeling the physics of the star's surface more accurately, at the expense of violating the SEC.

\section{Conclusions}\label{sec:Conclusion}

In this work, we have studied the properties of ECOs in relation to their tidal deformability, both from a theoretical and a phenomenological perspective. Below, we aim to outline and summarize the key points of our analysis and the results we have obtained, particularly emphasizing the original aspects.
\begin{itemize}
    \item[$\circ$] In section\,\ref{sec:Stiffness}, we considered the constraint of relativistic causality for both neutron stars and ECOs supported by a linear EoS, motivated by the fact that, in the latter, the speed of sound saturates the speed of light. 
    Specifically, we analyzed how the speed of sound regulates the stiffness or softness of the EoS and discussed how this property is reflected in the behavior of tidal deformability, see figs.\,\ref{fig:CompaLambdaNS},\,\ref{fig:NeutronStarMassRadius}.
    \item[$\circ$]
    In section\,\ref{sec:Tidal}, we used the linear EoS to derive a causality-based lower limit on the tidal 
    deformability $\Lambda$, see eq.\,(\ref{eq:CausalTidal}).
    In fig.\,\ref{fig:KeyPlotCLambda}, we identified the `compactness vs. tidal deformability' plane $(\mathcal{C},\Lambda)$ as a particularly useful parameter space for theoretically organizing the properties of ECOs in comparison to black holes and neutron stars.
    \item[$\circ$] 
    In section\,\ref{sec:Pheno}, 
    we investigated the possibility that a population of ECOs described by a linear EoS could occupy a mass range extending from the mass gap to the sub-solar region, see fig.\,\ref{fig:MassComparison}. Since a gravitational merger event in this range is believed to be a smoking-gun signature of primordial black holes, these ECOs could be considered as `primordial black hole mimickers,' distinguished, however, by their non-zero tidal deformability. We then sought to determine, using a Fisher matrix analysis, whether it is feasible to measure the tidal deformability in this mass range with the LIGO/Virgo/KAGRA interferometer network as well as with next-generation detectors such as the Einstein Telescope,
    see figs.\,\ref{fig:TidalPlotFisher},\,\ref{fig:FisherTest}.
    \item[$\circ$] 
    In section\,\ref{sec:BS}, we considered solitonic boson stars as an explicit model of ECOs whose equation of state approximates the properties of the linear EoS. 
    The calculation of the tidal deformability of solitonic boson stars showed that it is possible to further lower the causality bound obtained with the linear EoS, see eq.\,(\ref{eq:LimitLambdaSBS}), at the cost of violating the SEC. However, the property of relativistic causality is preserved, as solitonic boson stars satisfy the DEC.
\end{itemize}
These findings not only deepen our understanding of the interplay between tidal deformability and the properties of ECOs but also highlight the potential of these objects to serve as viable alternatives to primordial black holes in the ongoing quest to elucidate the mysteries of compact astrophysical objects.

\section*{Acknowledgments}

We thank Loris Del Grosso for useful   discussions. 
This work is partially supported by
ICSC - Centro Nazionale di Ricerca in High Performance Computing, Big Data and Quantum Computing, funded by European Union-NextGenerationEU and by the research grant number 20227S3M3B ``Bubble Dynamics in Cosmological
Phase Transitions'' under the program PRIN 2022 of the Italian Ministero dell’Università e Ricerca
(MUR).

\bibliography{draft}

\end{document}